\documentclass[review]{elsarticle}

\usepackage{framed}
\usepackage{multicol}
\usepackage{subcaption}
\usepackage{graphicx}
\usepackage{amsmath}
\usepackage{color,soul}
\usepackage{soul}
\usepackage{xcolor}
\usepackage{rotating}
\usepackage{comment}
\usepackage{lscape}
\usepackage{multirow}
\usepackage{array}

\usepackage{amssymb}

\usepackage{lineno}
\usepackage{amsmath}
\usepackage{geometry}
\makeatletter
\def\ps@pprintTitle{%
	\let\@oddhead\@empty
	\let\@evenhead\@empty
	\def\@oddfoot{}%
	\let\@evenfoot\@oddfoot}
\makeatother

\journal{Journal Name}

\begin{document}

\begin{frontmatter}


\title{Stability analysis of a single-phase rectangular Coupled Natural Circulation Loop system employing a Fourier series based 1-D model}



\author{Akhil Dass, Sateesh Gedupudi \footnote{Corresponding author}}

\address{Heat Transfer and Thermal Power Laboratory, Department of Mechanical Engineering, IIT Madras, Chennai, India }

\begin{abstract}
A linear stability analysis of a single-phase Coupled Natural Circulation Loop (CNCL) is carried out using a Fourier series based 1-D model. A 3-D CFD study is undertaken to assess the ability of the 1-D model to capture the non-periodic oscillatory behaviour exhibited by the CNCL system. After the model verification, the stability maps of the system are obtained from the eigenvalues of the steady-state. The CNCL has multiple steady states and the stability maps of consistently observed steady states are presented. A thorough parametric study is conducted to observe the influence of non-dimensional numbers on the CNCL system. Increase in the Fourier number and flow resistance coefficient lead to an increase in the domain of stability. A comparison of the linear stability map with the empirical stability map indicates that the non-linear terms do not significantly affect the stability boundary.
\end{abstract}

\begin{keyword}
Linear stability \sep Multiple steady states \sep Fourier series \sep Coupled Natural Circulation Loop \sep Grashof number \sep Fourier number 


\end{keyword}

\end{frontmatter}


\section{Introduction}
Study of buoyancy-driven heat exchange systems has been an area of active interest due to its applications in domains where safety is of paramount importance such as nuclear reactors. Such heat exchange systems are used primarily because of their inherent passivity and lack of moving components, which make them ideal devices for use in emergency situations. A Natural Circulation Loop (NCL) is such a buoyancy-driven heat exchange system which has been extensively studied during the last five decades and is still an active research area. An NCL comprises of a fluid-filled conduit which exhibits internal flow when thermally stimulated because of the buoyancy forces generated. The NCL acts as a chaotic system under very high loads (\cite{welander1967oscillatory}, \cite{cammarata2003stability}). Employing a simple NCL system is adequate for preliminary investigation of the chaotic oscillation phenomena, but the systems used in practice which employ natural circulation to satisfy their heat load requirements are more complex in nature (Passive Residual Heat Removal System (PRHRS) and Liquid Metal Fast Breeder Reactor (LMFBR)). 
Systems such as PRHRS and LMFBR are multiloop systems which employ buoyancy forces to ensure fluid flow and avoid critical emergency situations when the power-grid and backup power systems fail. The multiloop systems are employed in practice to accomplish the task of heat transfer due to the following reasons \cite{todreas1992nuclear} :

\begin{itemize}
    \item isolate the fluid which is in direct contact with the radioactive nuclear core.
    \item divide the heat load.
    \item provide redundant backup capability in case of component failure.
\end{itemize}

Thus, even though valuable insights are obtained from the study of an NCL system, it is inadequate for understanding multiloop systems. Duffey and Hughes \cite{duffey2016safety} have identified a CNCL system as an ideal system which links an NCL to a complex system such as a PRHRS. The CNCL system is the most elementary multiloop system which comprises of two-component NCLs thermally linked via the common heat exchange section. A detailed study of the transient modelling and analysis of single-phase two-loop CNCL systems has been carried out by Dass and Gedupudi \cite{dass2019}. A CNCL system can also exhibit chaotic behaviour; thus, it is of significant interest to determine when such conditions arise for the efficient design and operation of buoyancy-driven heat exchange systems. The observations obtained from the study of CNCL systems may be of greater relevance to understanding PRHRS and LMFBR systems.

\subsection{Chaotic oscillations and stability analysis of NCL systems}

The observation that the NCL systems exhibit chaotic oscillations at high heat loads was first reported and thoroughly investigated by Welander \cite{welander1967oscillatory}. The transient and steady-state behaviour of an NCL system was analysed, and the neutral stability curves of the NCL system were identified. The NCL system considered for the study had infinitesimal length horizontal legs with a point heat source and heat sink for thermal excitation. The phase difference between the growth rate of the buoyancy and the viscous forces encountered by the fluid within the NCL was identified to be the cause for exhibiting oscillatory and chaotic behaviour. Keller \cite{keller1966periodic} studied the periodic oscillations and stability of rectangular NCL systems with point heat source and sink. The criteria for obtaining periodic oscillations in such systems were proposed, and the interplay between the buoyancy and viscous forces was identified, with the inertia of the system having a negligible contribution. Creveling at al. \cite{creveling1975stability} studied the stability of NCL systems of finite heating and cooling sections with a toroidal geometry. The study considered a heat flux at the bottom half and a constant temperature condition on the top half of the toroidal loop. The study compared the stability predictions of the 1-D model with experiment, and a good agreement between them was reported. An extensive review of transient and steady-state behaviour of NCL systems and their stability was reported by Zvirin in 1981 \cite{zvirin1982review}. An approximated 1-D transient study of the NCL system with a point heat source and sink was conducted by Zvirin and Grief \cite{zvirin1979transient}. The approximated model was unable to capture the unstable behaviour of the NCL system. A more refined approach for 1-D modelling of toroidal NCL systems was presented by Hart \cite{hart1984new}. The approach utilised a Fourier series based modelling to reduce the nonlinear Partial Differential Equation (PDE) system, which represents the NCL, to a system of nonlinear coupled ordinary differential equations (ODE). This model was capable of predicting the unstable transient dynamics of the NCL system reported thus far. The chaotic oscillations of the NCL system were observed to have a Lorenz type attractor. Gorman et al. \cite{gorman1986nonlinear} utilised the 1-D model of toroidal NCL developed by Hart, and a thorough investigation of different transient regimes was carried out. Yorke et al. \cite{yorke1987lorenz} studied the effect of truncating the Fourier nodes on the point of Hopf bifurcation of the toroidal NCL system modelled using the Fourier series. It was identified that $4$ Fourier nodes were required to identify the location of Hopf bifurcation.

The stability analysis of toroidal NCL or NCL systems with a point heat source or sink has been considered so far, but NCL systems with rectangular geometry and finite-length heating/cooling sections are of greater relevance. The stability analysis of an NCL system with heating and cooling sections on the horizontal legs was studied by Nayak et al. \cite{nayak1995mathematical}. The stability map of the NCL was obtained using both the linear and non-linear stability approach, and they reported a good match between the stability maps obtained from both the approaches. Ambrosini and Ferreri \cite{ambrosini1998effect} investigated the effect of numerical schemes on the prediction of stability maps using Finite Difference Methods (FDM). A second-order explicit (Lax-Wendroff) scheme was identified to provide a good estimation of the stability boundary. The effect of ratio of  total loop length to hydraulic diameter ($L_t/D_h$) on the stability boundary was investigated by Vijayan \cite{vijayan2002experimental} and an increase in $L_t/D_h$ was found to push the stability boundary upward in the $Gr$-$St$ domain. Fichera and Pagano \cite{fichera2003modelling} employed the Fourier series based method used by Hart \cite{hart1984new} to model rectangular NCL systems and both uni-directional and bi-directional oscillations in the NCL system were reported. The rectangular NCL system was observed to have a Lorenz like attractor for bi-directional oscillations. Cammarata et al. \cite{cammarata2003stability} employed the Fourier series method for linear stability analysis of the NCL. The stability map was developed in the $Gr$-$L/L1$ domain. The linear stability of the NCL system was based on the identification of eigenvalues of the truncated ODE system at steady state. If the eigenvalue is `+Ve' then the system is unstable and if it is `-Ve' the system is categorised as stable. The stability map of the NCL system is generally plotted in the $Gr$-$St$ domain, but it can also be plotted in the $Gr$-$Re$ domain. Wu and Sienicki \cite{wu2003stability} obtained the stability boundary of an NCL system in the $Gr$-$Re$ domain. The effect of heater-cooler position on the stability maps was extensively studied by Pilkhwal et al. \cite{pilkhwal2007analysis}, Vijayan et al. \cite{vijayan2007steady} and Ruiz et al. \cite{ruiz2015dynamic}. Pilkhwal et al. \cite{pilkhwal2007analysis} also conducted a 3-D CFD study of chaotic oscillations and observed thermal stratification across the cross-sectional diameter of the NCL system. The effect of these stratifications are not incorporated in the 1-D models and thus may contribute to the deviation of 1-D results from CFD. The Fourier series based stability analysis described by Cammarata et al. \cite{cammarata2003stability} was used by Lu et al. \cite{lu2014stability} to obtain the stability map of an NCL system with heat flux (heat source) and constant temperature (heat sink) condition. A detailed 3-D CFD study of chaotic oscillations was conducted by Kudariyawar et al. \cite{kudariyawar2016computational}. The physics of the uni-directional and bi-directional oscillations were presented, and a good agreement with the previous experimental work was reported. The low Reynolds number $k-\epsilon$ turbulence model was identified to capture the laminar to turbulent flow regime transition. Luzzi et al. \cite{luzzi2017assessment}, Pini et al. \cite{pini2016analytical}, Nadella et al. \cite{nadella2018semi} and Cammi et al. \cite{cammi2016influence} carried out a detailed and extensive assessment of the 1-D numerical and semi-analytical models against experimental literature and observed good agreement between the 1-D model predictions and the experimental literature data. The thermal inertia introduced due to the piping material was also observed to influence the stability of NCL systems. Cammi et al. \cite{cammi2017stability} utilised information entropy to obtain the stability map of NCL system. The stability map obtained using the information entropy is in good agreement with the stability map predicted using linear stability analysis. The stability map was plotted in the $Re$-$Pr$ domain. Pilehvar et al. \cite{pilehvar2020stability} conducted the linear stability analysis of integrated self-pressurized water reactor. The indirect Lyapunov approach and the Routh-Hurwitz criteria is employed to determined the stability boundary. They observed that the system remained stable in the considered power range as long as the system remained in the single phase flow at the reactor core. Elton et al. \cite{elton2020investigations} carried out an experimental study to determine the instability threshold of the NCL systems with larger diameters. They reported the influence of operating procedures (start-up from rest, power raising from stable steady state, etc.) on the instability threshold. Goyal et al. \cite{goyal2020non} conducted the linear stability analysis of the NCL system and identifies the regions of unstable limit cyles and chaos. The methods for analysing chaotic single phase NCL system and obtaining their stability map have been summarized by Mukhopadhyay at al. \cite{mukhopadhyay2019dynamics}. Lu and Rizwan-uddin \cite{lu2020stability} carried out an extensive stability analysis of natural circulation in lead cooled fast reactor accounting for both the thermal-hydaulics and neutronics using a 1-D model. Saha et al. \cite{saha2020flow} conducted the symbolic time series analysis to predict the flow reversal in NCL systems employing a 1-D NCL model accounting for the wall effects. The symbolic state histograms are utilised to identify how far the system is from the chaotic regime.

\subsection{Chaotic oscillations and stability analysis of CNCL systems}

A CNCL system is the simplest of the multiloop systems. Davis and Roppo \cite{davis1987coupled} studied a CNCL system with toroidal loops which have a point contact heat exchange section between them. A Fourier series based approach was used to model the system and identify the steady states. The toroidal CNCL system has multiple steady states corresponding to conduction, counter-flow and parallel flow conditions at the common heat exchange section. Ehrhard \cite{ehrhard1988dynamisches} utilised the model developed by Davis and Roppo \cite{davis1987coupled} and experimentally validated it employing the bifurcation map. The point contact coupling and toroidal geometry of the component NCL restrict the generalisation of the results. The study of CNCL systems with rectangular component loops and area contact was first studied by Salazar et al. \cite{salazar1988flow}. The steady-state analysis of the system was performed, and the existence of multiple steady states was demonstrated. Zhang et al. \cite{zhang2015analysis} performed a 2-D transient study of chaotic oscillations in CNCL system with offset coupling and then validated it with the 1-D model. The focus of the study was primarily to determine the influence of the rate of power addition to the system on the oscillatory behaviour of the CNCL system. The CNCL systems were reported to have lesser stability relative to NCL systems and decreasing the thermal resistance at the common heat exchange section can lead to better stability of the CNCL system. Dass and Gedupudi \cite{dass2019} studied CNCL systems of different orientations and counter and parallel flow configurations at the common heat exchange section using 3-D CFD study. All the considered systems had stable steady-state solutions. A Fourier series based 1-D semi-analytical model of the CNCL system was developed and was thoroughly validated with all the considered cases. A good agreement was observed between the 1-D model and 3-D CFD results indicating the suitability of the 1-D model for the stable transient dynamic predictions of CNCL systems.

The present study utilises the 1-D model developed and verified by Dass and Gedupudi \cite{dass2019} for non-chaotic transient CNCL behaviour and explores its capability to model chaotic oscillatory behaviour, identify the steady states of the CNCL system and obtain stability maps of such systems. Thus, the objectives of the present paper can be listed as follows:

\begin{enumerate}
	\item Demonstrate the ability of the Fourier series based 1-D CNCL model to adequately capture the chaotic oscillations via comparison with 3-D CFD simulations.
	\item Study of the multiple steady states of the CNCL system, and the development of stability maps of the CNCL systems.
	\item Stability assessment of the counterflow vs parallel flow configurations for the CNCL systems.
	\item Study the effect of non-dimensional parameters on the stability map.
	\item Determine the accuracy of the stability map via comparison with transient simulations.
\end{enumerate}

\section{3-D CFD study of Chaotic oscillations in CNCL system}

The 3-D CFD study of chaotic oscillations in CNCL system is undertaken in the present study and ANSYS FLUENT 16.1 software is employed for the CFD investigation of the CNCL system. The 3-D transient study of CNCL systems with non-chaotic dynamics was extensively studied by Dass and Gedupudi \cite{dass2019}, and the present work focuses on the study of CNCL system with chaotic dynamics.

\subsection{Geometry, dimensions and meshing}

\begin{figure}[!htb]
	\centering
	\includegraphics[width=0.9\linewidth]{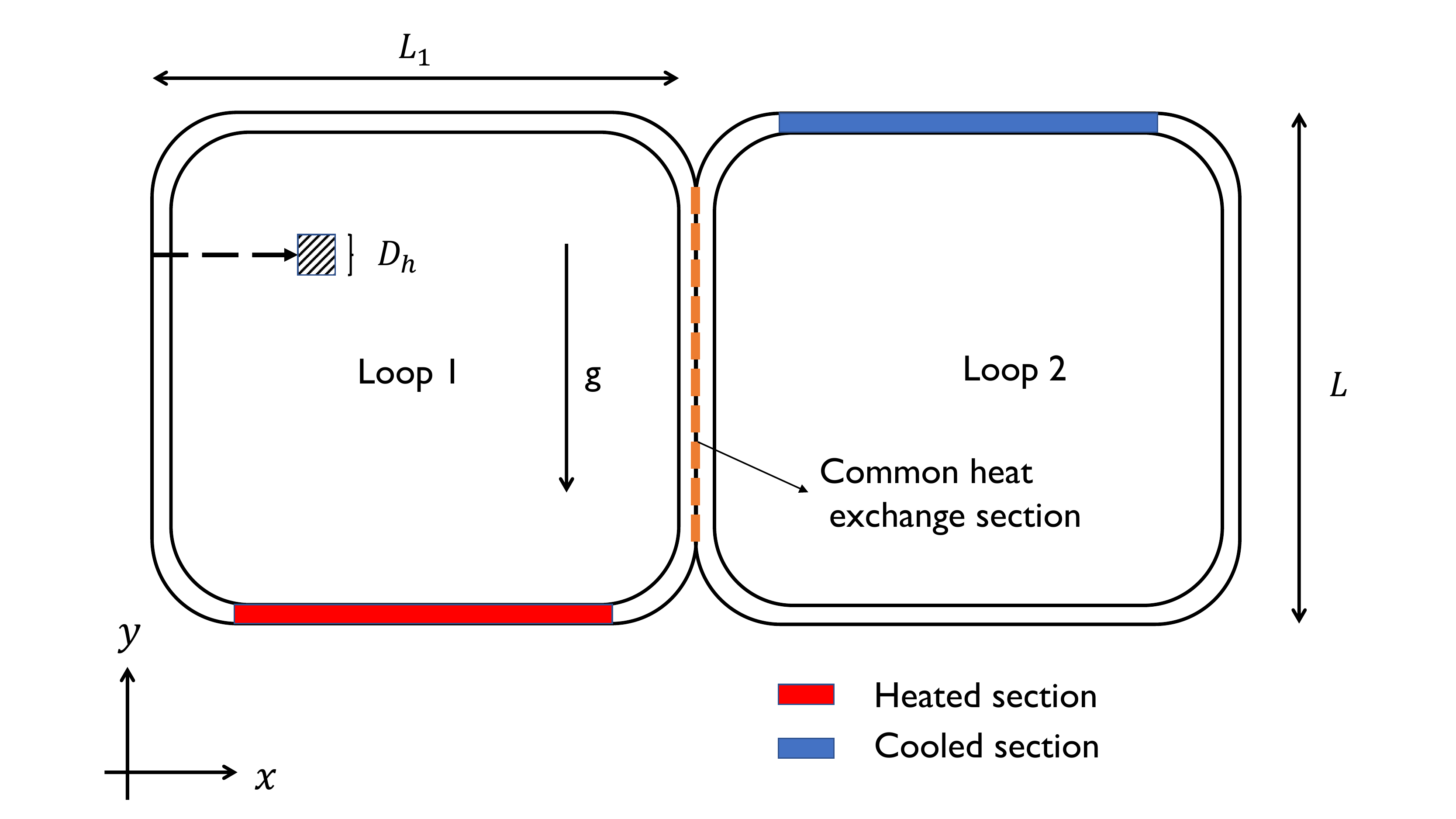}
	\caption{Schematic of the geometry used for 3-D CFD study.}
	\label{Fig1}
\end{figure}

\begin{figure}[!htb]
    \centering
    \includegraphics[width=\linewidth]{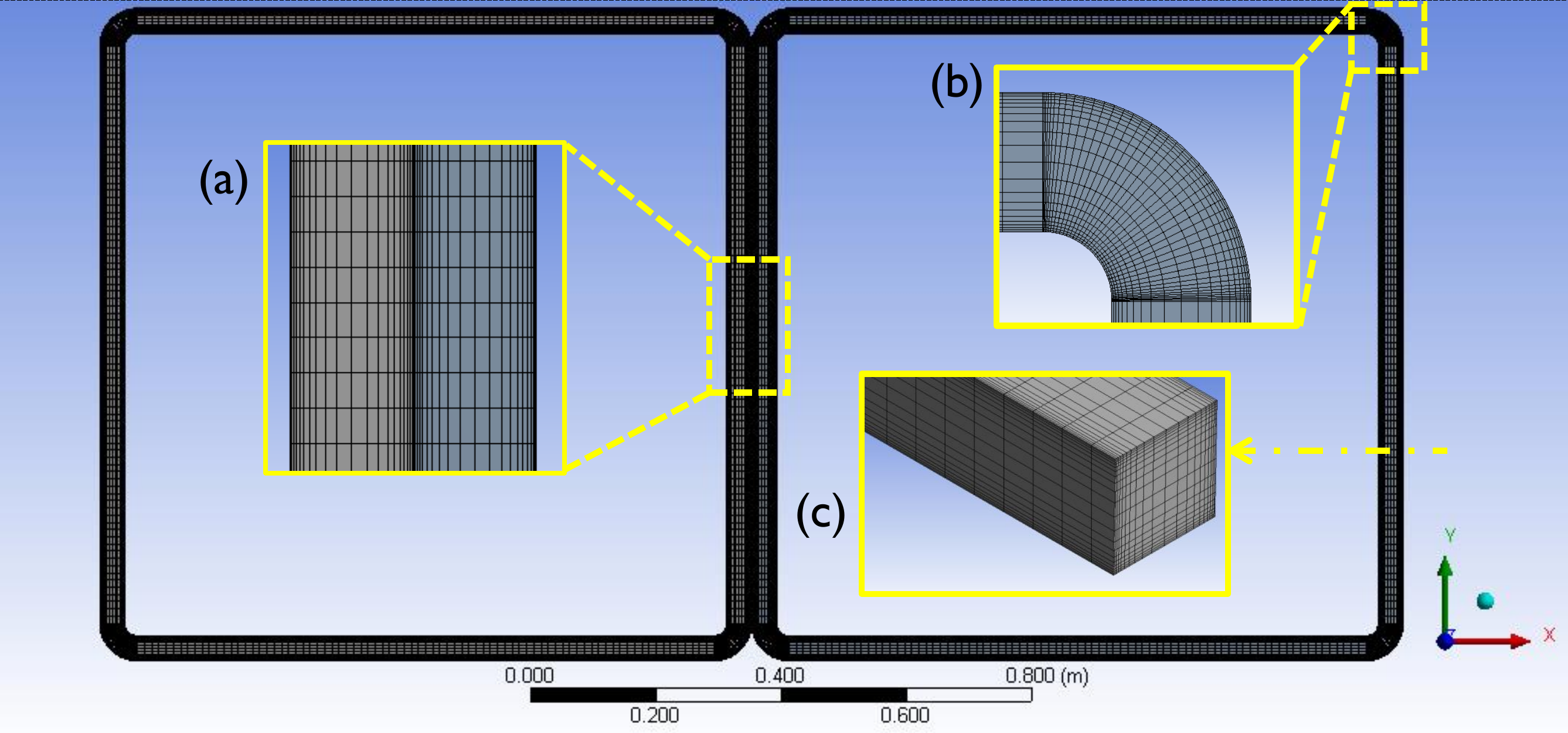}
    \caption{Schematic of the coarse mesh used for the 3-D CFD study. (a) Mesh at the common heat exchange section, (b) Mesh at the elbow, (c) Cross-sectional view of the mesh.}
    \label{Fig2}
\end{figure}

Figure \ref{Fig1} represents the schematic of the geometry employed for 3-D CFD investigation of chaotic oscillations in CNCL system. The CNCL consists of two thermally coupled NCL systems with an area contact at the common heat exchange section. The component NCLs have a square cross-section with a hydraulic diameter $D_h=2\; cm$, height $L=1\;m$ and width $L1=1\;m$. The bottom horizontal leg of Loop 1 is the heated section, and the top horizontal leg of Loop 2 is the cooled section. The corners of the component NCLs are rounded with a radius of curvature ($R_d$) having the same magnitude as the cross-section of the component NCL, $R_d=D_h$. Figure \ref{Fig2} represents the schematic of the mesh used for the CFD study of the CNCL system. A structured mesh is generated with a finer mesh resolution near the walls to account for the boundary layer phenomena. The mesh at the common heat exchange section, bend and the cross-section of the component NCL are clearly depicted in figure \ref{Fig2}.

\clearpage

\subsection{Case setup}
The case settings are chosen such that the flow is always in the laminar regime; this is intentionally dove to avoid the complications arising from the flow regime shift from laminar to turbulent (which occurs as the velocity of the system increases). Ensuring that the flow is within the laminar regime also enables us to focus specifically on the physics instead of dealing with multiple turbulent models which only complicates the study further.

A laminar flow model is thus used with a pressure-based solver to obtain the transient behaviour of the CNCL system. Second-order upwind schemes are utilised for the energy and momentum schemes along with the PISO scheme for the pressure and velocity coupling. The magnitude of acceleration due to gravity ($g$) is set to $9.81\; \mathrm{m/s^2}$ and fluid whose properties are listed in Table 1 is selected. Both the loops of the CNCL system contain the same fluid. The heat flux boundary condition is set to $Q^{\prime \prime}= 2000\; \mathrm{W/m^2}$ at the heated and cooled sections. The zero-velocity flow field conditions with initial temperature ($T_0$) set to $300\;\mathrm{K}$ and initial pressure set to 1 $\mathrm{atm}$ are used as the initial conditions. The transient behaviour of the system is captured by using a second-order implicit temporal discretisation. The buoyancy forces generated are modelled employing the Boussinesq hypothesis. The number of iterations taken per time step is set to 400, and the scaled residuals are set to $10^{-5}$ for continuity and momentum and $10^{-6}$ for energy to test temporal convergence. The transient simulation is performed from $t=0\;\mathrm{s}$ to $t=1200\;\mathrm{s}$.

\begin{table}[!htb]
\centering
\caption{ Thermophysical properties of the fluid used for simulation.}
\scalebox{0.9}{
\begin{tabular}{|l|l|l|}
\hline
\textbf{Property}                       & \textbf{Unit}                 & \textbf{Value}            \\ \hline
Density ($\rho$)                        & \multicolumn{1}{c|}{$\mathrm{kg/m^3}$} & \multicolumn{1}{c|}{1000} \\ \hline
Dynamic viscosity ($\mu$)               & $\mathrm{Pa\;s}$                       & 0.01                      \\ \hline
Specific heat ($C_p$)          & $\mathrm{J/(kg \;K)}$                    & 500                       \\ \hline
Thermal expansion coefficient ($\beta$) & $\mathrm{1/K}$                         & 0.01                      \\ \hline
Thermal diffusivity ($\alpha$)          & $\mathrm{m^2/s}$                       & $10^{-7}$                     \\ \hline
\end{tabular}}
\end{table}

\subsection{Grid and time step independence tests}

\begin{figure}[!htb]
	\centering
	\begin{subfigure}[b]{0.49\textwidth}
		\includegraphics[width=1\linewidth]{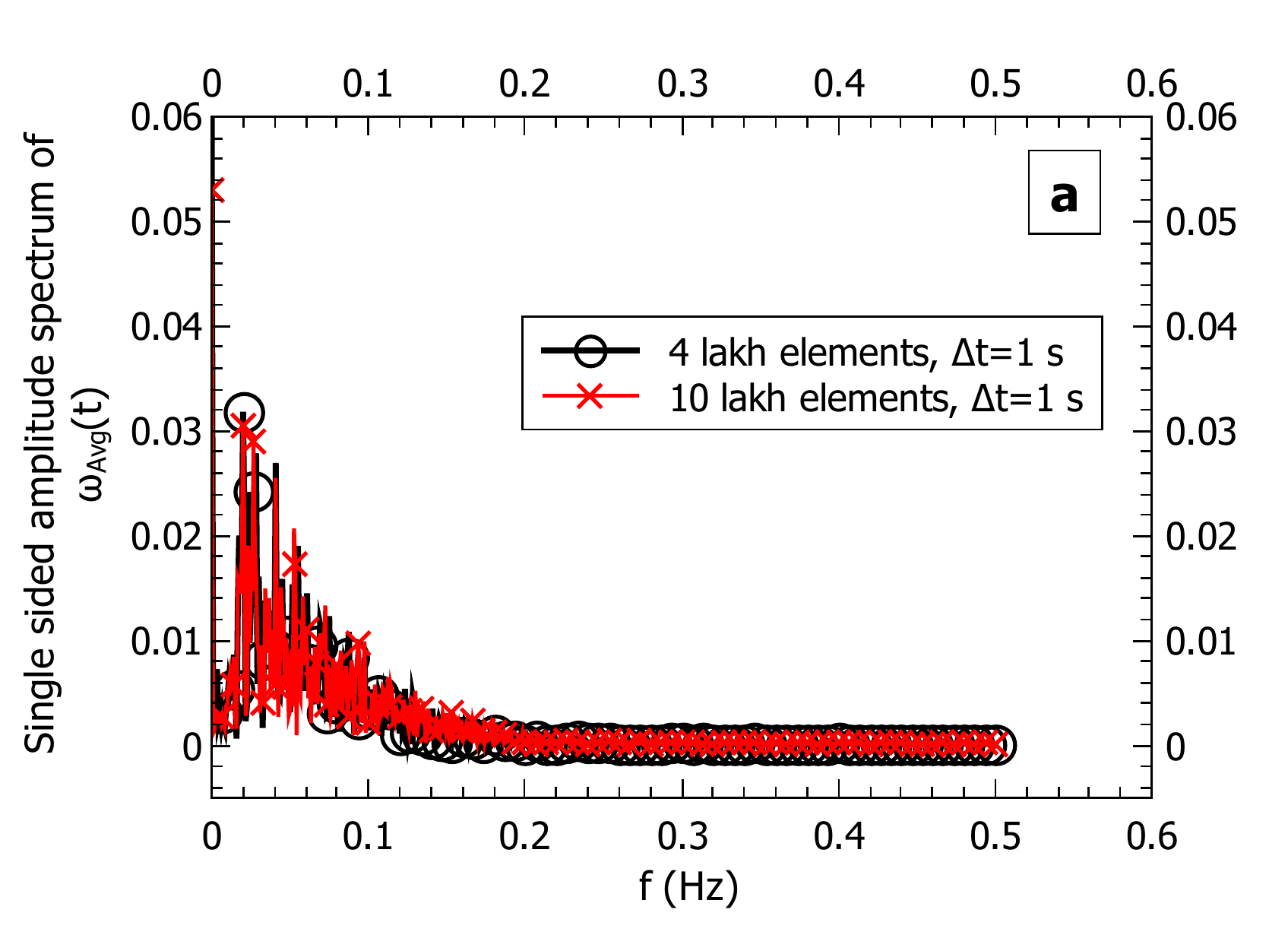}
	\end{subfigure}
	\hspace{\fill}
	\begin{subfigure}[b]{0.49\textwidth}
		\includegraphics[width=1\linewidth]{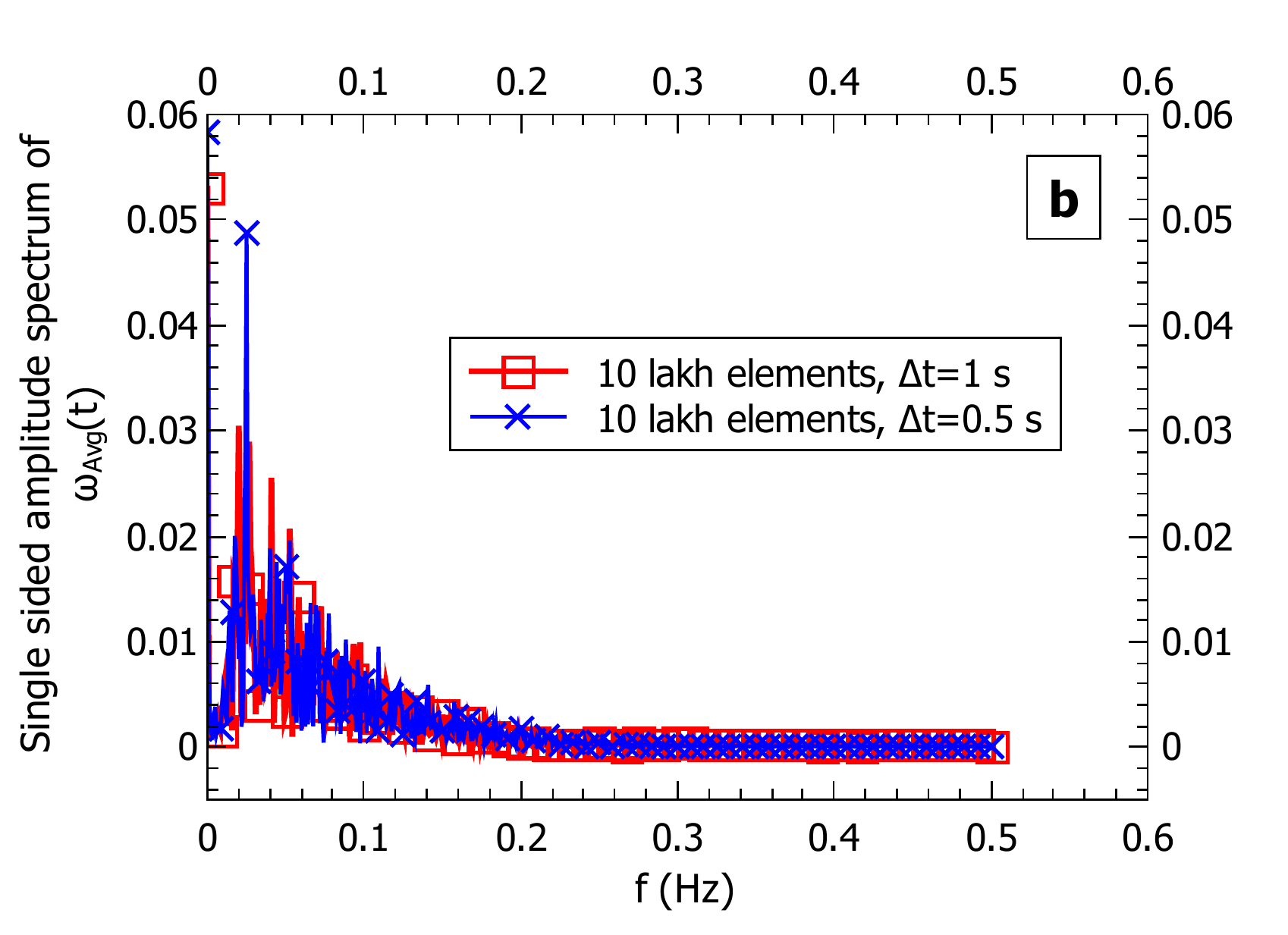}
	\end{subfigure}
	\caption{Reliability evaluation of 3-D CFD study: (a) Grid independence test, (b) Time-step independence test.}
	\label{Fig3}
\end{figure}

 To ensure the discretisation errors are minimised, the grid and time-step independence study are undertaken. The amplitude spectrum of velocity ($\omega_{Avg}$) is used as the parameter of choice for the grid and time-step independence tests. Figure \ref{Fig3} presents the grid and time-step independence tests of the CNCL system undertaken, and we can infer that a mesh with 10 lakh elements and a time step of $1\;\mathrm{s}$ is adequate for the 3-D CFD study.

\subsection{Validation of 3-D CFD methodology}

\begin{figure}[!htb]
    \centering
    \includegraphics[width=0.55\linewidth]{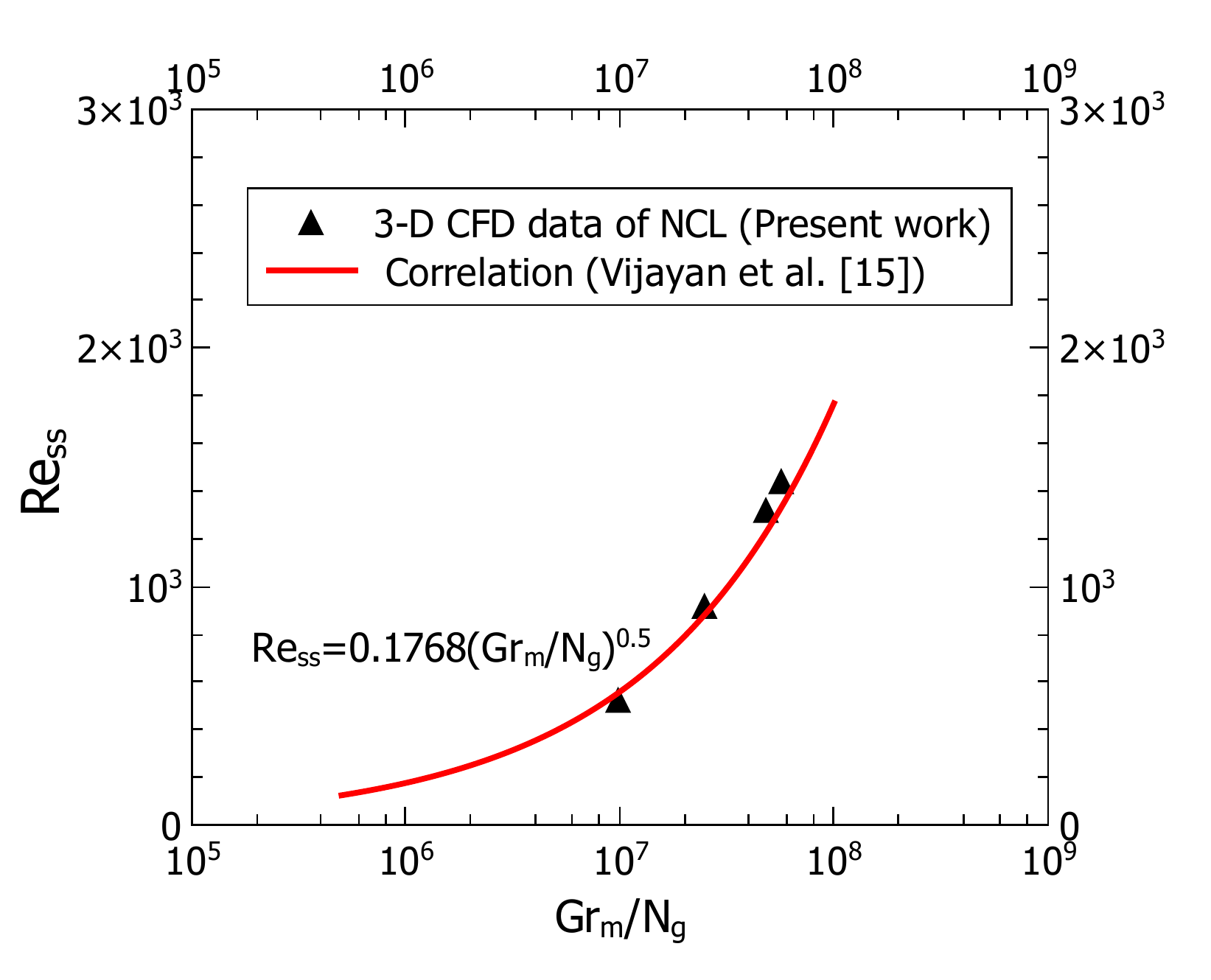}
    \caption{Validation of the methodology and settings employed for the present 3-D CFD study.}
    \label{Fig4}
\end{figure}

The settings used for the 3-D CFD simulation needs to be validated to ensure accurate description of flow physics. To accomplish this task, we compare the predictions of the 3-D CFD simulation of an NCL loop having the same settings described in the previous section with the available literature data. Figure \ref{Fig4} represents the comparison of the 3-D CFD predictions with Vijayan's correlation \cite{vijayan2002experimental}. A good match is observed between the 3-D CFD predictions and the available literature data indicating that the settings used for the CFD simulations accurately capture the physics of natural circulation systems.

\subsection{3-D CFD results}

\begin{figure}[!htb]
	\centering
	\begin{subfigure}[b]{0.49\textwidth}
		\includegraphics[width=1\linewidth]{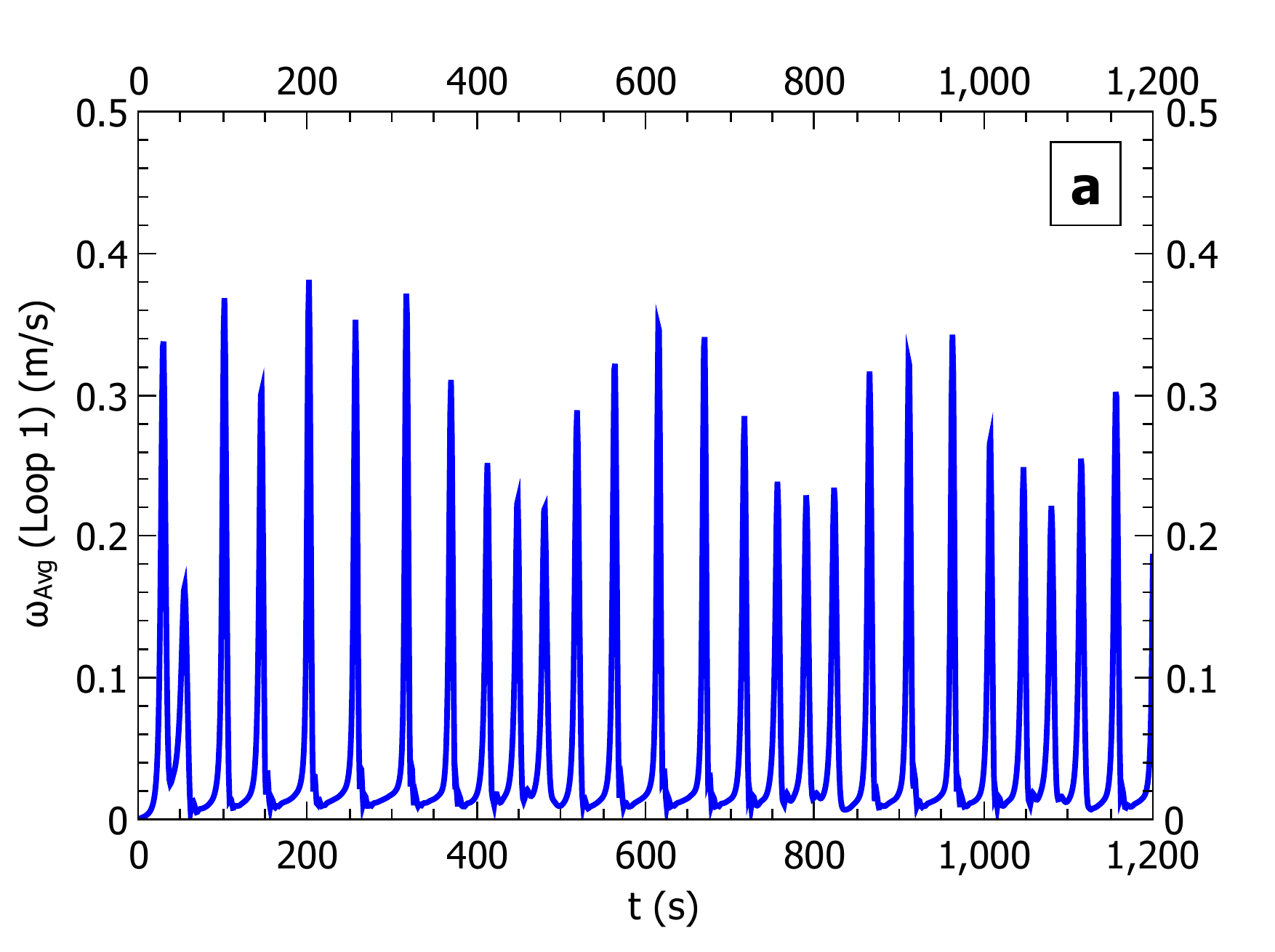}
	\end{subfigure}
	\hspace{\fill}
	\begin{subfigure}[b]{0.49\textwidth}
		\includegraphics[width=1\linewidth]{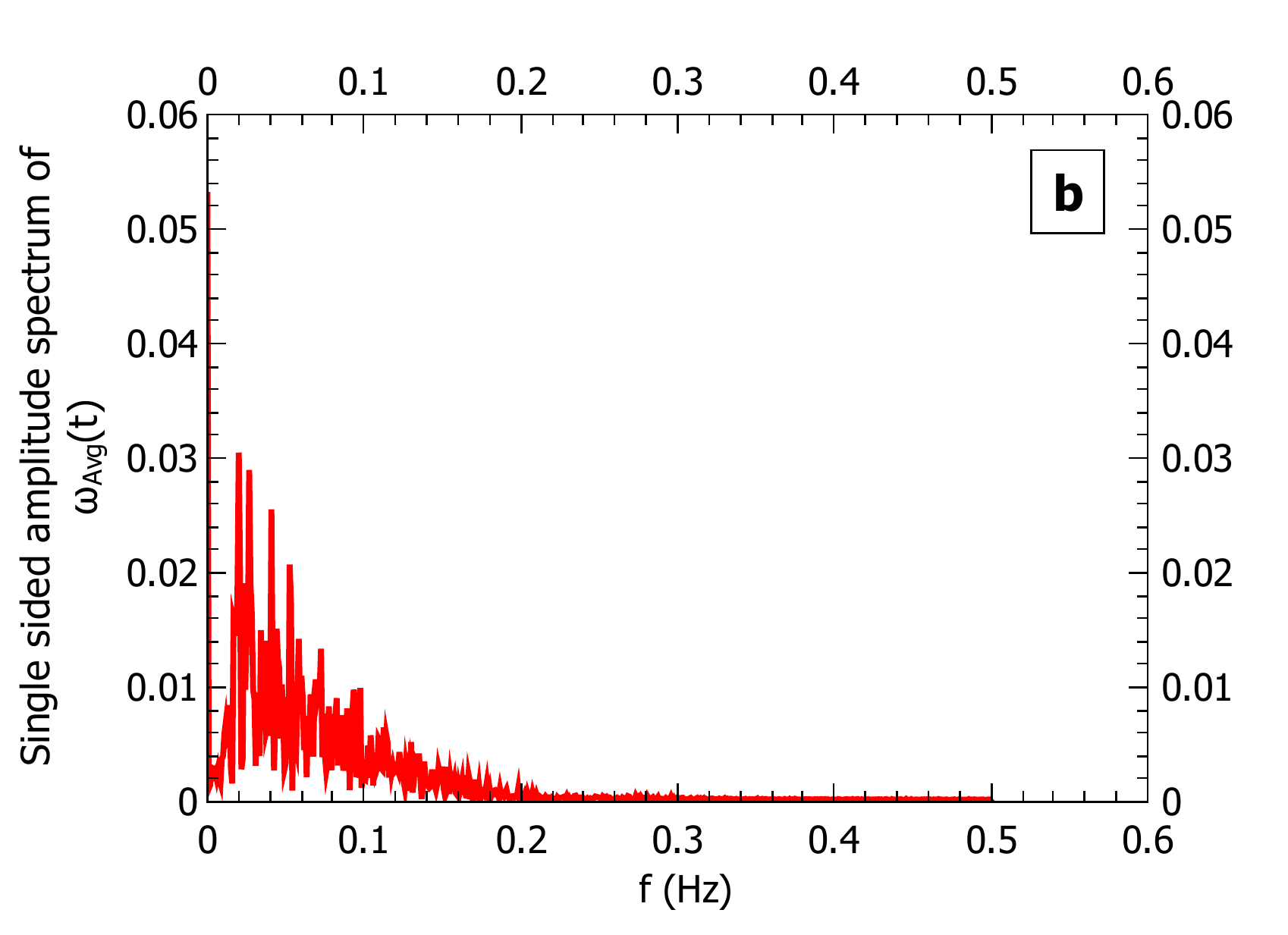}
	\end{subfigure}
	\hspace{\fill}
	\begin{subfigure}[b]{0.49\textwidth}
		\includegraphics[width=1\linewidth]{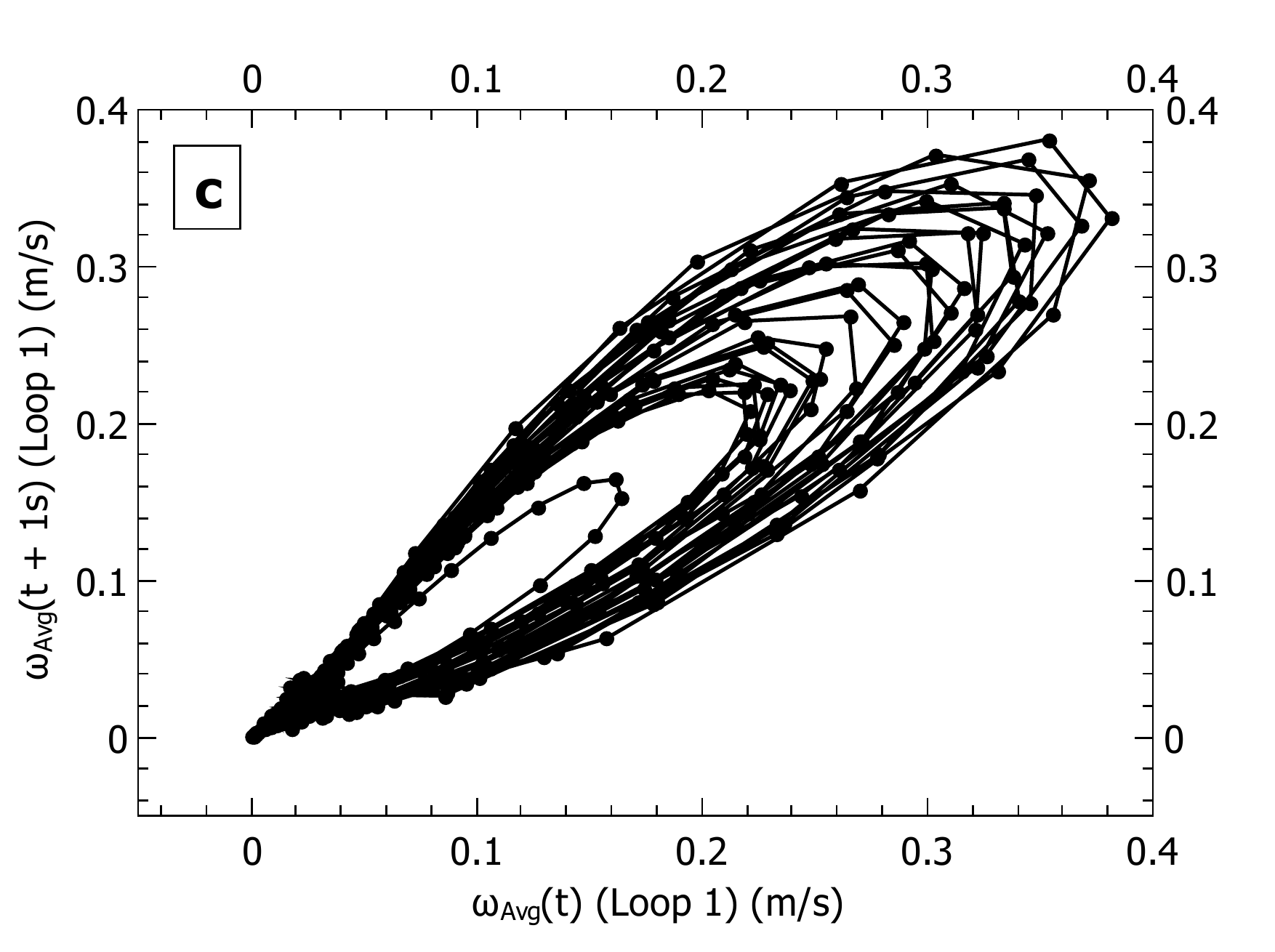}
	\end{subfigure}
	\caption{Results from the 3-D CFD study of chaotic oscillations in CNCL system, (a) Transient trend of average velocity, (b) Single sided amplitude spectrum of  average velocity, and (c) Attractor of the CNCL system for non-periodic unidirectional oscillations.}
	\label{Fig5}
\end{figure}

\begin{figure}[!htb]
	\centering
	\includegraphics{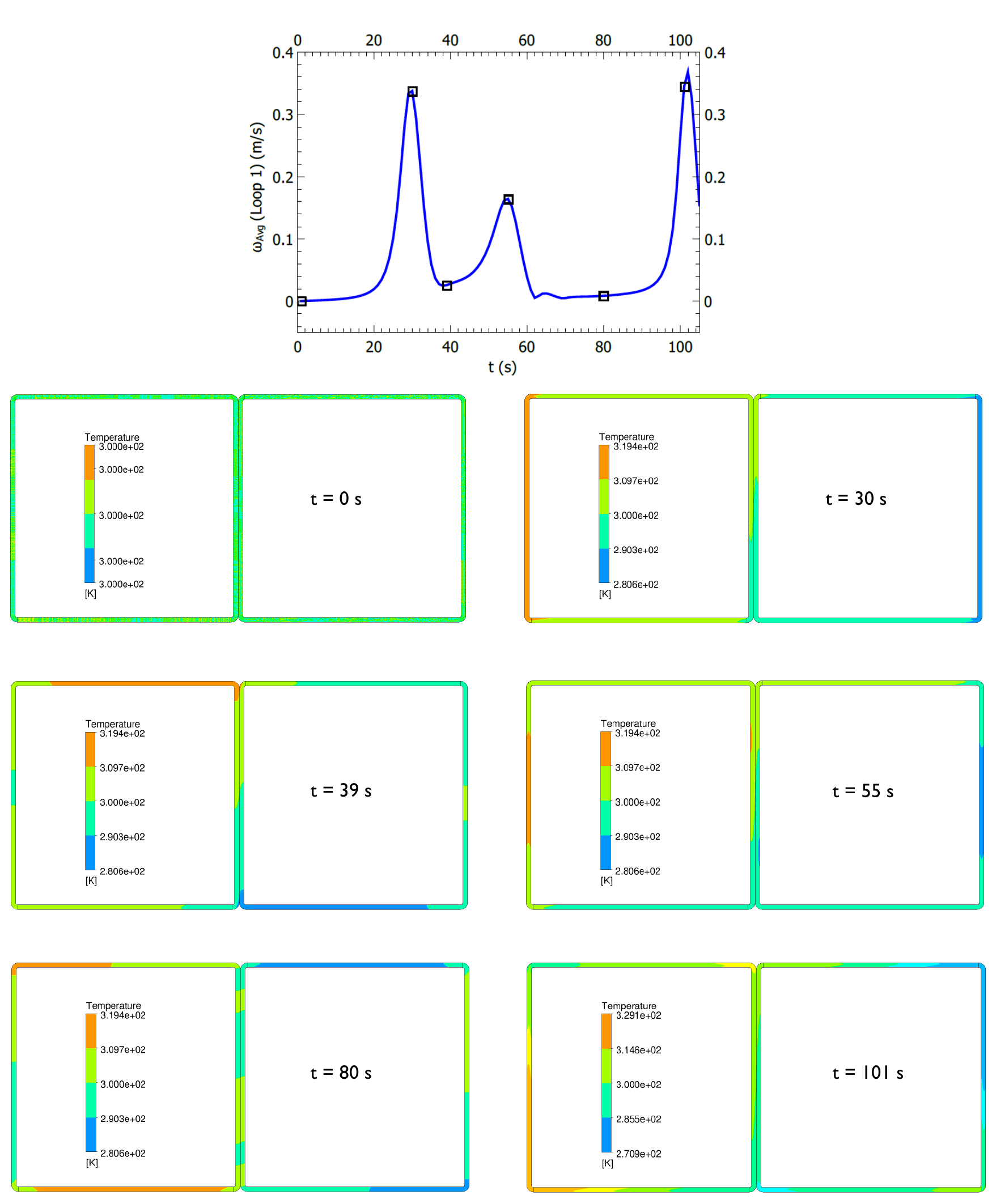}
	\caption{The temperature contours of the mid-plane of the CNCL system used for the 3-D CFD study corresponding to the peaks and valleys of the transient average velocity.}
	\label{Fig6}
\end{figure}

The present section describes in detail the results obtained from the transient 3-D CFD study. For the described conditions of the CNCL mentioned in the previous sections, it is observed that the system exhibits chaotic unidirectional oscillations. The chaotic unidirectional oscillations can be represented using the average velocity of Loop 1 versus time plot, as shown in figure \ref{Fig5}(a). Since both the loops of the CNCL contain the same fluid, they exhibit symmetric transient behaviour for the chosen heater-cooler arrangement. Thus, it is adequate to study one of the loops of the CNCL to characterise the CNCL system completely. The chaotic oscillations exhibited by the CNCL indicate that CNCL is a dynamical system. A dynamical system, by definition, implies a system with sensitive dependence on the initial conditions \cite{liu2010chaotic}, and it is necessary to employ other parameters to represent such dynamical systems. A detailed list of tools to study chaotic time series is given by Liu \cite{liu2010chaotic}. For the present case, we employ Fourier transform and attractor reconstruction to study the CNCL system further. The Fast Fourier Transform (FFT) is applied to the transient CFD data represented in figure \ref{Fig5}(a) to obtain the single-sided amplitude spectrum of the average velocity of Loop 1, which is represented in figure \ref{Fig5}(b). MATLAB software is employed to determine the FFT of the average velocity of the CNCL system and generate its single-sided amplitude spectrum. A single peak in the single-sided amplitude spectrum would indicate a periodic oscillation, but we note that there are multiple peaks observed in figure \ref{Fig5}(b), indicating that it is not a periodic flow. Another method employed to study chaotic time series is using attractor reconstruction. For the present case, the attractor is constructed using the average velocity of Loop 1 of the CNCL system. The delay time is computed to be 1s from the average mutual information algorithm of the phaseSpaceReconstruction function in MATLAB. The delay time is calculated, and the constructed attractor is represented in figure \ref{Fig5}(c).

Figure \ref{Fig6} represents the mid-plane temperature contours of the CNCL system used for the 3-D CFD study at different time instants which correspond to the peaks and valleys in the transient average velocity plot. It is observed from figure \ref{Fig6} that with progress in time there is a creation of hot and cold fluid packets within Loop 1 and Loop 2 of the CNCL system, respectively. The presence of the hot and cold packets within the respective loops leads to the development of buoyancy forces which propel the fluid resulting in the observed unidirectional chaotic oscillations. The peaks in the transient average velocity plot of Loop 1 occur when the hot packet of fluid is in the vertical limb of Loop 1 of the CNCL system. The valleys of the transient average velocity plot of Loop 1 of the CNCL correspond to the location of the hot fluid packet in the horizontal limb of Loop 1 of the CNCL system. The vertical limbs of the CNCL accelerate the hot fluid packet, and the horizontal limbs slow it down.  It can also be noted that with progress in time ($t= 80\;\mathrm{s}$) there are two hot fluid packets in Loop 1 of the system. The individual hot fluid packets try to dominate the transient dynamics of the CNCL system, and the presence of multiple heated packets may be the reason for the occurrence of non-periodic flow in the CNCL system. The symmetry of the locations of the hot (in Loop 1) and cold (in Loop 2) packets within the CNCL system about the common heat exchange section is observed.

\section{Mathematical model of 1-D CNCL system}
A detailed discussion on the modelling approach of a CNCL system with rectangular component CNCL loops and having a flat plate heat exchanger at the point of coupling was given by Dass and Gedupudi \cite{dass2019}. The model was validated via comparison with 3-D CFD results for stable convective flows. The present study utilises the same model for the stability analysis of the CNCL system. 

\subsection{Governing equations of the CNCL system}

Equations 1-4 denote the governing equations of the CNCL system, and a detailed description of the derivation of these equations is available in \cite{dass2019}.

\begin{equation}
    \rho_1\frac{d\omega_{1}(t)}{dt}+\frac{4\tau_{1}}{D_h}=\frac{\rho_1 g\beta_1}{2(L+L1)}(\oint\!(T_{1}-T_{0})f(x)dx\,) - \frac{n K\rho_1 \omega_1 ^2}{4(L+L1)}    
\end{equation}
 
 \begin{equation}
      \frac{\partial T_{1}}{\partial t}+\omega_{1}\frac{\partial T_{1}}{\partial x}= \frac{4Q^{\prime \prime} h_1(x)}{\rho_1 C_{p,1} D_h}-\frac{U}{\rho_1 C_{p,1}D_h}\lambda(x)(T_{1}-T_{2}) +\alpha_1\frac{\partial^{2}T_{1}}{\partial x^{2}}
 \end{equation}
 
 \begin{equation}
       \rho_2\frac{d\omega_{2}(t)}{dt}+\frac{4\tau_{2}}{D_h}=\frac{\rho_2 g\beta_2}{2(L+L1)}(\oint\!(T_{2}-T_{0})f(x)dx\,) - \frac{n K\rho_2 \omega_2 ^2}{4(L+L1)}
 \end{equation}
   
  \begin{equation}
       \frac{\partial T_{2}}{\partial t}+\omega_{2}\frac{\partial T_{2}}{\partial x}=\frac{-4Q^{\prime \prime} h_2(x)}{\rho_2 C_{p,2} D_h}+\frac{U}{\rho_2 C_{p,2}D_h}\lambda(x)(T_{1}-T_{2})+\alpha_2\frac{\partial^{2}T_{2}}{\partial x^{2}}
  \end{equation}
   
where, 

\begin{equation}
    \tau_i=\frac{\rho_i \omega_i^2}{2}\bigg( \frac{b}{Re_i} \bigg)^d
\end{equation}

\begin{equation}
    K=\frac{800}{Re_i} +0.14 \bigg(1 + \frac{4}{(D_h)^{0.3}} \bigg)
\end{equation}

\subsection{Non-dimensional governing equations}
 The non-dimensionalised governing equations which represent the system behaviour are:

\begin{equation}
\frac{d Re_1}{d\zeta}=\big [ Gr_1 \big ] \oint(\theta_1) f(s) ds - \big [ Co_1 \big] (Re_1)^{2-d} -\frac{nK}{4}(Re_1)^2
\end{equation}

\begin{equation}
\frac{\partial \theta_1}{\partial \zeta} +  Re_1\frac{\partial \theta_1}{\partial s} =
\big[ Fo_1 \big]\frac{\partial^2\theta_1}{\partial s^2} + h_1(s) - \big[     St_1 \big]\lambda(s)\bigg(\theta_1- \frac{\theta_2}{Co_2}\bigg)
\end{equation}

\begin{equation}
\frac{d Re_2}{d\zeta}=\big [ Gr_2 \big ] \oint(\theta_2) f(s) ds - \bigg [\frac{\nu_2}{\nu_1}\bigg]\big[ Co_1 \big] (Re_2)^{2-d} -\bigg[\frac{\nu_2}{\nu_1}\bigg]\frac{nK}{4}(Re_2)^2
\end{equation}

\begin{equation}
\frac{\partial \theta_2}{\partial \zeta} +  \bigg[\frac{\nu_2}{\nu_1}\bigg]Re_2\frac{\partial \theta_2}{\partial s}  =
\big[ Fo_2\big]\frac{\partial^2\theta_2}{\partial s^2} + h_2(s)  + \big[     St_2 \big]\lambda(s)\big(Co_2 \theta_1- \theta_2\big)
\end{equation}

where $\theta_i={T_i-T_0}/{\Delta T_i}$; $\zeta={t}/{t_0}$; $s={x}/{x_0}$; $t_0={x_0D_h}/{\nu_1} $; $\Delta T_i=(4Q^{''} t_0)/(\rho Cp D_h) $; $x_0=(L+L1)$  and the non-dimensional parameters are defined as follows:

\begin{equation}	
Co_1=\frac{2bx_0}{D_h}
\end{equation}

\begin{equation}	
Co_2=\frac{\Delta T_1}{\Delta T_2}
\end{equation}

\begin{equation}	
Fo_1=\frac{\alpha_i t_{0}}{x_0^2}
\end{equation}

\begin{equation}	
St_i=\frac{Ut_{0}}{\rho_i Cp_i D_h}
\end{equation}

\begin{equation}		
Re_i=\frac{\omega_i D_h}{\nu}
\end{equation}

\begin{equation}	
Gr_i=\frac{g \beta_i \Delta T_i x_0 D_h t_{0}}{2(L+L1) \nu_i}
\end{equation}

with,

\begin{equation}
f(s)=\left\{ \global\long\def\arraystretch{1.2}
\begin{array}{@{}c@{\quad}l@{}}
1 & {0<s<\frac{L}{L+L1}}\\
0 & {\frac{L}{L+L1}<s<1}\\
-1 & {1<s<\frac{2L+L1}{L+L1}}\\
0 & {\frac{2L+L1}{L+L1}<s<2}
\end{array}\right.
\end{equation}

\begin{equation}
\lambda (s)=\left\{ \global\long\def\arraystretch{1.2}
\begin{array}{@{}c@{\quad}l@{}}
1 & {0<s<\frac{L}{L+L1}}\\
0 & {\frac{L}{L+L1}<s<1}\\
0 & {1<s<\frac{2L+L1}{L+L1}}\\
0 & {\frac{2L+L1}{L+L1}<s<2}
\end{array}\right.
\end{equation}

\begin{equation}
h_1 (s)=\left\{ \global\long\def\arraystretch{1.2}
\begin{array}{@{}c@{\quad}l@{}}
0 & {0<s<\frac{L}{L+L1}}\\
0 & {\frac{L}{L+L1}<s<1}\\
0 & {1<s<\frac{2L+L1}{L+L1}}\\
1 & {\frac{2L+L1}{L+L1}<s<2}
\end{array}\right.
\end{equation}

\begin{equation}
h_2 (s)=\left\{ \global\long\def\arraystretch{1.2}
\begin{array}{@{}c@{\quad}l@{}}
0 & {0<s<\frac{L}{L+L1}}\\
1 & {\frac{L}{L+L1}<s<1}\\
0 & {1<s<\frac{2L+L1}{L+L1}}\\
0 & {\frac{2L+L1}{L+L1}<s<2}
\end{array}\right.
\end{equation}

When same fluid is considered within the constituent loops of the CNCL system, it results in the following simplifications: $Gr_1=Gr_2=Gr$, $Fo_1=Fo_2=Fo$, $St_1=St_2=St$, $\Delta T_1=\Delta T_2\; \implies Co_2=1$. 
Equations (7) and (9) represent the non-dimensional momentum equation and equations (8) and (10) represent the non-dimensional energy equation of the CNCL system. These equations represent the coupled nature of the CNCL system. In the above equations, $f(s)$ represents the orientation of the component loops of the CNCL system w.r.t. gravity, $\lambda (s)$ represents the location of the thermal coupling. The functions $h_1(s)$ and $h_2(s)$ represent the locations of the heater and cooler on the CNCL system, respectively for the  heater cooler configuration represented in figure \ref{Fig1}. Details of the location of origin used for mathematical representation of the aforementioned can be found in \cite{dass2019}.

\subsection{Initial conditions}

The initial conditions of the CNCL system are:

\begin{equation}
Re_1(\zeta=0)=0
\end{equation}

\begin{equation}
Re_2(\zeta=0)=0
\end{equation}

\begin{equation}
\theta_1 (s,\zeta=0)=0
\end{equation}

\begin{equation}
\quad \theta_2 (s,\zeta=0)=0
\end{equation}

Equations (21) and (22) represent the initial conditions of the non-dimensional momentum equations and equations (23) and (24) represent the initial conditions of the non-dimensional energy equations.

\subsection{Non-dimensional stencil of the CNCL system}

To obtain the transient solution of the CNCL system behaviour, the PDEs (Partial Differential Equations) represented by equations (7) to (10) are converted into system of ODEs (Ordinary Differential Equation). This is achieved by substituting the Fourier series of non-dimensional temperature and boundary conditions in the PDEs.
The Fourier series of the non-dimensional temperature and boundary conditions are:

\begin{equation}
\theta_1(s,\zeta)= \sum \limits_{k = -\infty }^\infty \theta_{1,k} (\zeta) e^{ik\pi s}
\end{equation}

\begin{equation}
\theta_2(s,\zeta)= \sum \limits_{k = -\infty }^\infty \theta_{2,k} (\zeta) e^{ik\pi s}
\end{equation}

\begin{equation}
h_1(s) = \sum \limits_{k = -\infty }^\infty h_{1,k} e^{ik\pi s} 
\end{equation}

\begin{equation}
h_2(s) = \sum \limits_{k = -\infty }^\infty h_{2,k} e^{ik\pi s}
\end{equation}

\begin{equation}
\lambda(s) = \sum \limits_{k = -\infty }^\infty \lambda_k e^{ik\pi s}
\end{equation}

\begin{equation}
f(s) = \sum \limits_{k = -\infty }^\infty f_k e^{ik\pi s} 
\end{equation}

and the representative ODE stencil of the CNCL system when same fluids are used in both loops, is given as:

\begin{equation}
\frac{d Re_1}{d\zeta}=\big [ Gr \big ] \sum_{k=-\infty}^{\infty}(\theta_{1,k}) f_{-k}  - \big [ Co_1 \big] (Re_1)^{2-d} -\frac{nK}{4}(Re_1)^2
\end{equation}

\begin{equation}
\frac{\partial \theta_{1,k}}{\partial \zeta} = -i k \pi Re_1\theta_{1,k} 
-k^2 {\pi}^2 \big[ Fo \big] \theta_{1,k} + h_{1,k} - \big[     St \big] \sum_{l=-\infty}^{\infty} \lambda_{k-l} \big(\theta_{1,l}- \theta_{2,l} \big)
\end{equation}

\begin{equation}
\frac{d Re_2}{d\zeta}=\big [ Gr \big ] \sum_{k=-\infty}^{\infty}(\theta_{2,k}) f_{-k} - \big [ Co_1 \big] (Re_2)^{2-d} -\frac{nK}{4}(Re_2)^2
\end{equation}

\begin{equation}
\frac{\partial \theta_{2,k}}{\partial \zeta} =   -i k \pi Re_2\theta_{2,k} 
-k^2 {\pi}^2 \big[ Fo \big] \theta_{2,k} + h_{2,k} + \big[     St \big] \sum_{l=-\infty}^{\infty} \lambda_{k-l} \big(\theta_{1,l}- \theta_{2,l} \big)
\end{equation}

with initial conditions of $\theta_{1,k}$ and $\theta_{2,k}$ calculated using:

\begin{equation}
	\theta_{i,k}(\zeta=0)= \frac{\int_{0}^{2} \theta_i(s,\zeta=0)e^{-ik\pi s}ds } {2}
\end{equation}

but $\theta_i (s,\zeta=0)=0$ (from equations (23) and (24)), and hence $\theta_{i,k} (\zeta=0)=0$ for all $k$.

\subsection{Numerical integration of the ODE system}

To estimate the transient dynamics of the CNCL system, equations (31) to (34) are truncated by restricting the number of nodes (i.e. $k=0,1,2,...,N$), separating the real and imaginary components and then the ODE system is integrated with the appropriate initial conditions using a numerical solver. In the current paper, we have employed the MATLAB solver $ode15s$ for numerical integration of the dimensional and non-dimensional CNCL equations.

\subsection{Fourier node independence test}

\begin{figure}[!htb]
	\centering
	\begin{subfigure}[b]{0.49\textwidth}
		\includegraphics[width=1\linewidth]{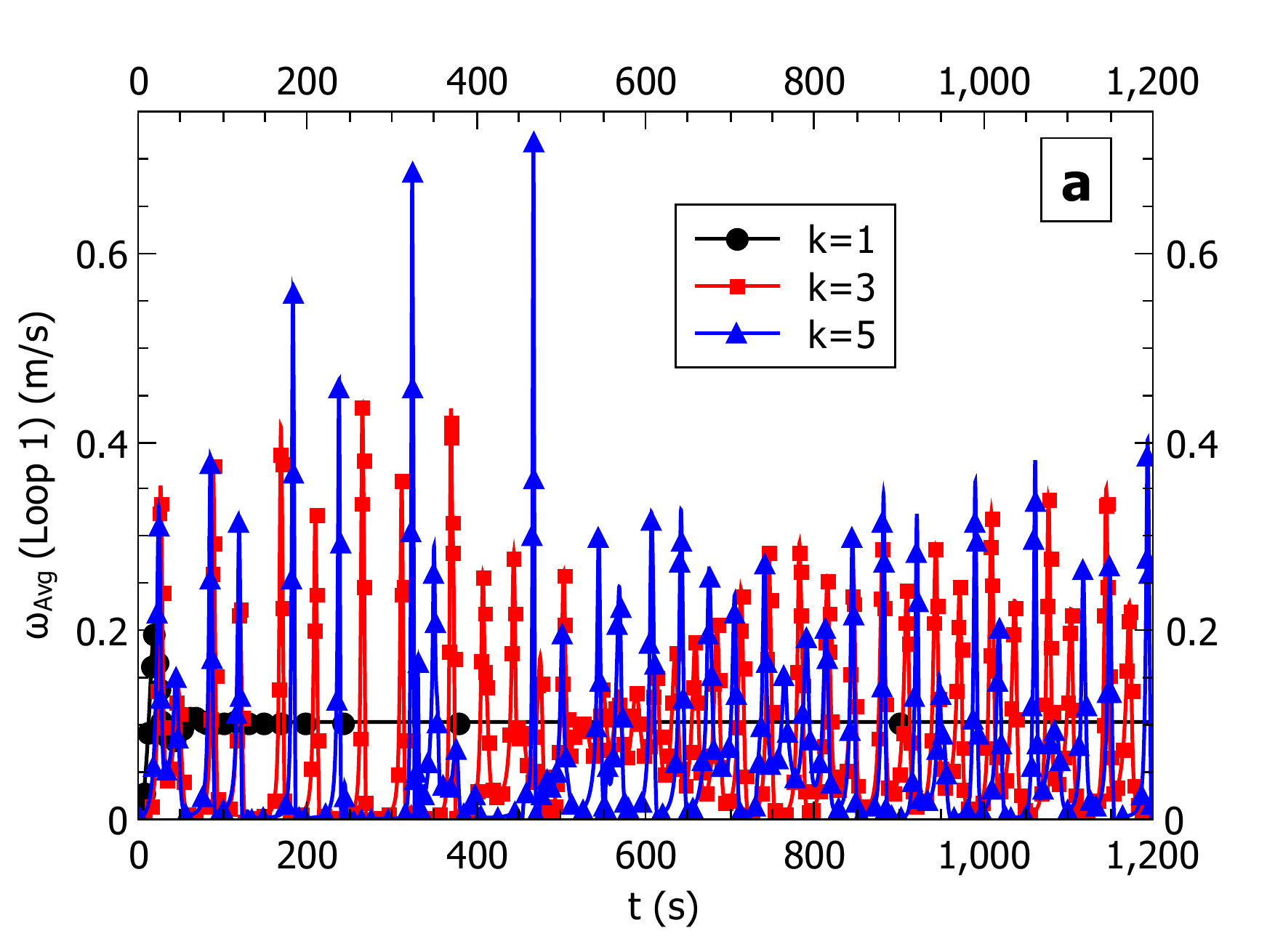}
	\end{subfigure}
	\hspace{\fill}
	\begin{subfigure}[b]{0.49\textwidth}
		\includegraphics[width=1\linewidth]{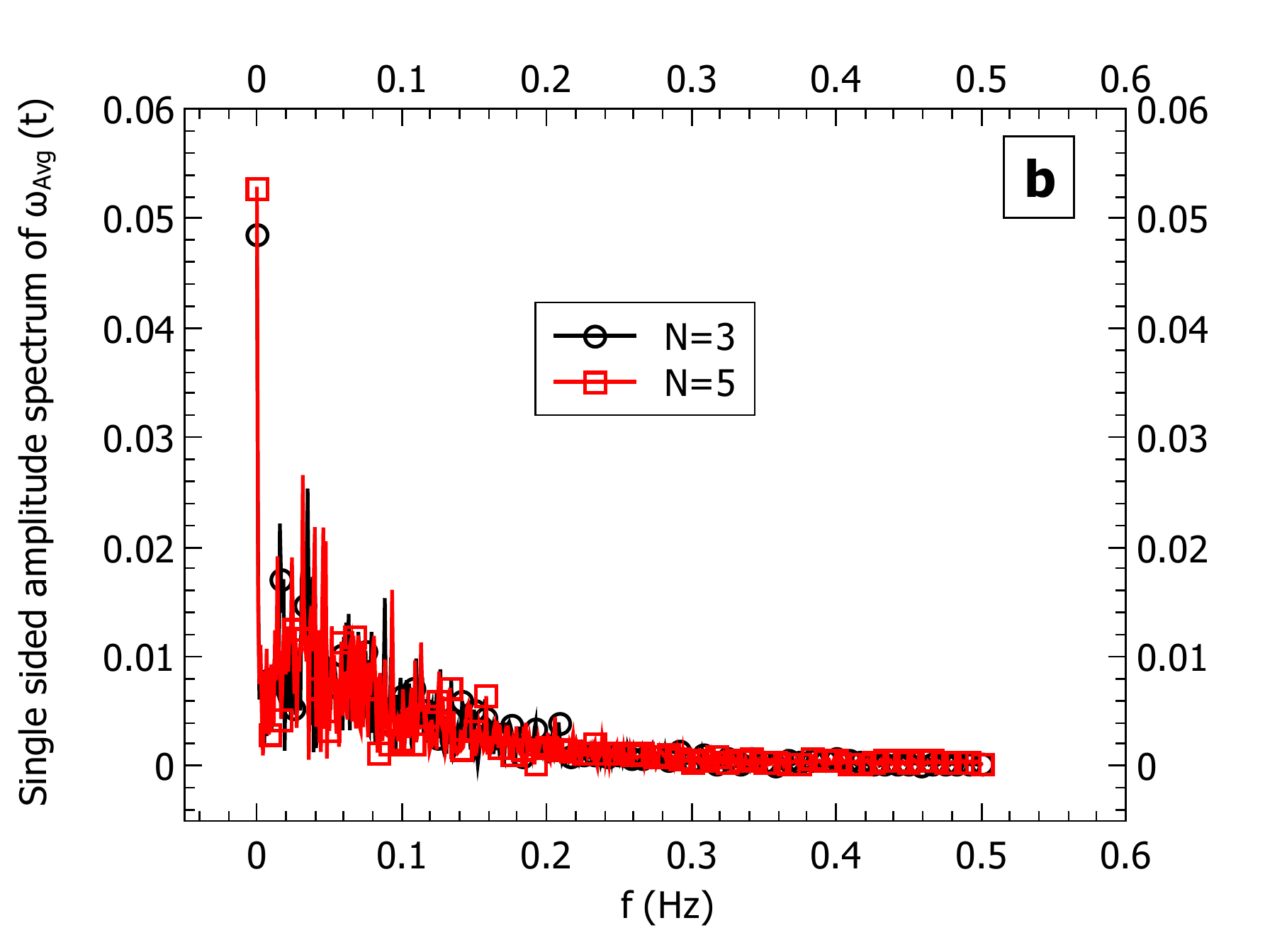}
	\end{subfigure}
	\caption{Fourier node independence test of the truncated ODE system, (a) Average velocity versus time, (b) Single sided amplitude spectrum of average velocity, for parameters considered for the 3-D CFD case.}
     \label{Fig7}
\end{figure}

 To identify the number of Fourier nodes required to adequately represent the phenomena being studied, a Fourier node independence test is carried out. It can be noted from figure \ref{Fig7}(a) that $N=1$ truncated ODE system is unable to predict the non-periodic unidirectional oscillations, but $N=3,5$ capture this behaviour. There is a phase difference and slight variation in the prediction of average velocity using $N=3,5$ truncated ODE systems. Thus, an FFT (single sided amplitude spectrum) of the predictions of the average velocity by  $N=3,5$ truncated ODE systems is compared in figure \ref{Fig7}(b). A good agreement is witnessed in the FFT of both  $N=3,5$ truncated ODE systems. Hence, from figure \ref{Fig7}, we can conclude that truncating the ODE system to $N=3$ is adequate to capture the non-periodic unidirectional oscillation of the considered 3-D CFD case.
 
 \subsection{Choosing heat transfer coefficient ($U$) for the 1-D model}
 
 
 From the 3-D CFD study it is observed that the heat transfer coefficient of the CNCL system considered for the present study also exhibits an oscillatory behaviour and has an average magnitude of $U \approx 4200\;\mathrm{W/m^2K}$. Since the current 1-D model employs a constant heat transfer coefficient for determination of the CNCL system transience, we utilise the average heat transfer coefficient magnitude obtained from the 3-D CFD study.
 
 
\section{Assessment of 1-D CNCL model to capture chaotic oscillations }

\subsection{ Sensitivity of the CNCL system to initial conditions}

\begin{figure}[!htb]
    \centering
    \begin{subfigure}[b]{0.49\textwidth}
    	\includegraphics[width=1\linewidth]{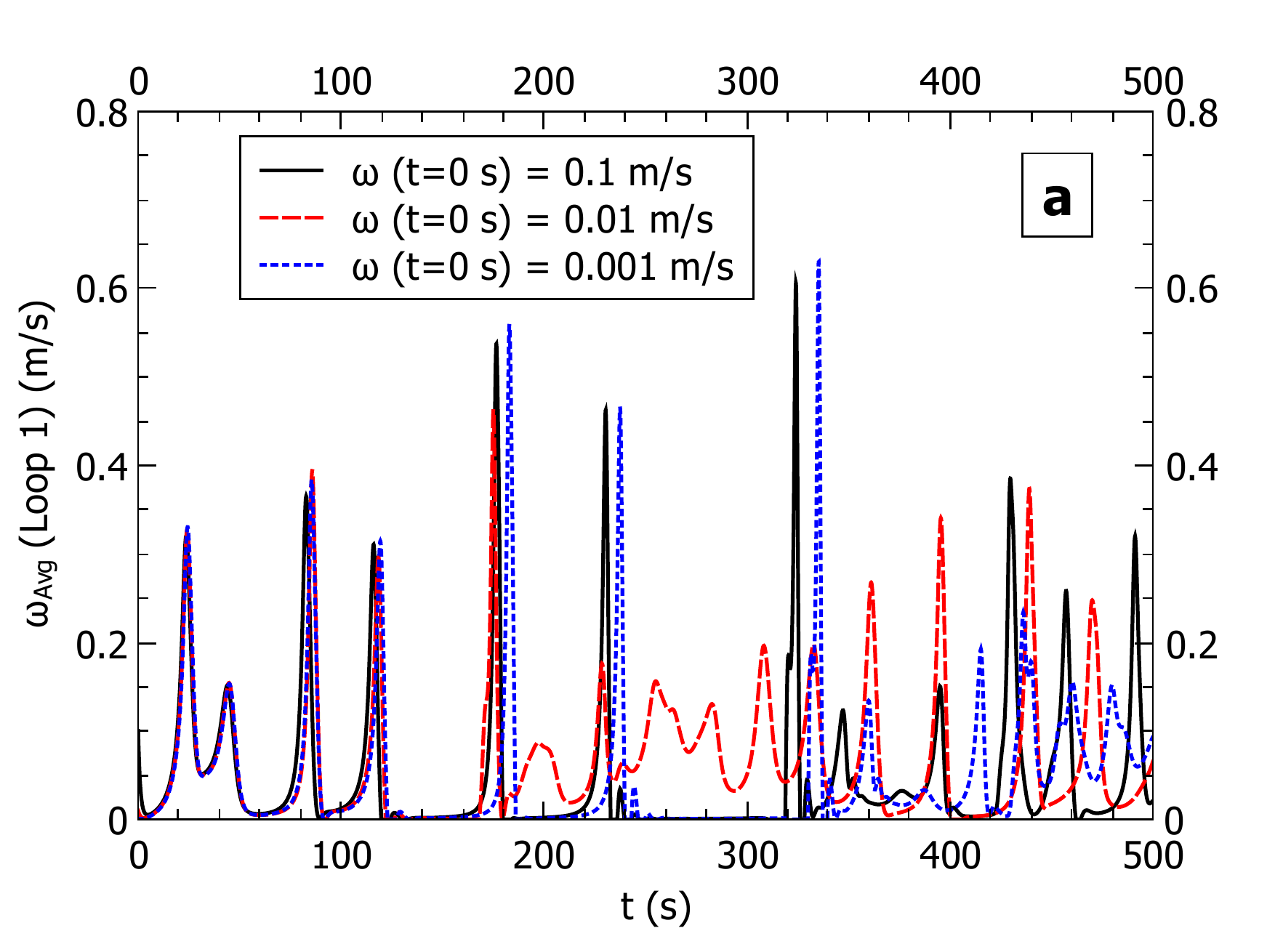}
    \end{subfigure}
    \hspace{\fill}
    \begin{subfigure}[b]{0.49\textwidth}
    	\includegraphics[width=1\linewidth]{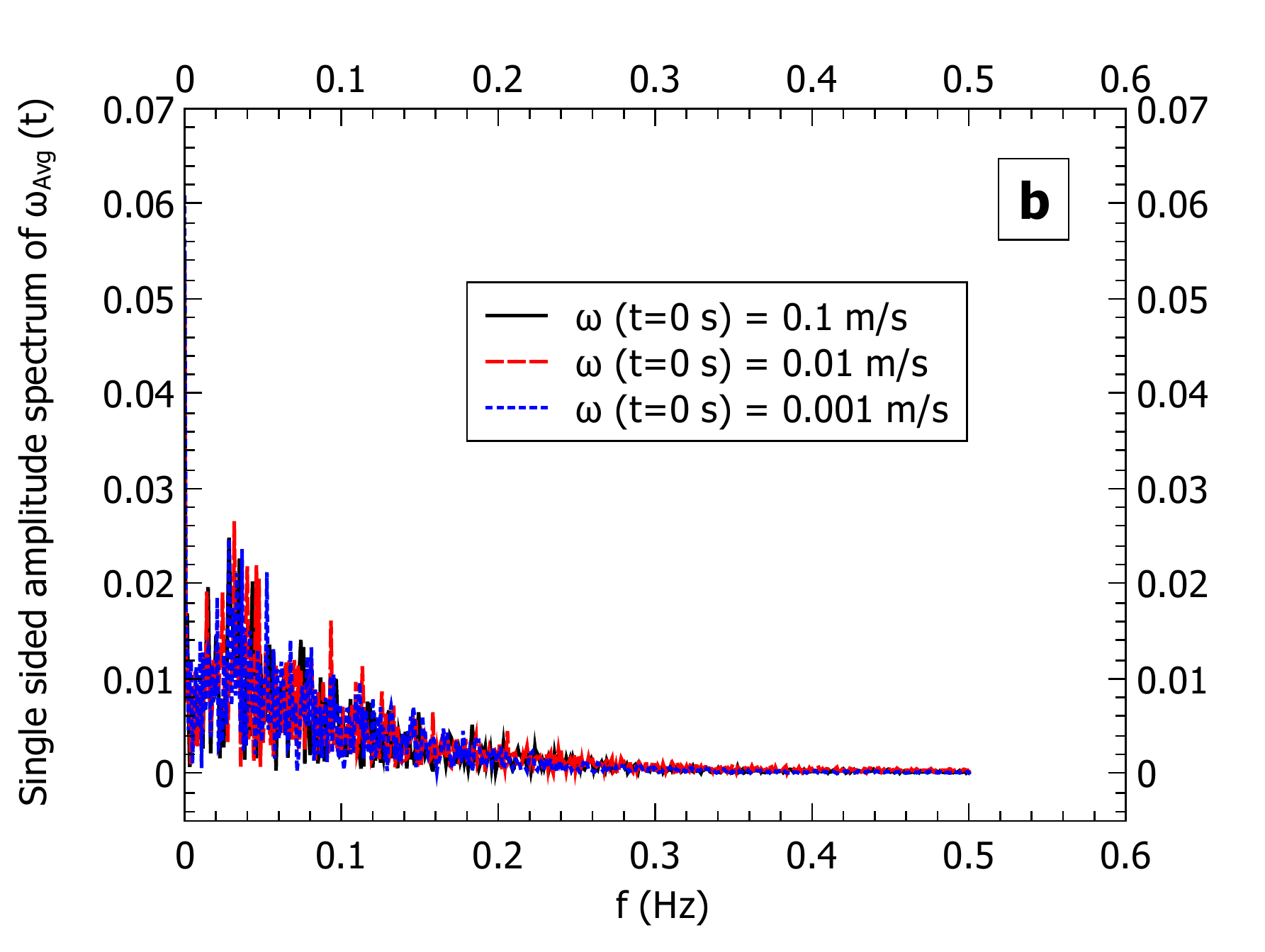}
    \end{subfigure}
    \caption{Effect of initial conditions on the transient behaviour of the CNCL exhibiting uni-directional oscillations, (a) Average velocity versus time, (b) Single sided amplitude spectrum of average velocity, for parameters considered for the 3-D CFD case.}
    \label{Fig8}
\end{figure}

Figure \ref{Fig8} demonstrates the sensitivity of the 1-D modelled CNCL system exhibiting unidirectional oscillation to the initial conditions for the parameters representing the considered 3-D CFD case. It is noted that even a small deviation in the initial condition leads to a difference in the observed transience, as shown in figure \ref{Fig8}(a), though the single sided amplitude spectrum is identical for different initial conditions as shown in figure \ref{Fig8}(b). It may be noted that non-chaotic dynamics of the CNCL system  studied by Dass and Gedupudi \cite{dass2019} converges to a steady state irrespective of the initial conditions. The present study  corroborates the fact that the CNCL is a dynamical system for specific conditions. Thus, the transient behaviour of parameters should not be employed to compare dynamical systems and hence, various tools (FFT (single sided amplitude spectrum) and attractor reconstruction) listed by Liu \cite{liu2010chaotic} need to be used for CNCL system comparison and validation.

\subsection{Verification of the 1-D model with 3-D CFD}

\begin{figure}[!htb]
	\centering
	\begin{subfigure}[b]{0.49\textwidth}
		\includegraphics[width=1\linewidth]{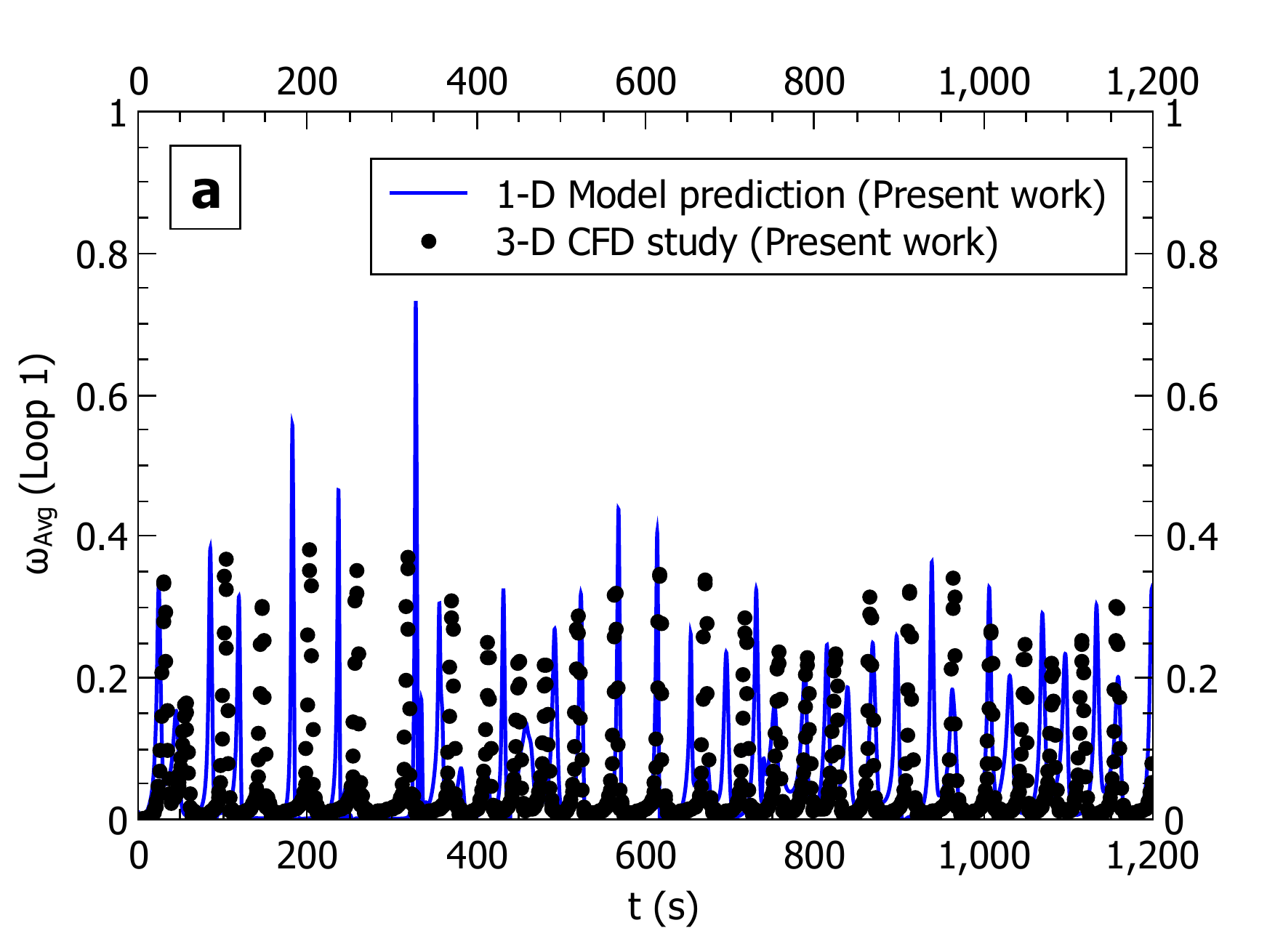}
	\end{subfigure}
	\hspace{\fill}
	\begin{subfigure}[b]{0.49\textwidth}
		\includegraphics[width=1\linewidth]{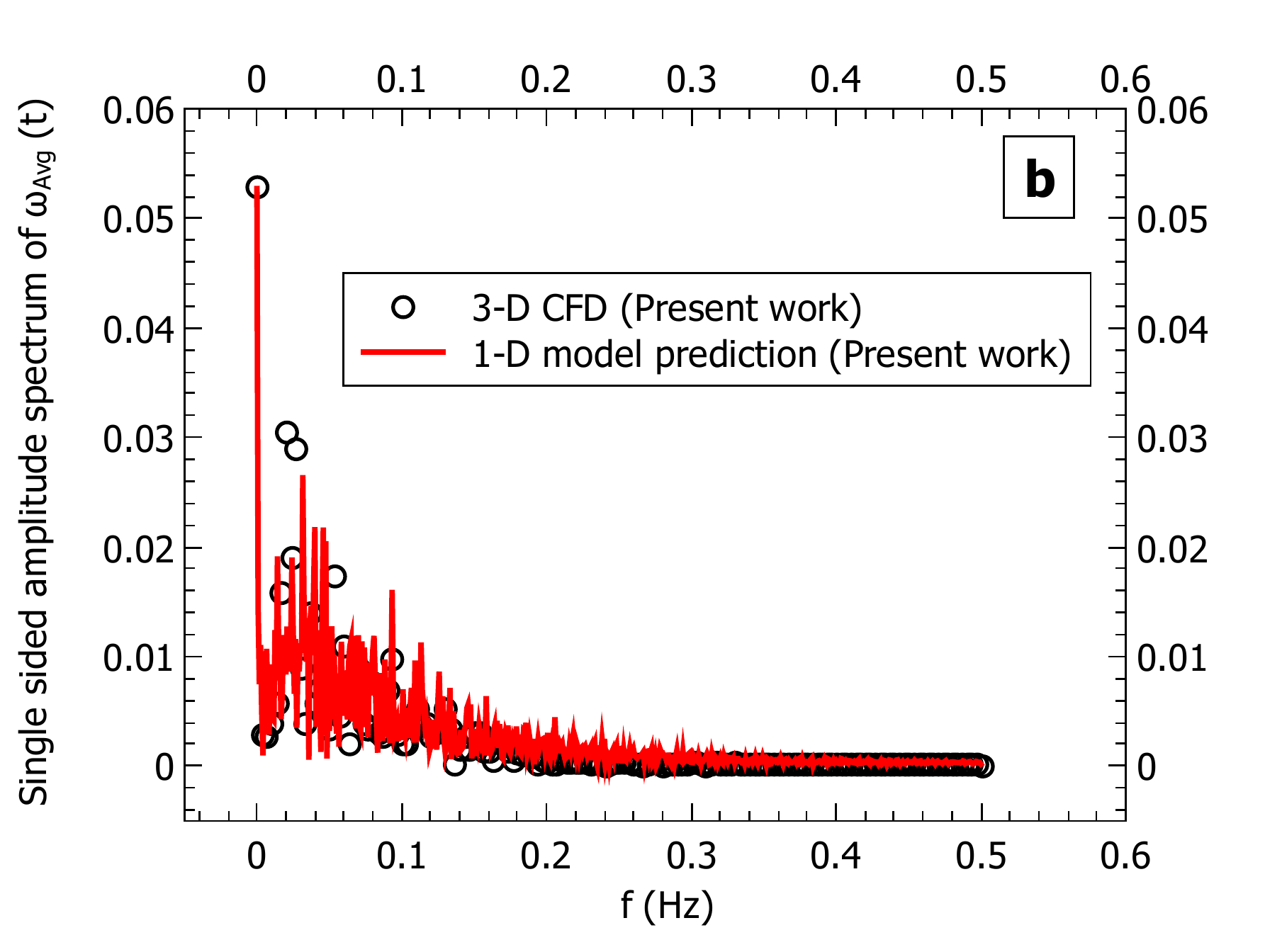}
	\end{subfigure}
		\hspace{\fill}
	\begin{subfigure}[b]{0.49\textwidth}
		\includegraphics[width=1\linewidth]{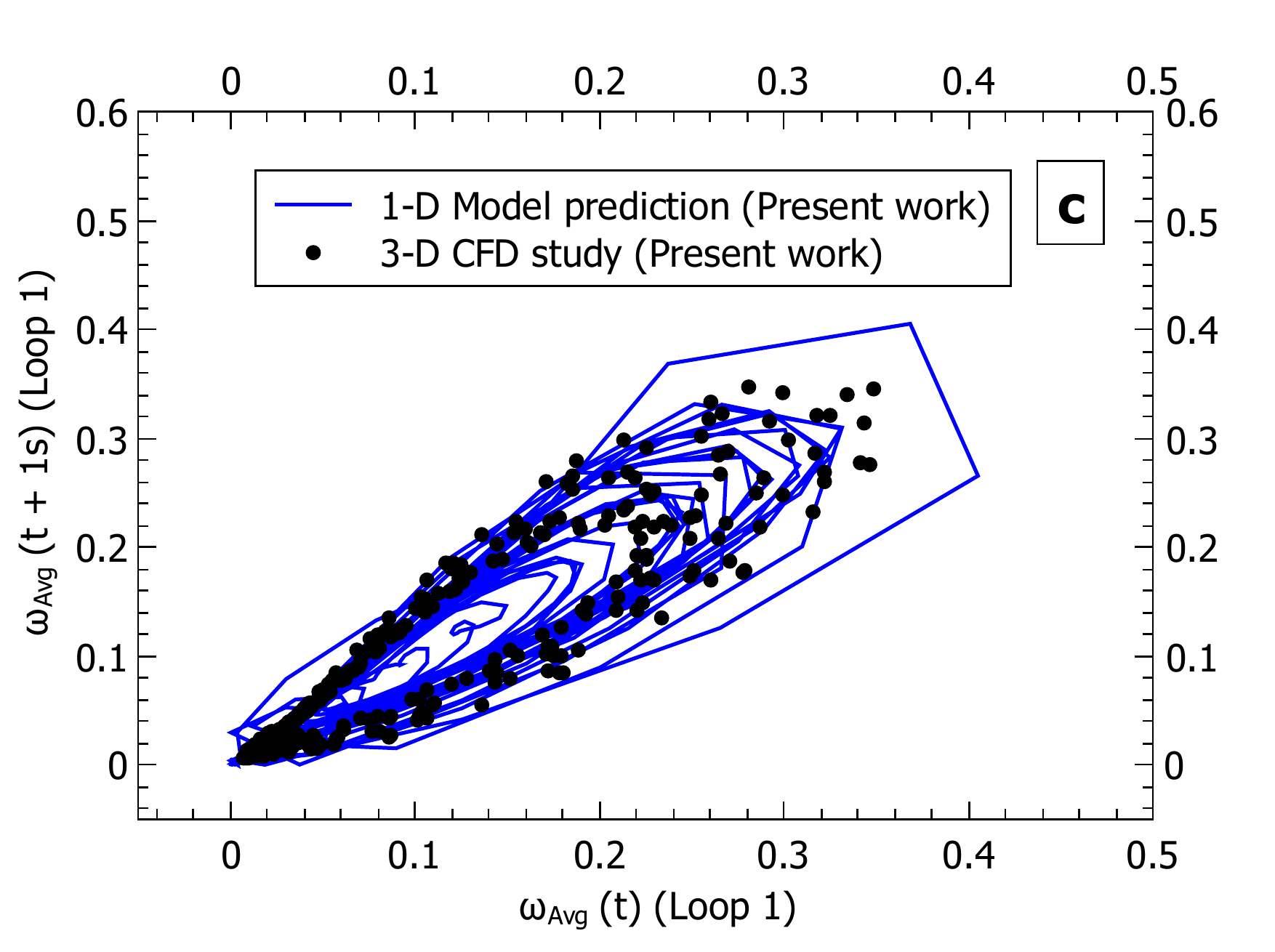}
	\end{subfigure}
	\caption{Validation of the 1-D CNCL with 3-D CFD data for chaotic oscillations in CNCL system employing, (a) Transient absolute averaged velocity, (b) Single sided amplitude spectrum of transient absolute averaged velocity, and (c) Attractor of the CNCL system, for parameters considered for the 3-D CFD case. }
	\label{Fig9}
\end{figure}

Figure \ref{Fig9} represents the comparison of predictions of the 1-D model against the results of the 3-D CFD case presented in the current paper. The transient dynamics, the single-sided amplitude spectrum and the attractor \footnote{ The attractor is constructed considering the CNCL system transience from $t=400\;\mathrm{s}.$} of the CNCL system are used as parameters to assess the ability of the 1-D CNCL model to capture the unidirectional non-periodic oscillation. A good agreement is observed for all the three parameters considered, indicating the suitability of the 1-D model to capture chaotic CNCL behaviour and thus can be utilised as a tool to predict when the system exhibits chaotic transient behaviour.

\section{Linear stability analysis of the CNCL system}

\begin{figure}[!htb]
    \centering
    \includegraphics[width=0.625\linewidth]{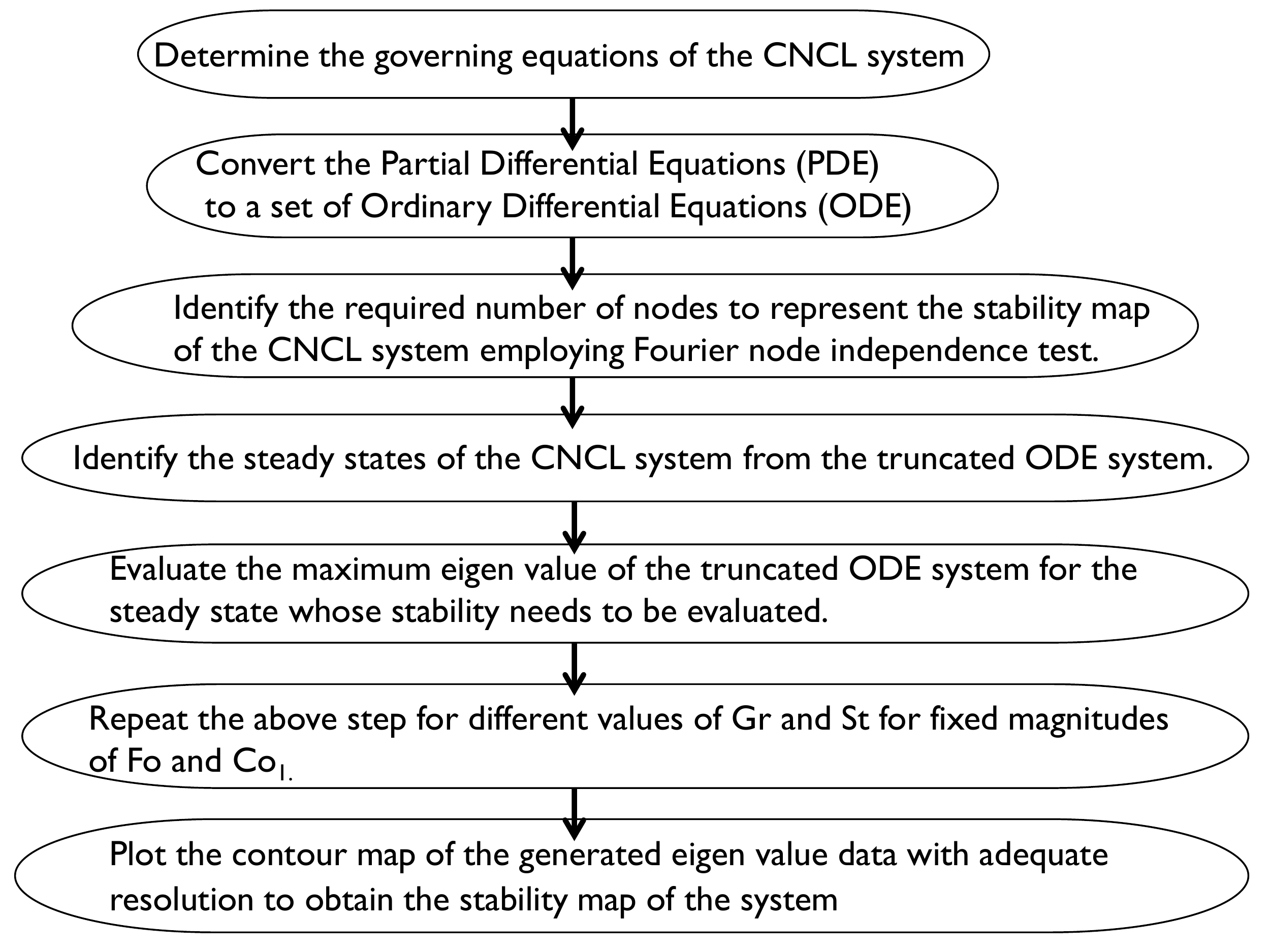}
    \caption{Flow chart representing the procedure used to obtain the linear stability map of the CNCL system.}
    \label{Fig10}
\end{figure}

The linear stability analysis of CNCL systems is presented in this section. The method to determine the linear stability analysis of an ODE system is well established. As the partial differential equations of the CNCL system can be reduced to an ODE stencil, the same approach can be utilised to determine the stability of the CNCL system. In the current study, the linear stability analysis is used to generate stability maps which demarcate the regions of stable convective flow and chaotic flows of the CNCL system. The flow chart representing the steps followed to obtain the stability map of the CNCL system is presented in figure \ref{Fig10}.

\subsection{Steady state analysis of the CNCL system}

The steady state ODE stencil of the CNCL system is :

\begin{equation}
\big [ Gr \big ] \sum_{k=-\infty}^{\infty}(\theta_{1,k}) f_{-k}  - \big [ Co_1 \big] (Re_1)^{2-d} -\frac{nK}{4}(Re_1)^2 = 0
\end{equation}

\begin{equation}
-k^2 {\pi}^2 \big[ Fo \big] \theta_{1,k} + h_{1,k} - \big[     St \big] \sum_{l=-\infty}^{\infty} \lambda_{k-l} \big(\theta_{1,l}- \theta_{2,l} \big) - i k \pi Re_1\theta_{1,k} = 0
\end{equation}

\begin{equation}
\big [ Gr \big ] \sum_{k=-\infty}^{\infty}(\theta_{2,k}) f_{-k} - \big [ Co_1 \big] (Re_2)^{2-d} -\frac{nK}{4}(Re_2)^2 =0
\end{equation}

\begin{equation}
-k^2 {\pi}^2 \big[ Fo \big] \theta_{2,k} + h_{2,k} + \big[     St \big] \sum_{l=-\infty}^{\infty} \lambda_{k-l} \big(\theta_{1,l}- \theta_{2,l} \big) - i k \pi Re_2\theta_{2,k} = 0
\end{equation}

\begin{figure}[!htb]
	\centering
	\begin{subfigure}[b]{0.49\textwidth}
		\includegraphics[width=1\linewidth]{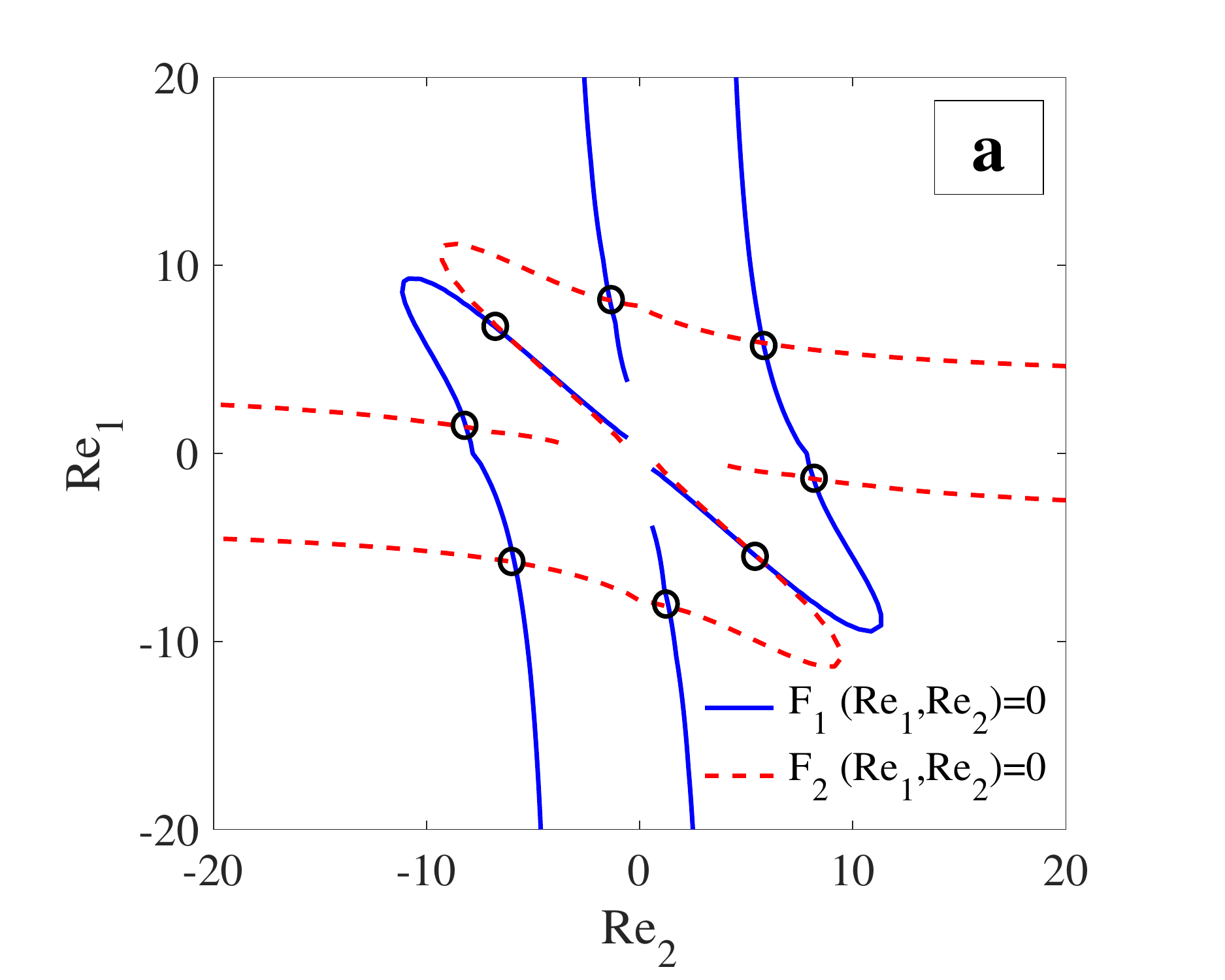}
	\end{subfigure}
	\hspace{\fill}
	\begin{subfigure}[b]{0.49\textwidth}
		\includegraphics[width=1\linewidth]{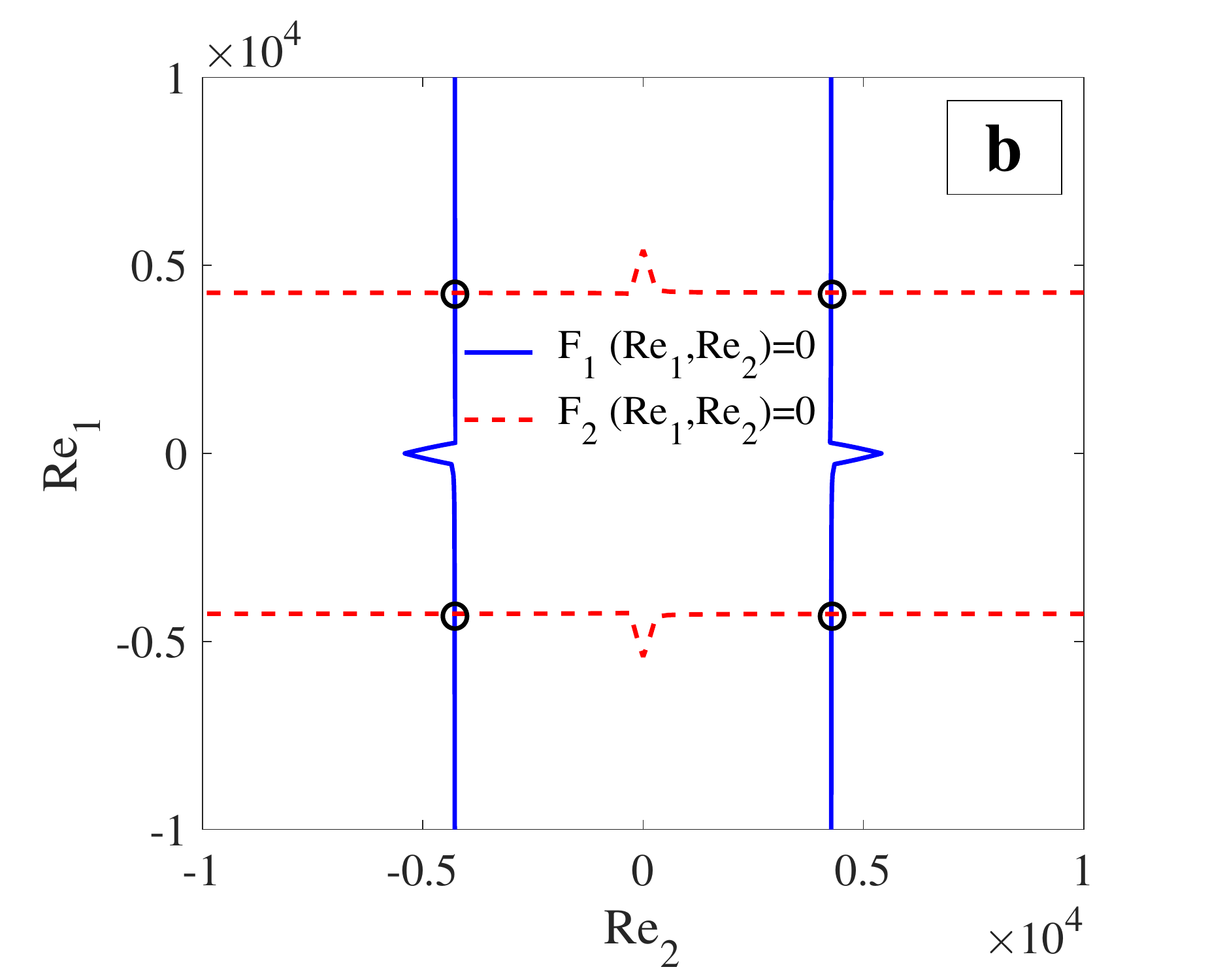}
	\end{subfigure}
	\caption{Multiple steady-states of the CNCL system considered for the present 3-D CFD study evaluated from the 1-D model after truncating the ODE system upto three Fourier nodes for (a) $Gr=10^6$ and $St=400$, (b) $Gr=10^{12}$ and $St=400$, for $Fo=0.0001$ and $As=1$. The points of intersection of $F_1$ and $F_2$ are encircled.}
	\label{Fig11}
\end{figure}

It can be observed from the above set of equations that equations (37) and (39) are linear, thus after expanding the ODE stencil represented by equations (37) and (39), all the Fourier nodes of the non-dimensional temperature of Loop 1 and Loop 2 can be expressed completely as functions of $Re_1$ and $Re_2$. The non-dimensional steady-state magnitudes of the Fourier nodes of the non-dimensional temperatures of both loops are substituted in equations (36) and (38) respectively to obtain two implicit equations $F_1 (Re_1,Re_2)=0$ and $F_2 (Re_1,Re_2)=0$. The points of intersection of $F_1$ and $F_2$ provide us with the possible steady-state magnitudes of $Re_1$ and $Re_2$ for a given set of non-dimensional numbers. Figure \ref{Fig11} represents the points of intersection of $F_1$ and $F_2$ for $St=400$ for $Gr=10^6$ and $Gr=10^{12}$. It can be inferred from figure \ref{Fig11} that the CNCL system has multiple steady states and that the number of steady states possible depends on the magnitudes of the non-dimensional numbers $Gr$, $St$, $Fo$ and $As$. The sign associated with $Re_1$ and $Re_2$ indicates the direction of flow within the CNCL system. A positive value of $Re_1$ indicates anti-clockwise flow in Loop 1 from the readers perspective, and a positive value of $Re_2$ indicates clockwise flow in Loop 2. The negative magnitudes indicate the opposite directions in each of the loops, respectively.

The CNCL system has multiple steady states, and thus for a given set of non-dimensional numbers, it is necessary to determine which of these steady states are exhibited by the CNCL system and which of the steady states are not. The stability analysis of the system provides an answer to this question.

\subsection{Classification of steady-state solutions}

\begin{figure}[!htb]
    \centering
    \includegraphics[width=0.6\linewidth]{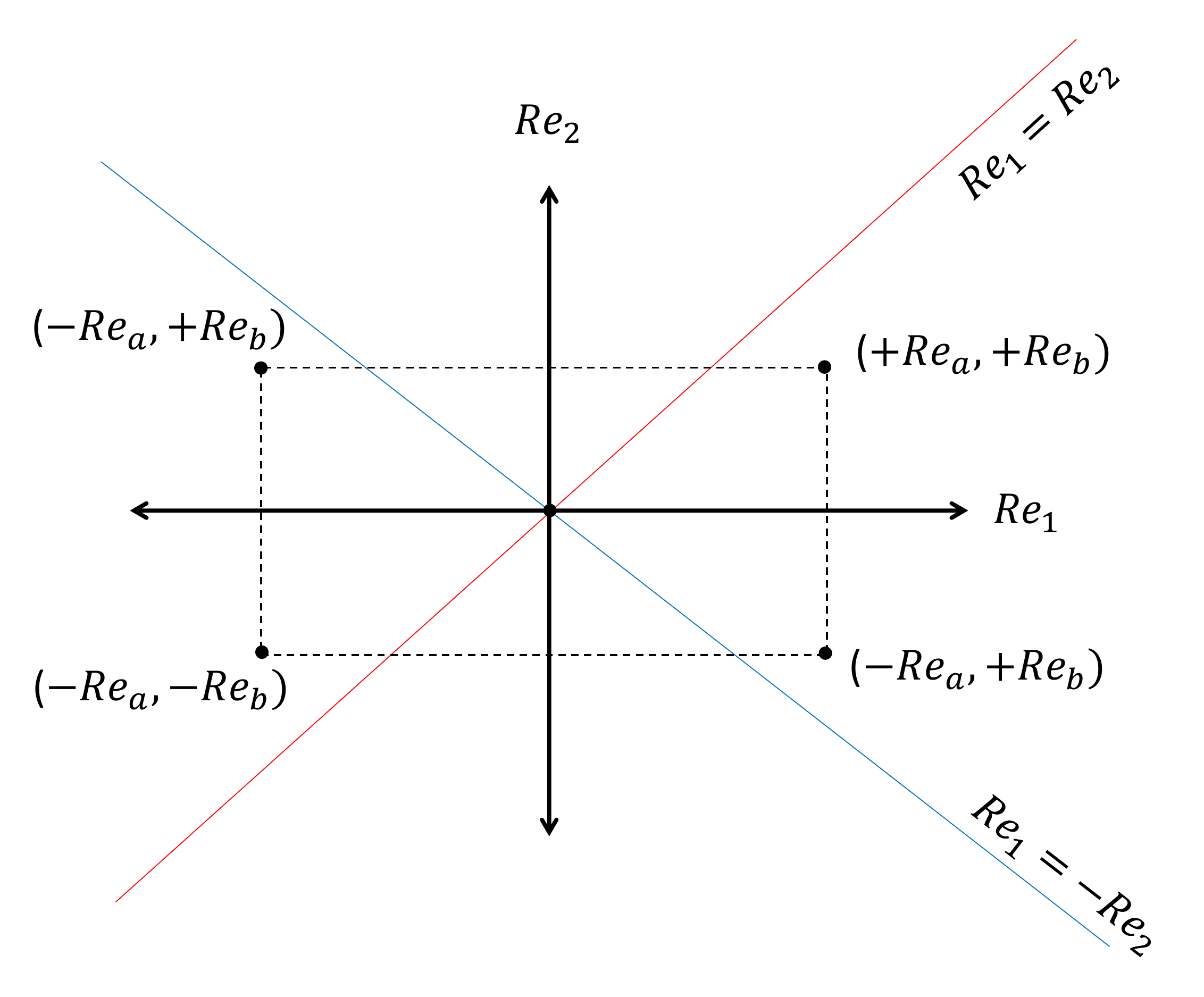}
    \caption{Classification of steady-states of the CNCL system based on the quadrant they lie in.}
    \label{Fig12}
\end{figure}

The CNCL systems have multiple steady states, and depending on the magnitude of the non-dimensional numbers, the number of steady states varies as observed in section 5.1. In order to plot the stability maps, it is necessary to find the stability of steady states which are consistently observed irrespective of the non-dimensional numbers chosen and the steady states which are observed in practice, so that the stability map can be used as an appropriate design tool for CNCL systems. From section 5.1, it can be observed that a CNCL system has multiple steady states, and they are classified as represented in figure \ref{Fig12} and can be in any of the four quadrants of the $Re_1 Re_2$ plane.


Here, the signs attached to $Re_1$ and $Re_2$ indicate the flow directions in each of the CNCL loops (refer to section 5.1 for more details). Knowing the values of $Re_1$ and $Re_2$, the steady state magnitudes of other parameters can be estimated. From section 5.1, we observe that the steady states corresponding to $Re_1=Re_2$ and $Re_1=-Re_2$ in the $Re_1 Re_2$ plane are consistently observed. Figure \ref{Fig11} presents the steady state magnitudes of $Re_1$ and $Re_2$ with varying $Gr$, which indicates that only the steady state values corresponding to $Re_1=-Re_2$ or $Re_1/Re_2=-1$ i.e. the steady states $(Re_1,Re_2)=(-Re_a,+Re_b)$, $(Re_1,Re_2)=(+Re_a,-Re_b)$ which lie on the line $Re_1=-Re_2$ are observed consistently. Thus, the stability of the steady states which lie on the line $Re_1=-Re_2$ in the $Re_1 Re_2$ plane are evaluated and their stability map is plotted. $Re_1/Re_2=-1$ implies a counterflow configuration at the common heat exchange section. This observation is consistent with the observation that a CNCL with the heater cooler configuration chosen for the present study always exhibits a counterflow configuration at the common heat exchange section \cite{dass2019}.

\subsection{Linearisation of the CNCL system}

The governing equations of the CNCL system are represented by a set of non-linear and coupled set of ODE as described by equations (31)-(34). The ODE stencil of the CNCL system represented by equations (31)-(34) can be represented in matrix format as follows:

\begin{equation}
    \frac{d [A]}{dt} = [B]
\end{equation}

where

\begin{equation}
    [A]= \begin{bmatrix}
Re_1 \\
\theta_{1,k} \\
Re_2 \\
\theta_{2,k}
\end{bmatrix}
\end{equation}

and

\begin{equation}
    [B]= \begin{bmatrix}
\big [ Gr \big ] \sum_{k=-\infty}^{\infty}(\theta_{1,k}) f_{-k}  - \big [ Co_1 \big] (Re_1)^{2-d} -\frac{nK}{4}(Re_1)^2 \\
-k^2 {\pi}^2 \big[ Fo \big] \theta_{1,k} + h_{1,k} - \big[     St \big] \sum_{l=-\infty}^{\infty} \lambda_{k-l} \big(\theta_{1,l}- \theta_{2,l} \big) - i k \pi Re_1\theta_{1,k} \\
\big [ Gr \big ] \sum_{k=-\infty}^{\infty}(\theta_{2,k}) f_{-k} - \big [ Co_1 \big] (Re_2)^{2-d} -\frac{nK}{4}(Re_2)^2 \\ 
-k^2 {\pi}^2 \big[ Fo \big] \theta_{2,k} + h_{2,k} + \big[     St \big] \sum_{l=-\infty}^{\infty} \lambda_{k-l} \big(\theta_{1,l}- \theta_{2,l} \big) - i k \pi Re_2\theta_{2,k} 
\end{bmatrix}
\end{equation}

After expanding the stencil for $N$ Fourier nodes, i.e. $k=0,1,2,...,N$ and separating the real and imaginary components, reorganising the equations and linearisation, we obtain:

\begin{equation}
    \frac{d [A]^{\prime}}{dt} = [B]^{\prime}_J 
\end{equation}

where $[B]^{\prime}_J$ is the Jacobian matrix of $[B]^{\prime}$ and

\begin{equation}
    [A]^{\prime}= \begin{bmatrix}
Re_1 \\
\theta_{1,0} + \theta_{2,0} \\
\theta_{1,{1_R}} \\
\theta_{1,{1_I}} \\
\vdots\\
\theta_{1,{N_R}} \\
\theta_{1,{N_I}} \\
Re_2 \\
\theta_{2,{1_R}} \\
\theta_{2,{1_I}} \\
\vdots\\
\theta_{2,{N_R}} \\
\theta_{2,{N_I}} \\
\end{bmatrix}
\end{equation}

\begin{equation}
    [B]^{\prime}= \begin{bmatrix}
\big [ Gr \big ] \sum_{k=-N}^{N}(\theta_{1,k}) f_{-k}  - \big [ Co_1 \big] (Re_1)^{2-d} -\frac{nK}{4}(Re_1)^2 \\
h_{1,0} - h_{2,0}  - \big[     2St \big] \big[ \sum_{l=-N}^{N} \lambda_{0-l} \big(\theta_{1,l}- \theta_{2,l} \big)\big]_R \\
- {\pi}^2 \big[ Fo \big] \theta_{1,1_R} + h_{1,1_R} - \big[     St \big] \big[ \sum_{l=-N}^{N} \lambda_{1-l} \big(\theta_{1,l}- \theta_{2,l} \big)\big]_R +  \pi Re_1\theta_{1,1_I} \\
- {\pi}^2 \big[ Fo \big] \theta_{1,1_I} + h_{1,1_I} - \big[     St \big] \big[ \sum_{l=-N}^{N} \lambda_{1-l} \big(\theta_{1,l}- \theta_{2,l} \big)\big]_I -  \pi Re_1\theta_{1,1_R} \\
\vdots\\
- N^2{\pi}^2 \big[ Fo \big] \theta_{1,N_R} + h_{1,N_R} - \big[     St \big] \big[ \sum_{l=-N}^{N} \lambda_{N-l} \big(\theta_{1,l}- \theta_{2,l} \big)\big]_R +  N \pi Re_1\theta_{1,N_I} \\
- N^2{\pi}^2 \big[ Fo \big] \theta_{1,N_I} + h_{1,N_I} - \big[     St \big] \big[ \sum_{l=-N}^{N} \lambda_{N-l} \big(\theta_{1,l}- \theta_{2,l} \big)\big]_I - N \pi Re_1\theta_{1,N_R} \\
\big [ Gr \big ] \sum_{k=-N}^{N}(\theta_{2,k}) f_{-k} - \big [ Co_1 \big] (Re_2)^{2-d} -\frac{nK}{4}(Re_2)^2 \\ 
-{\pi}^2 \big[ Fo \big] \theta_{2,1_R} + h_{2,1_R} + \big[     St \big] \big[ \sum_{l=-N}^{N} \lambda_{1-l} \big(\theta_{1,l}- \theta_{2,l} \big)\big]_R + \pi Re_2\theta_{2,1_I} \\
-{\pi}^2 \big[ Fo \big] \theta_{2,1_I} + h_{2,1_I} + \big[     St \big] \big[ \sum_{l=-N}^{N} \lambda_{1-l} \big(\theta_{1,l}- \theta_{2,l} \big)\big]_I - \pi Re_2\theta_{2,1_R} \\
\vdots \\
-N^2{\pi}^2 \big[ Fo \big] \theta_{2,N_R} + h_{2,N_R} + \big[     St \big] \big[ \sum_{l=-N}^{N} \lambda_{N-l} \big(\theta_{1,l}- \theta_{2,l} \big)\big]_R + N \pi Re_2\theta_{2,N_I} \\
-N^2 {\pi}^2 \big[ Fo \big] \theta_{2,N_I} + h_{2,N_I} + \big[     St \big] \big[ \sum_{l=-N}^{N} \lambda_{N-l} \big(\theta_{1,l}- \theta_{2,l} \big)\big]_I - N \pi Re_2\theta_{2,N_R}
\end{bmatrix}
\end{equation}

  The eigenvalues of $[B]^{\prime}_J$ for a particular steady-state enable us to determine if the steady-state exists or not. If the value of the maximum of the real part of the eigenvalue of $[B]^{\prime}_J$ is negative, then the steady-state is stable. If the value of the maximum of the real part of the eigenvalue of $[B]^{\prime}_J$ is positive, then the steady-state is unstable.

\section{Validation of methodology used for generating Stability maps}

\begin{figure}[!htb]
	\centering
	\begin{subfigure}[b]{0.49\textwidth}
		\includegraphics[width=1\linewidth]{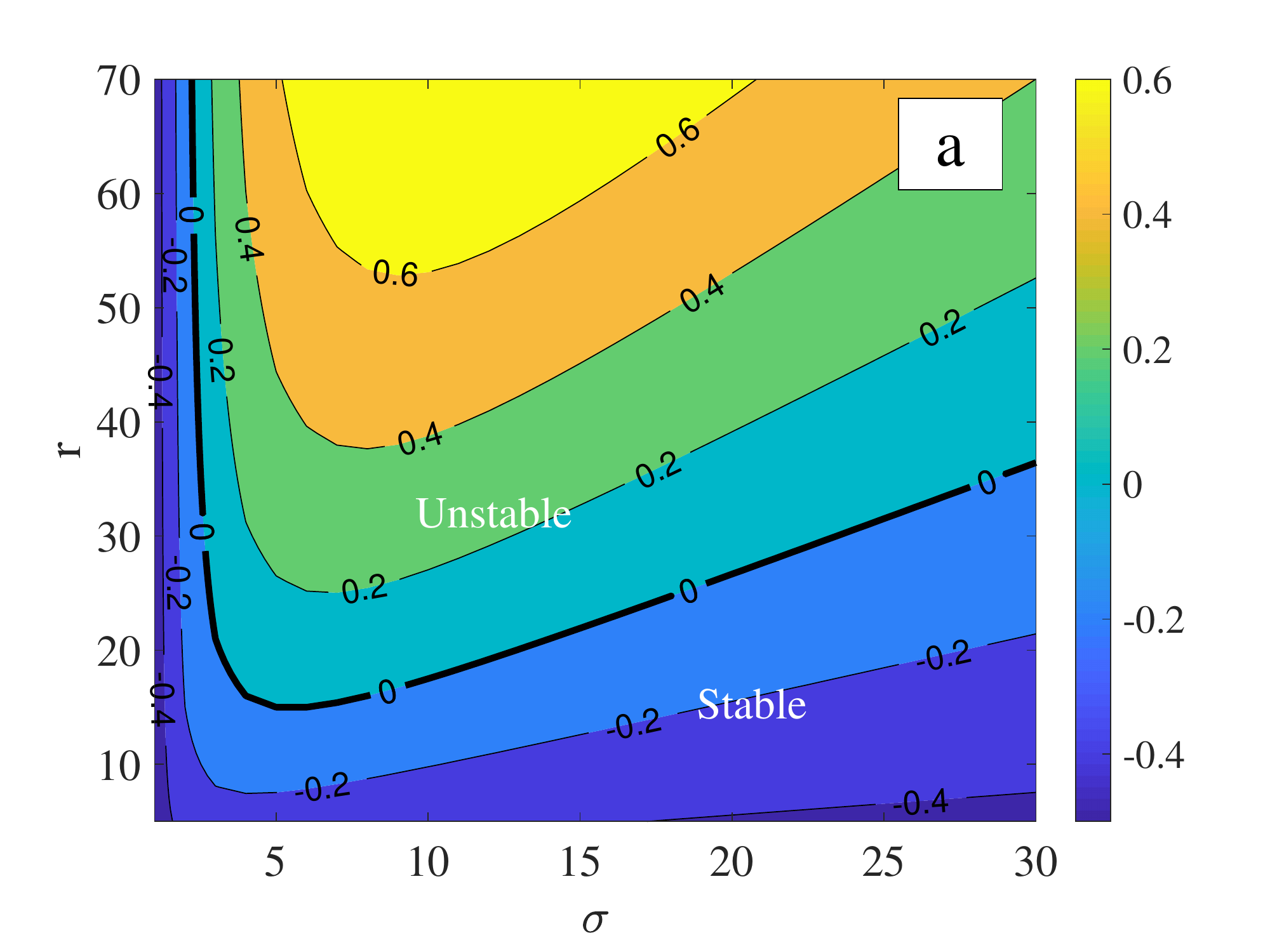}
	\end{subfigure}
	\hspace{\fill}
	\begin{subfigure}[b]{0.49\textwidth}
		\includegraphics[width=1\linewidth]{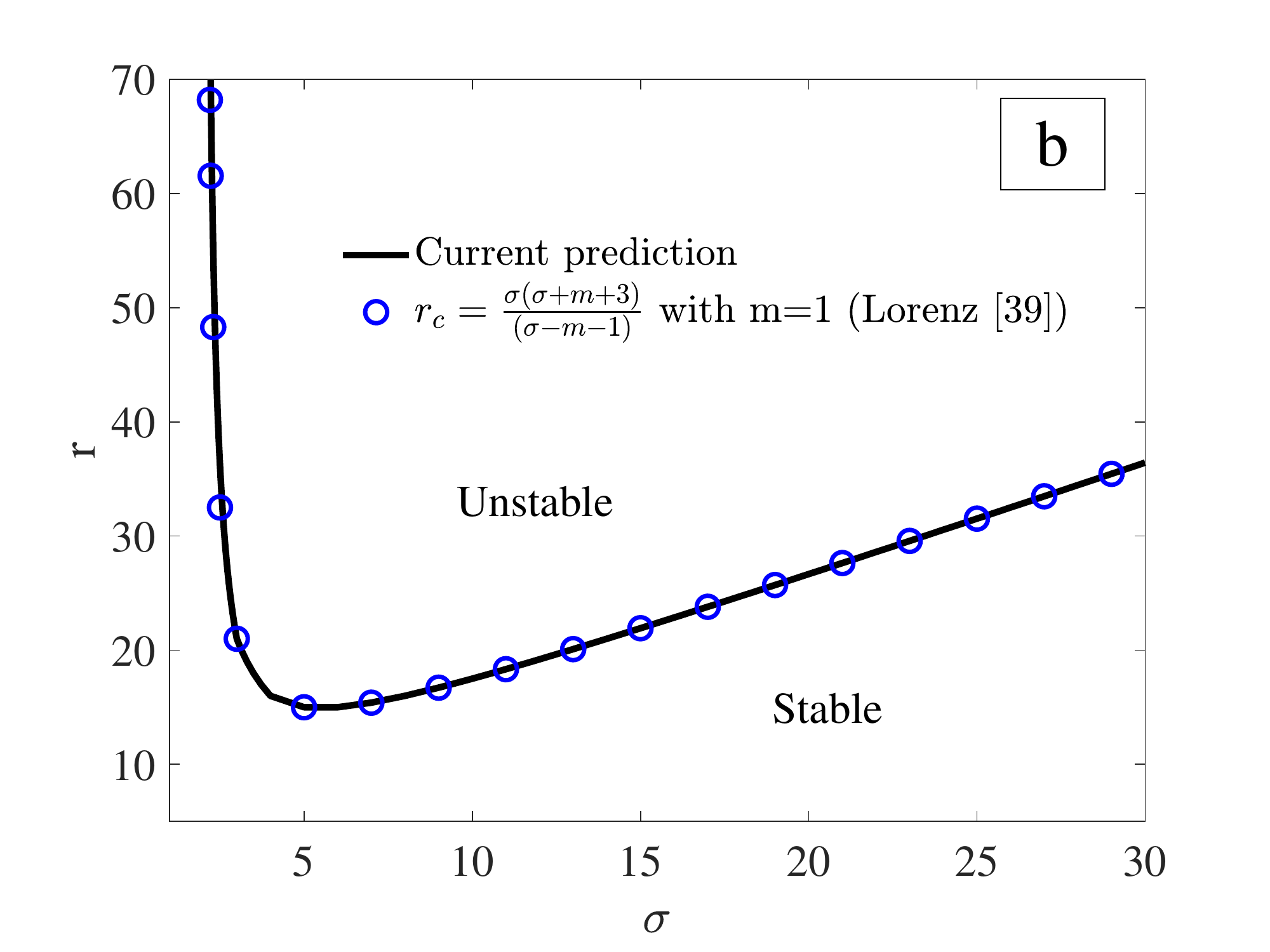}
	\end{subfigure}
	\caption{Assessment of the approach used to generate the stability map, (a) Stability map of the Lorenz system indicating stable and unstable regions, and (b) Validation of the stability boundary predicted using the current approach.}
	\label{Fig13}
\end{figure}

The current study utilises a contour map of the eigenvalues for a set of non-dimensional numbers to identify the regions of stability, instability and the stability boundary of the CNCL system.  The ability of this approach to capture the stability boundary is assessed by comparing the stability map generated using the contour maps with the known stability boundary for a particular ODE system. Thus, the stability map of the Lorenz system \cite{lorenz1963deterministic} is generated and compared against the prediction by Lorenz \cite{lorenz1963deterministic}. The results of the comparison are presented in figure \ref{Fig13}, and a good agreement is observed indicating the suitability of the methodology to determine the stability boundary accurately.

\section{Validation of 1-D CNCL model stability prediction with 3-D CFD results}

\begin{table}[!htb]
\centering
\caption{Validation of 1-D CNCL model with 3-D CFD results.}
\scalebox{0.85}{
\begin{tabular}{|c|c|c|c|c|c|c|c|c|c|c|}
\hline
Si.no & Description                                                                             & 3-D CFD case                                                      & \begin{tabular}[c]{@{}c@{}}Transient behaviour\\ exhibited by the system\end{tabular} & $Gr$                         & $Co_1$                & $Fo$                 & $As$              & \begin{tabular}[c]{@{}c@{}}Eigen value at \\ steady state from \\ the 1-D model\end{tabular} & \begin{tabular}[c]{@{}c@{}}Prediction from \\ the eigen value\end{tabular} \\ \hline
1     & \multirow{3}{*}{\begin{tabular}[c]{@{}c@{}}Dass and \\ Gedupudi \\ \cite{dass2019}\end{tabular}} & \begin{tabular}[c]{@{}c@{}}VCNCL\\ Counter flow\end{tabular}  & Attains steady state                                                                  & \multirow{3}{*}{$3.59 \times 10^{11}$} & \multirow{3}{*}{1423} & \multirow{3}{*}{800} & \multirow{4}{*}{1} & $-4.14 \times 10^{3}$                                                                     & Stable                                                                 \\ \cline{1-1} \cline{3-4} \cline{9-10} 
2     &                                                                                         & \begin{tabular}[c]{@{}c@{}}HCNCL\\ Parallel flow\end{tabular} & Attains steady state                                                                  &                              &                       &                      &                    & $-2.96 \times 10^{3}$                                                                   & Stable                                                                 \\ \cline{1-1} \cline{3-4} \cline{9-10} 
3     &                                                                                         & \begin{tabular}[c]{@{}c@{}}HCNCL\\ Counter flow\end{tabular}  & Attains steady state                                                                  &                              &                       &                      &                    & $-3.46 \times 10^{3}$                                                                    & Stable                                                                 \\ \cline{1-7} \cline{9-10} 
4     & \begin{tabular}[c]{@{}c@{}}Present\\  work\end{tabular}                                                                            & VCNCL                                                         & \begin{tabular}[c]{@{}c@{}}Exhibits chaotic\\  behaviour\end{tabular}                                                            &   $1.26 \times 10^{9}$
                    & 2846                  & 0.0001               &                    & $+47.69$                                                                    & Unstable                                                              \\ \hline
\end{tabular}}
\end{table}

Table 2 presents the validation of the 1-D CNCL model to predict the CNCL system stability from the existing literature and the present 3-D CFD study. The predicted eigenvalue from the 1-D model of the CNCL system represented in Table 2 accurately denotes whether the CFD case has a stable steady-state (eigenvalue is negative and a steady-state is attained) or unstable (eigenvalue is positive and a steady-state is not attained). Results from table 2 validate the ability of the linear stability analysis of the CNCL system from the 1-D model to predict the nature of long-term transient behaviour.

\section{Fourier node independence test employing stability map}

\begin{figure}[!htb]
	\centering
	\begin{subfigure}[b]{0.49\textwidth}
		\includegraphics[width=1\linewidth]{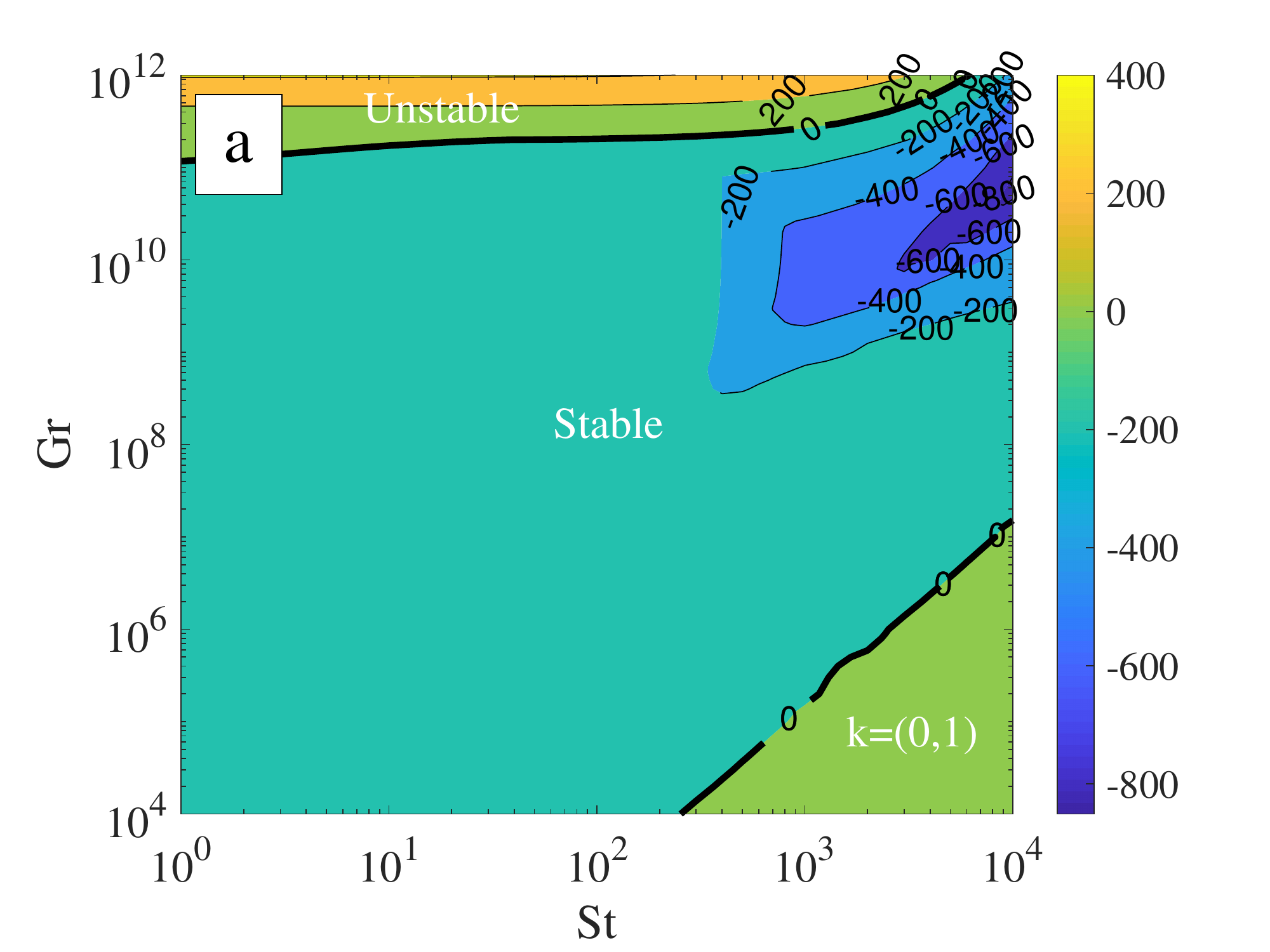}
	\end{subfigure}
	\hspace{\fill}
	\begin{subfigure}[b]{0.49\textwidth}
		\includegraphics[width=1\linewidth]{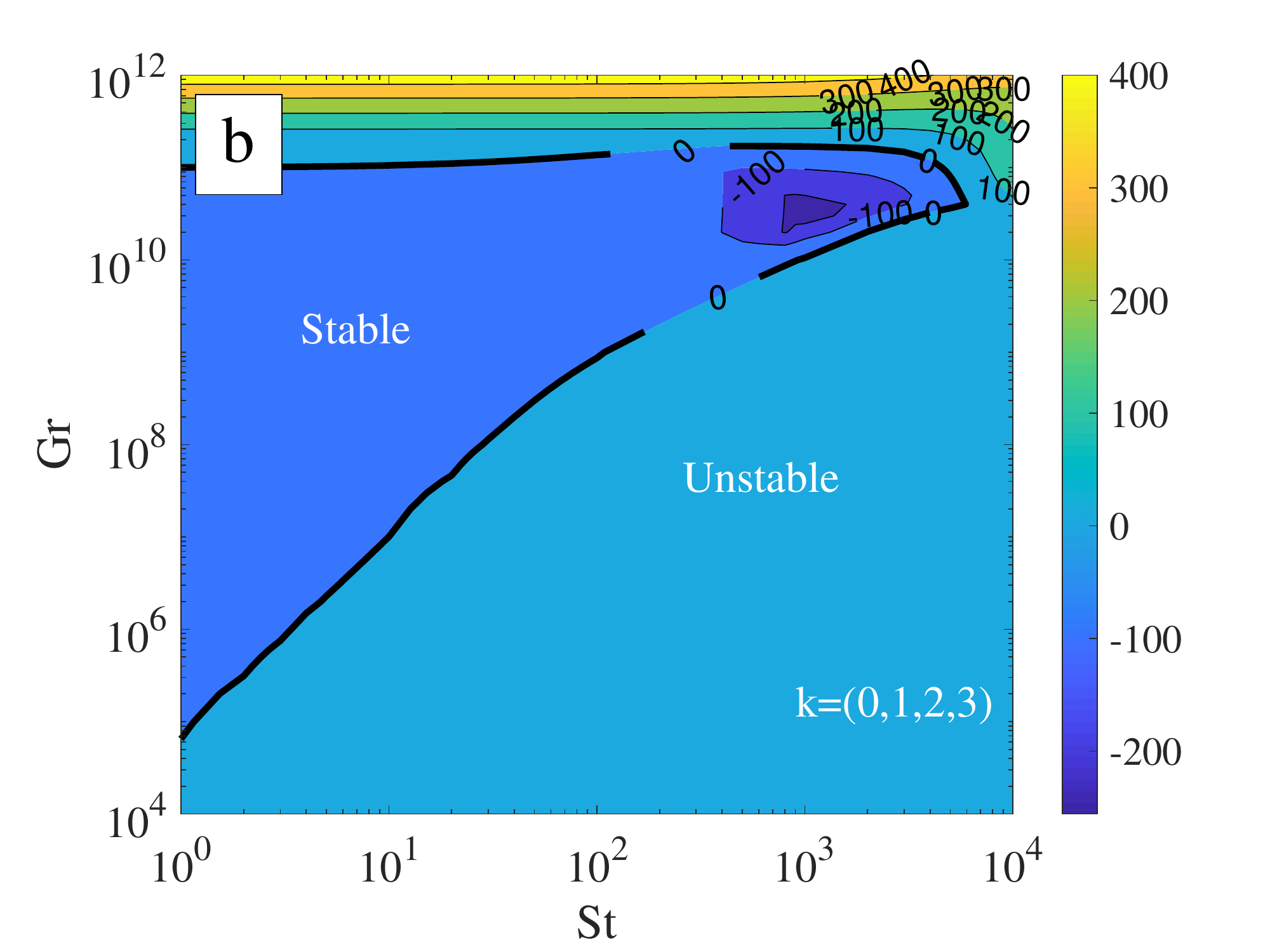}
	\end{subfigure}
	\hspace{\fill}
	\begin{subfigure}[b]{0.49\textwidth}
		\includegraphics[width=1\linewidth]{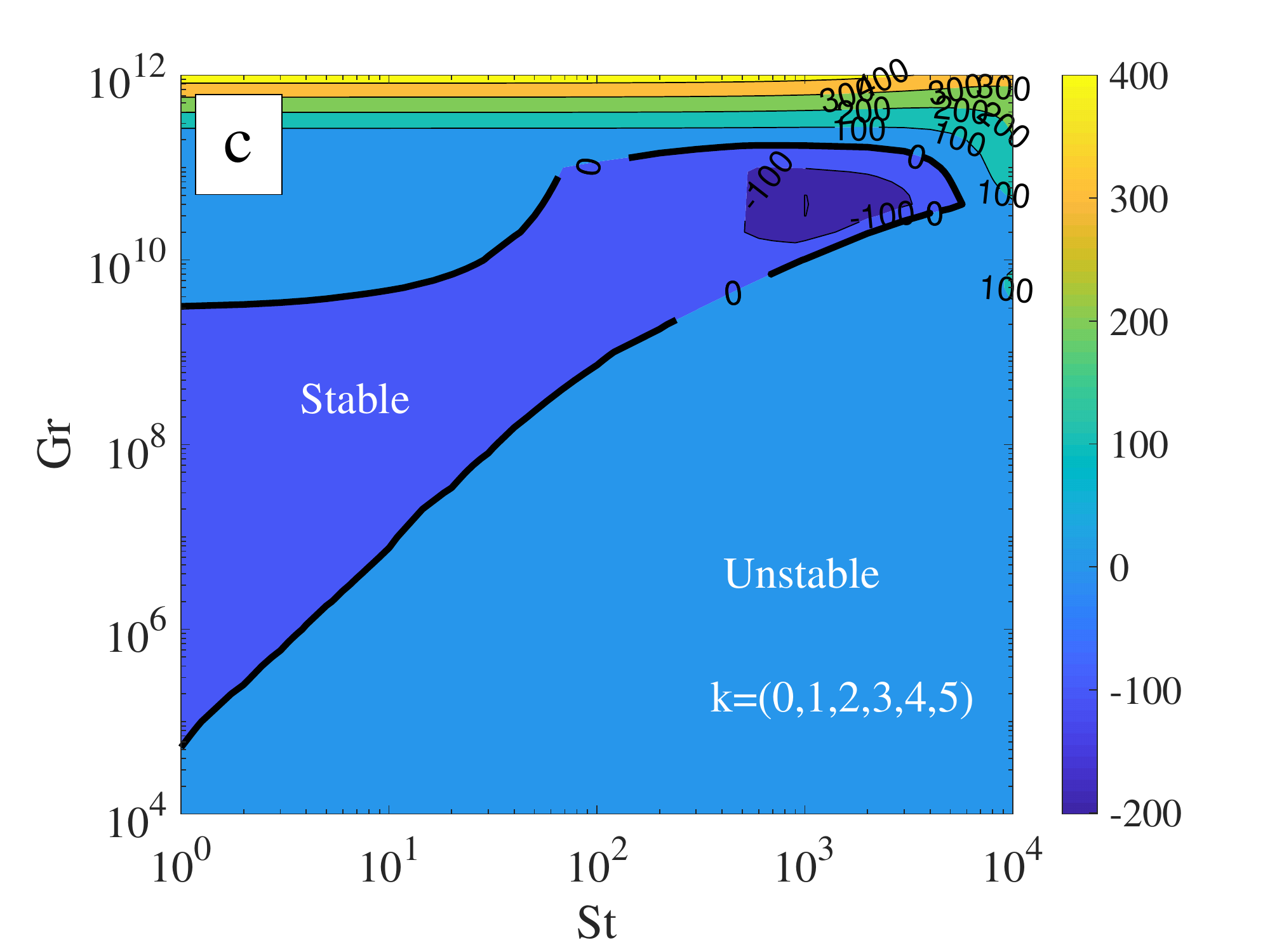}
	\end{subfigure}
	\hspace{\fill}
	\begin{subfigure}[b]{0.49\textwidth}
		\includegraphics[width=1\linewidth]{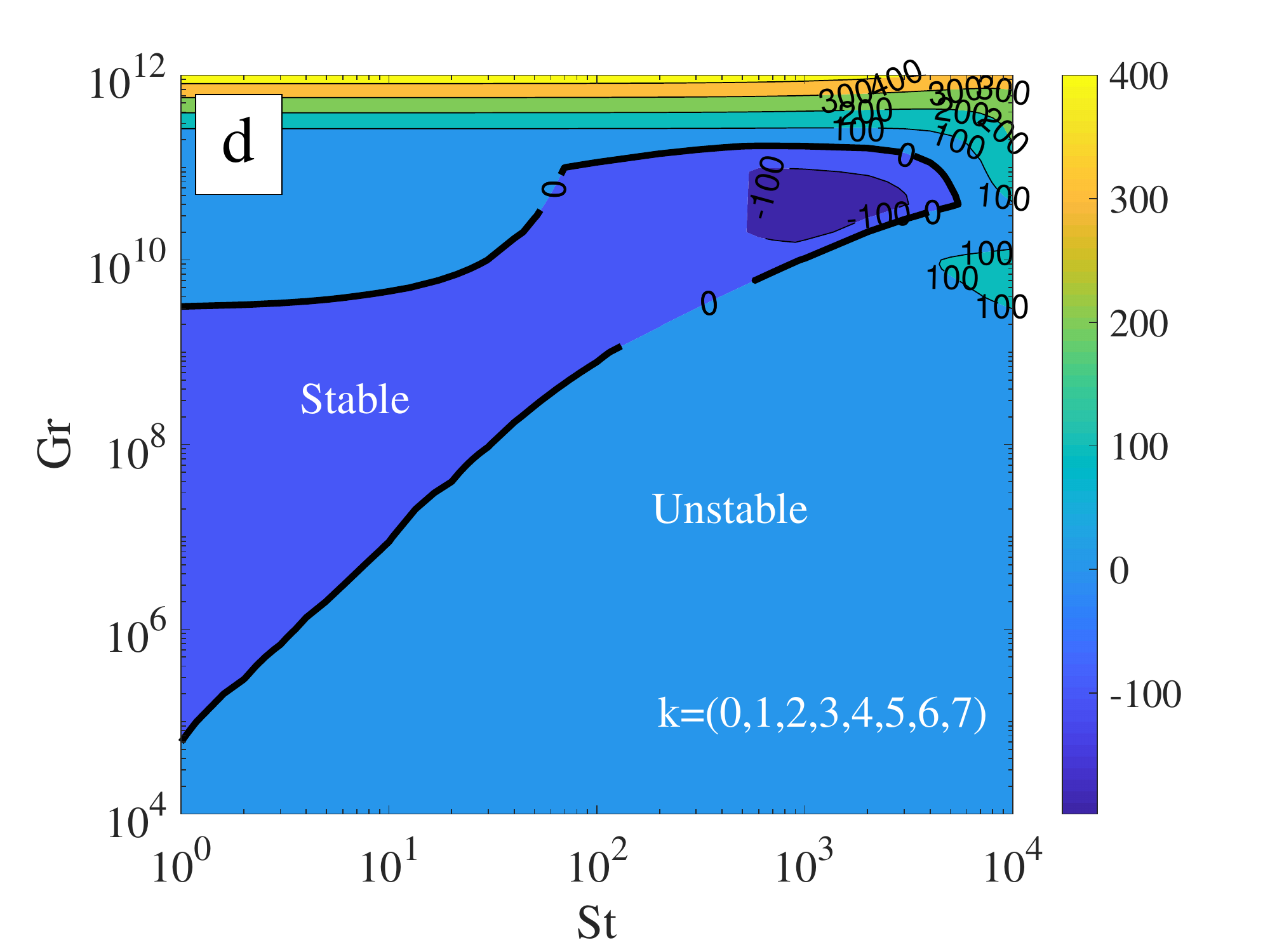}
	\end{subfigure}
	\caption{Stability map of the CNCL system considered for the 3-D CFD study for the steady state $(Re_1,Re_2)=(-Re_a,+Re_b)$ which lie on the line $Re_1=-Re_2$, (a) Stability map of CNCL employing  truncated ode system with $k=(0,1)$, (b) Stability map of CNCL employing  truncated ode system with $k=(0,1,2,3)$, (c) Stability map of CNCL employing  truncated ode system with $k=(0,1,2,3,4,5)$, and (d) Stability map of CNCL employing  truncated ode system with $k=(0,1,2,3,4,5,6,7)$, for $Fo=0.0001$, $As=1$, $Co_1=2846$ (see Table 3 for calculation).}
	\label{Fig14}
\end{figure}

The present section studies the influence of Fourier node truncation on the stability of CNCL systems. If adequate nodes are not considered, then the physics of the system is not accurately captured. A Fourier node independence test is carried out employing the stability map to determine the number of nodes necessary to capture the complete physics of the CNCL system. Figure \ref{Fig14} represents the Fourier node independence test employing a stability map for the 3-D CFD case of the current paper. It can be noted from figure 14 that a CNCL system with $N=5$ is adequate to fully represent the physics for $Gr$ varying from $10^4$ to $10^{12}$ and $St$ varying from $10^0$ to $10^4$. For the same range of $Gr$ if $St$ were considered only from $10^2$ to $10^4$, then CNCL system with $N=3$ would have been adequate as mentioned in section 3.6. However, as the CNCL is a simplified version of Passive Residual Heat Removal System (PRHRS), it is necessary to capture its system behaviour at low heat transfer conditions, and hence $St$ is considered from $10^0$ to $10^4$.

\section{Effect of resolution on the prediction of stability boundary}

\begin{figure}[!htb]
	\centering
	\begin{subfigure}[b]{0.49\textwidth}
		\includegraphics[width=1\linewidth]{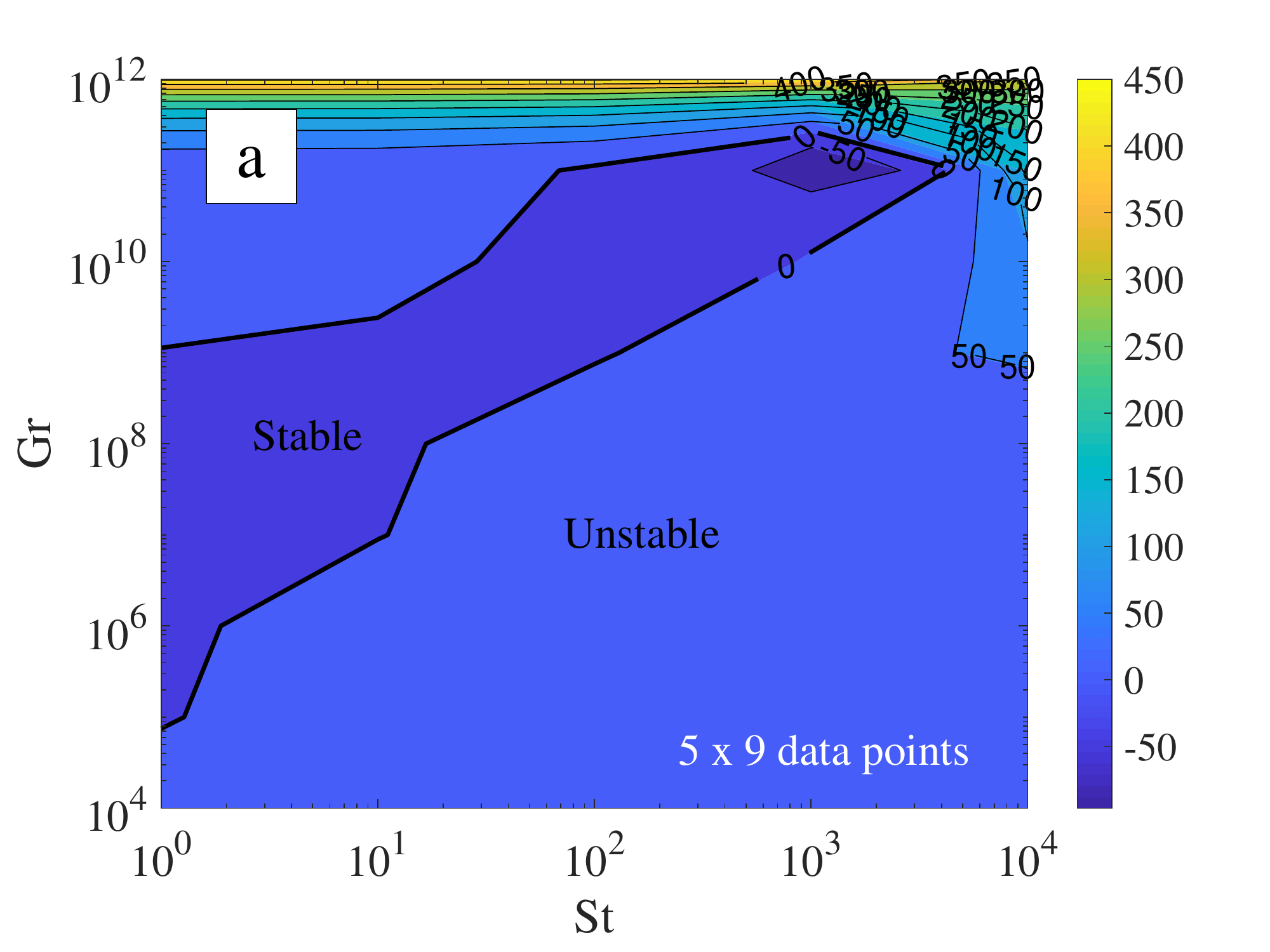}
	\end{subfigure}
	\hspace{\fill}
	\begin{subfigure}[b]{0.49\textwidth}
		\includegraphics[width=1\linewidth]{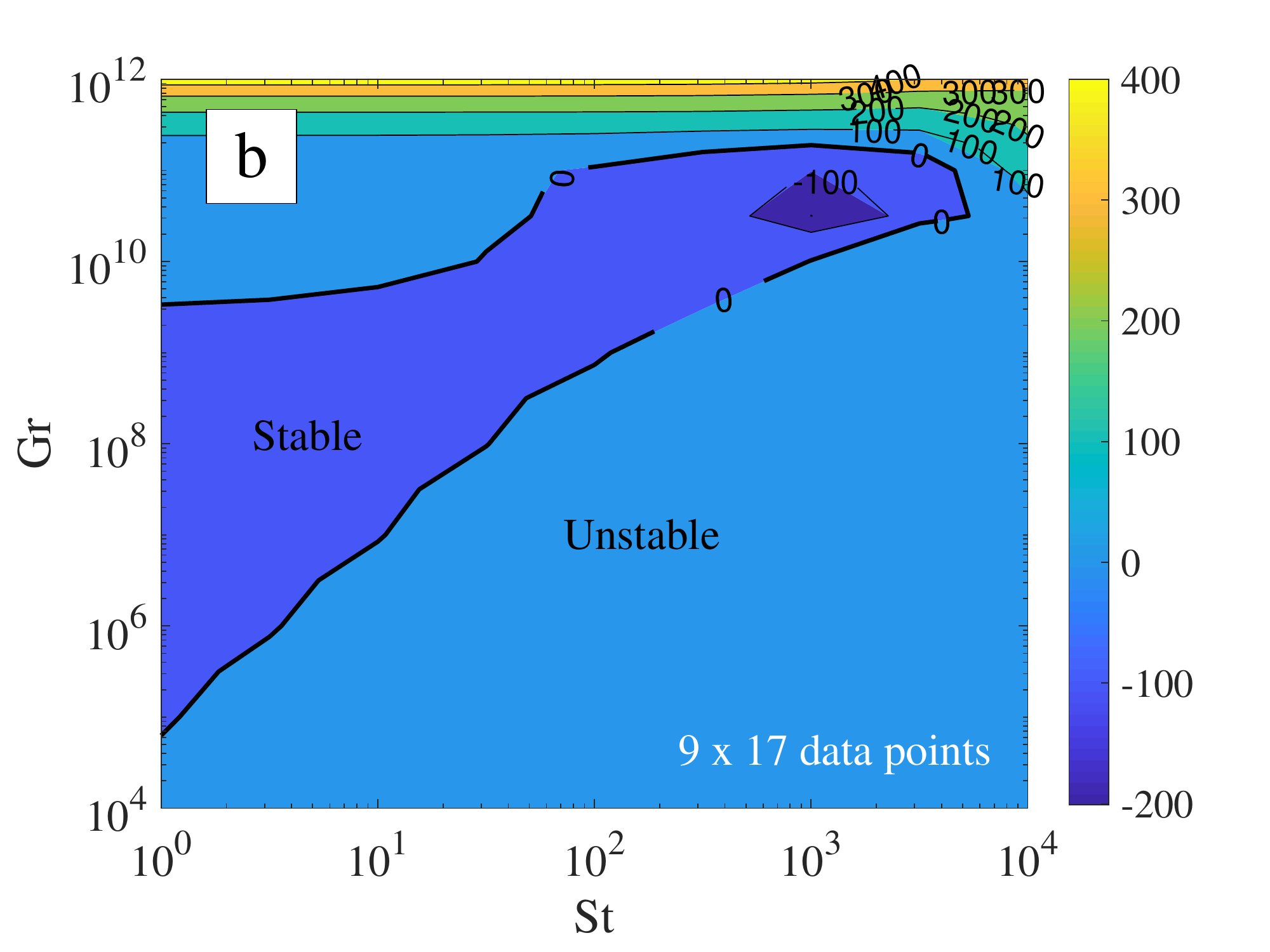}
	\end{subfigure}
	\hspace{\fill}
	\begin{subfigure}[b]{0.49\textwidth}
		\includegraphics[width=1\linewidth]{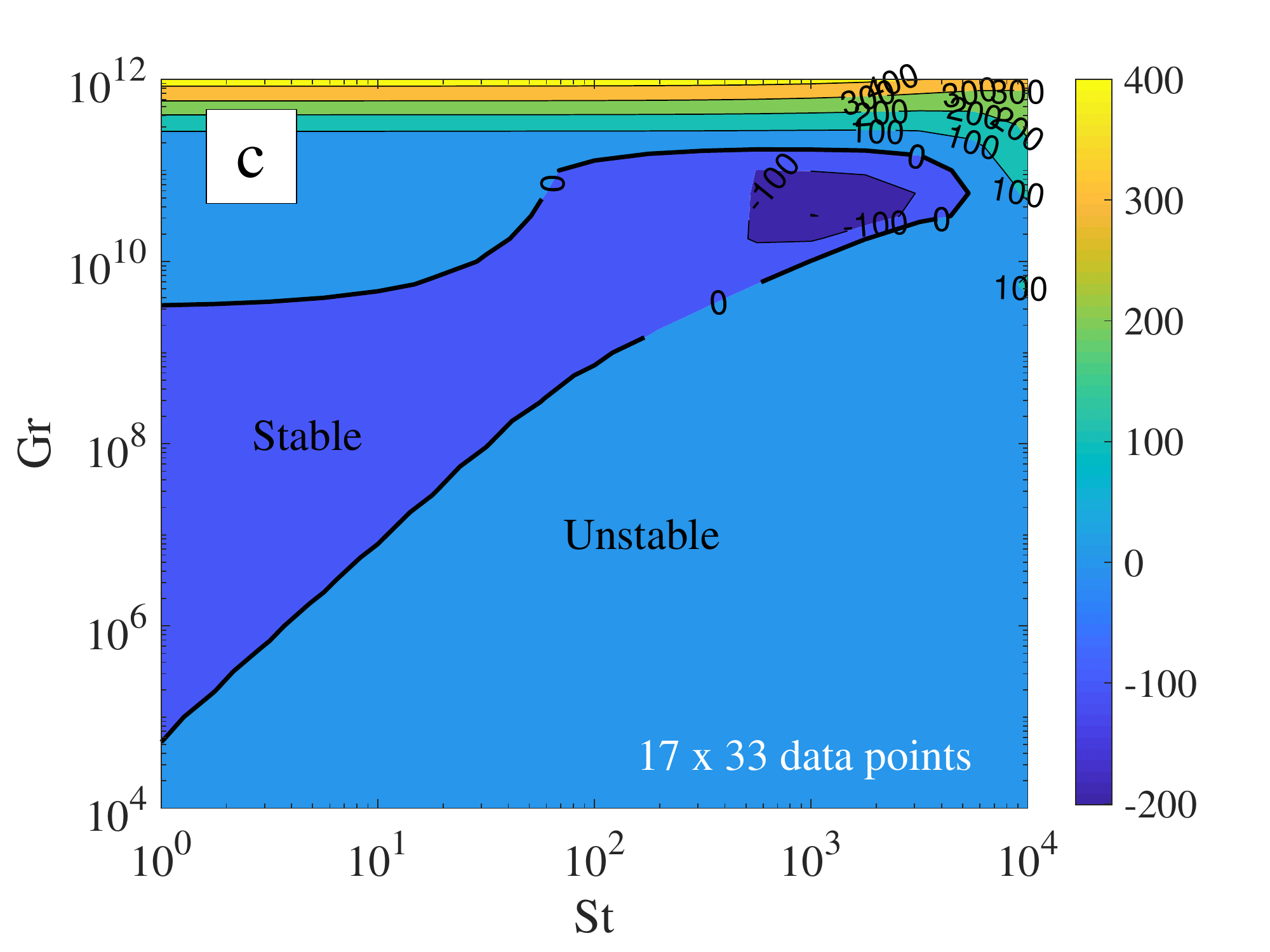}
	\end{subfigure}
	\hspace{\fill}
	\begin{subfigure}[b]{0.49\textwidth}
		\includegraphics[width=1\linewidth]{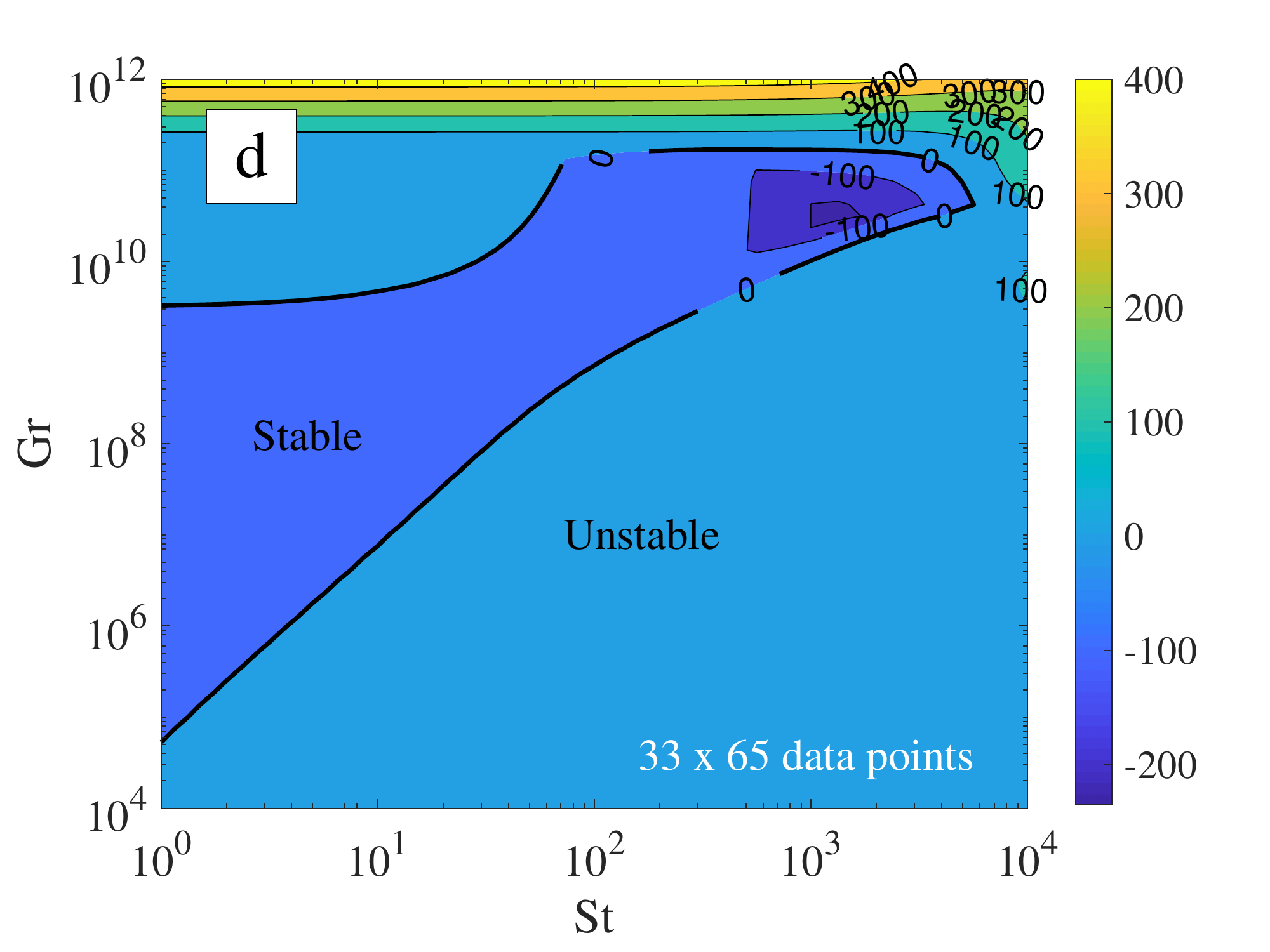}
	\end{subfigure}
	\hspace{\fill}
	\begin{subfigure}[b]{0.49\textwidth}
		\includegraphics[width=1\linewidth]{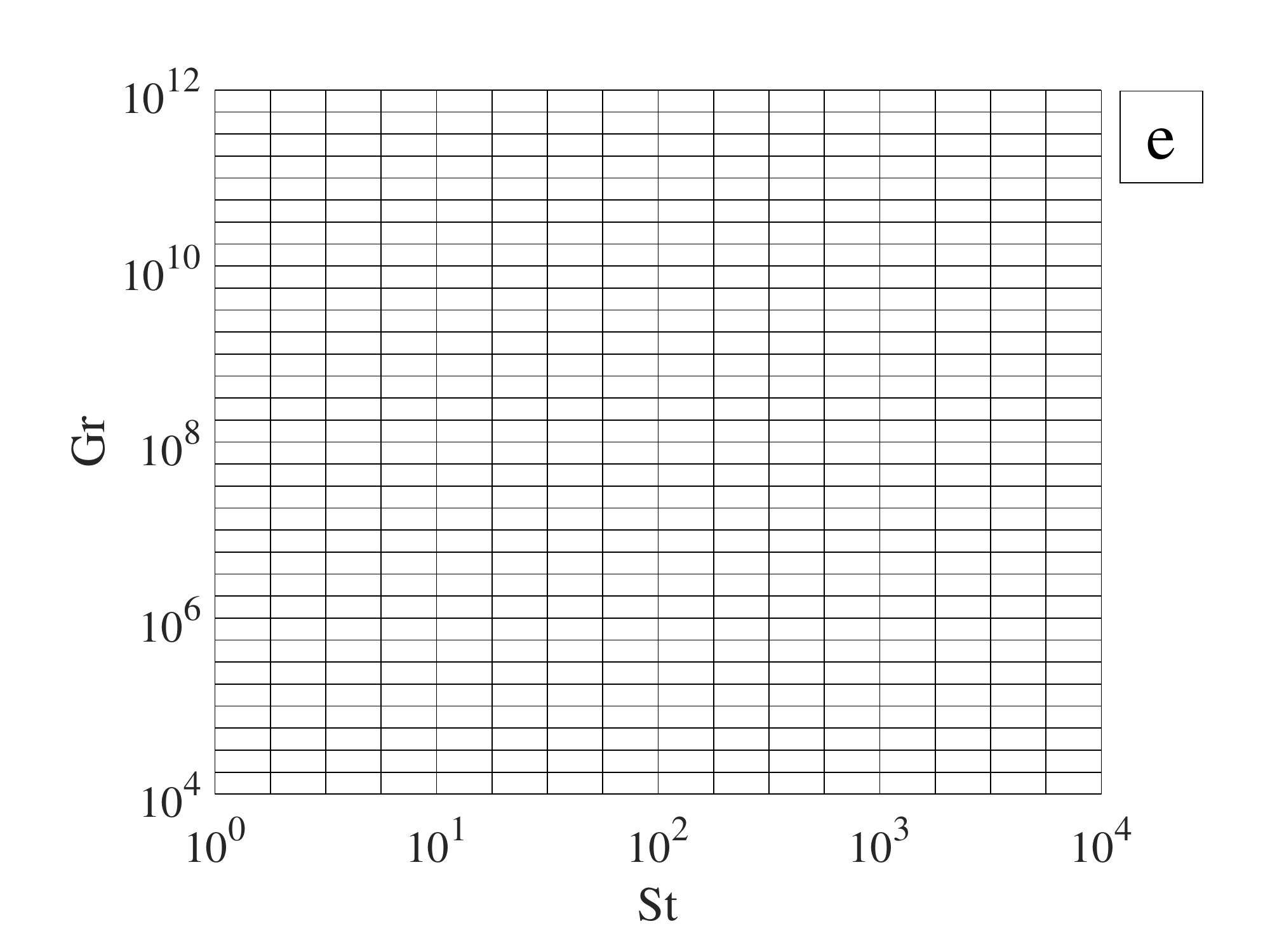}
	\end{subfigure}
	\caption{Effect of contour map resolution on the prediction of stability boundary of the CNCL system considered for the 3-D CFD study for the steady state $(Re_1,Re_2)=(-Re_a,+Re_b)$ which lie on the line $Re_1=-Re_2$, (a) Contour plot with $5 \times 9$ data points, (b) Contour plot with $9 \times 17$ data points, (c) Contour plot with $17 \times 33$ data points, (d) Contour plot with $33 \times 65$ data points, and (e) Gridded mesh with $17 \times 33$ data points used to obtain the contour plot, for $Fo=0.0001$, $As=1$, $Co_1=2846$ (see Table 3 for calculation).}
	\label{Fig15}
\end{figure}

As the current approach utilises a contour map to obtain the stability boundary of the CNCL system, the number of data points used to obtain the contour map also becomes relevant. From figure \ref{Fig15}, it can be observed that a stability map obtained with $17 \times 33$ data points and that with $33 \times 65$ data points are identical, implying that $17 \times 33$ data points is the minimum resolution required to identify the stability boundary accurately. All the stability maps plotted for the analysis in the upcoming sections have a minimum resolution of $17 \times 33$ data points. Figure 15(e) represents the uniformly spaced data points in the log-log graph used to obtain the stability map with 17 data points for $St$ and $33$ data points for $Gr$.

\section{Results and discussion}

The CNCL systems may be classified as vertical or horizontal depending on the orientation of the common heat exchange section with respect to gravity. If the common heat exchange section is parallel w.r.t. gravity, then it is termed as a Vertical CNCL (VCNCL). The CNCL utilised for the 3-D CFD study is a VCNCL system. If the common heat exchange section is perpendicular w.r.t. gravity then it is referred to as a Horizontal CNCL (HCNCL). The present section discusses in detail the stability of VCNCL and HCNCL systems and then a thorough parametric study is conducted employing the VCNCL system to investigate the influence of non-dimensional numbers $Fo$, $As$ and $Co_1$ on the stability of such systems. The stability maps of the CNCL system presented in the current paper are the $Gr-St$ stability maps with $Fo$, $As$ and $Co_1$ kept constant.

\subsection{Stability analysis of VCNCL system}

The present section focuses on the stability analysis of VCNCL systems and presents the stability maps of its various steady states.

\subsubsection{Steady state solutions of VCNCL systems}

\begin{figure}[!htb]
	\centering
	\begin{subfigure}[b]{0.49\textwidth}
		\includegraphics[width=1\linewidth]{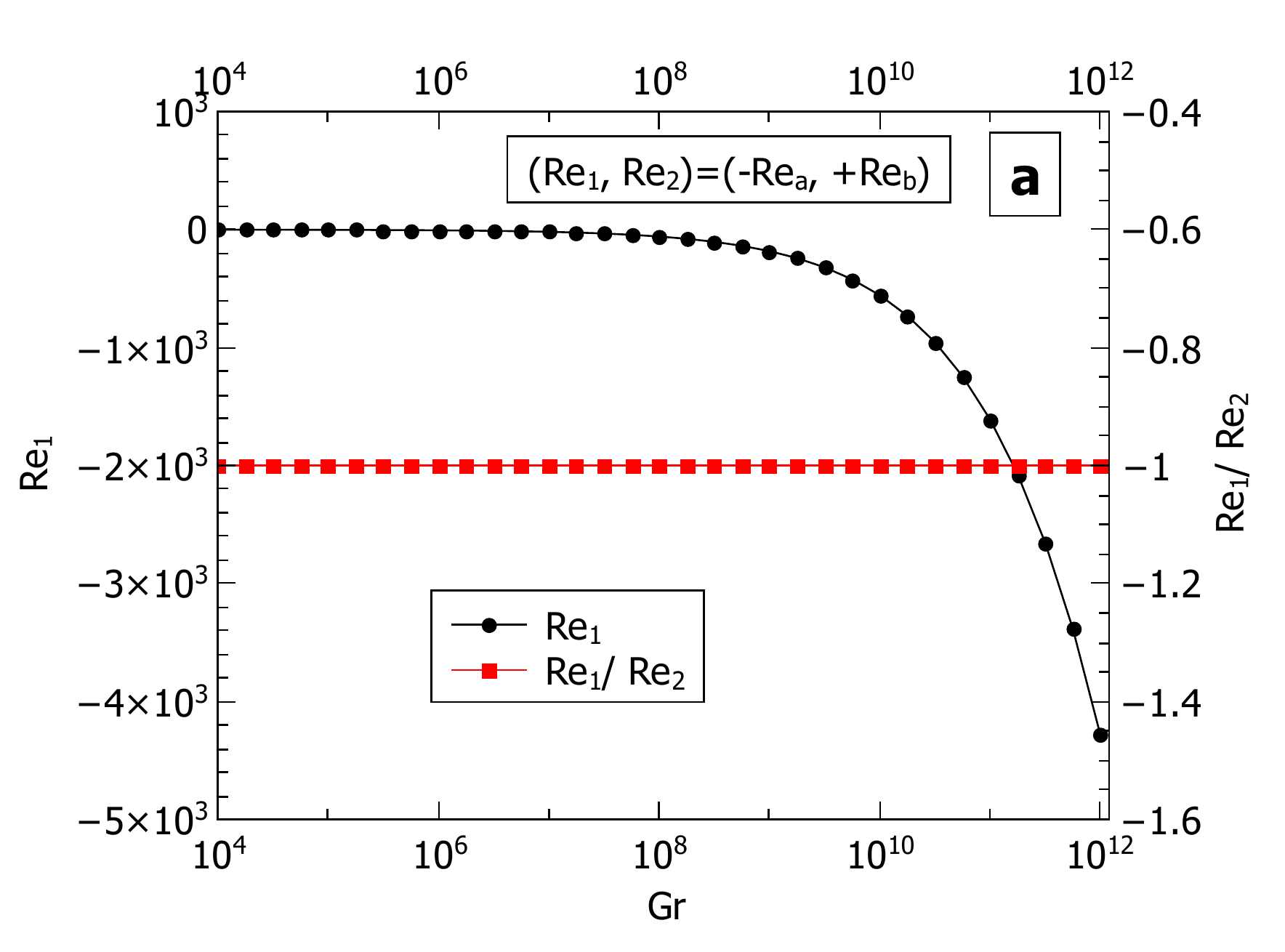}
	\end{subfigure}
	\hspace{\fill}
	\begin{subfigure}[b]{0.49\textwidth}
		\includegraphics[width=1\linewidth]{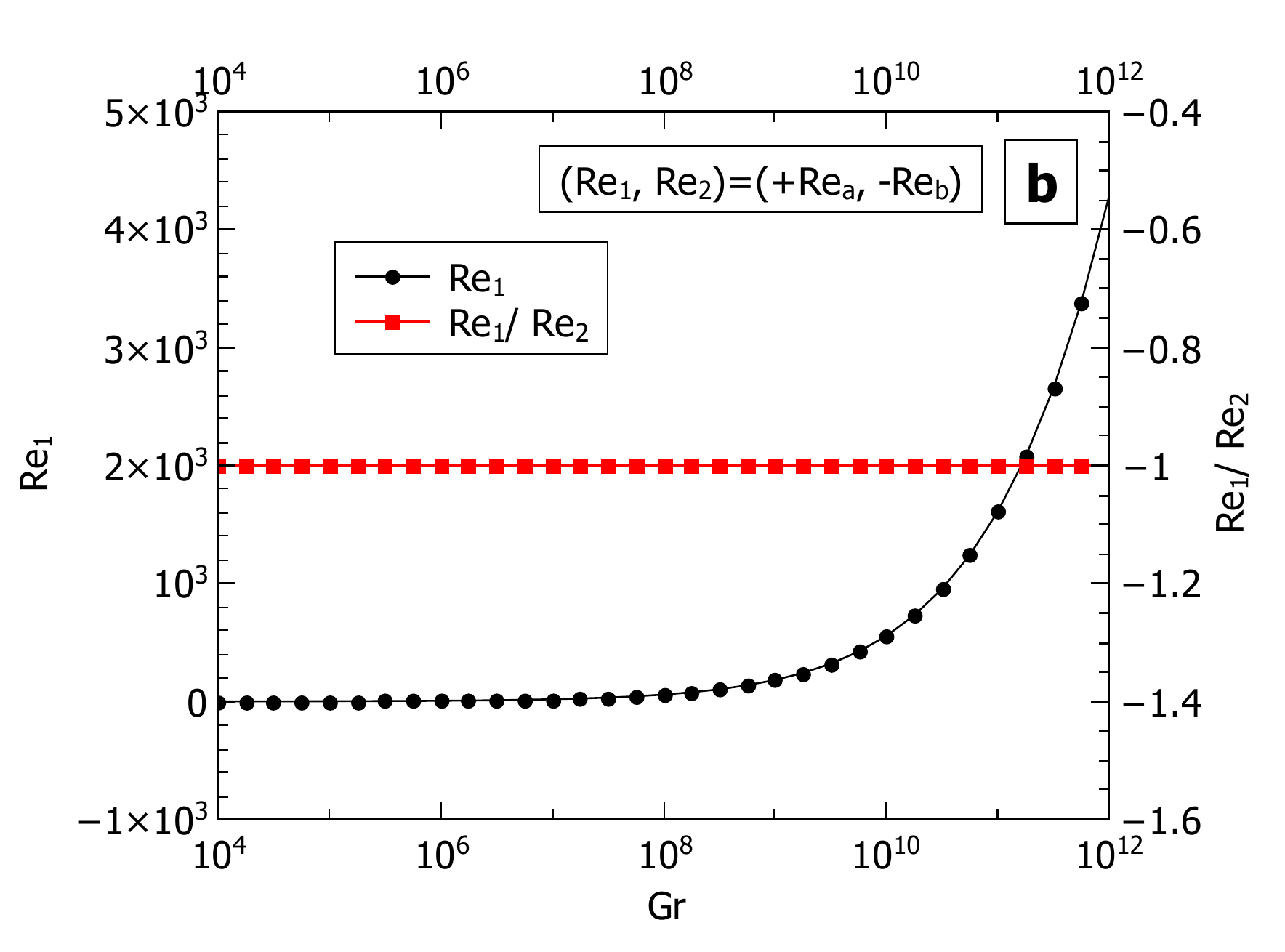}
	\end{subfigure}
	\hspace{\fill}
	\begin{subfigure}[b]{0.49\textwidth}
		\includegraphics[width=1\linewidth]{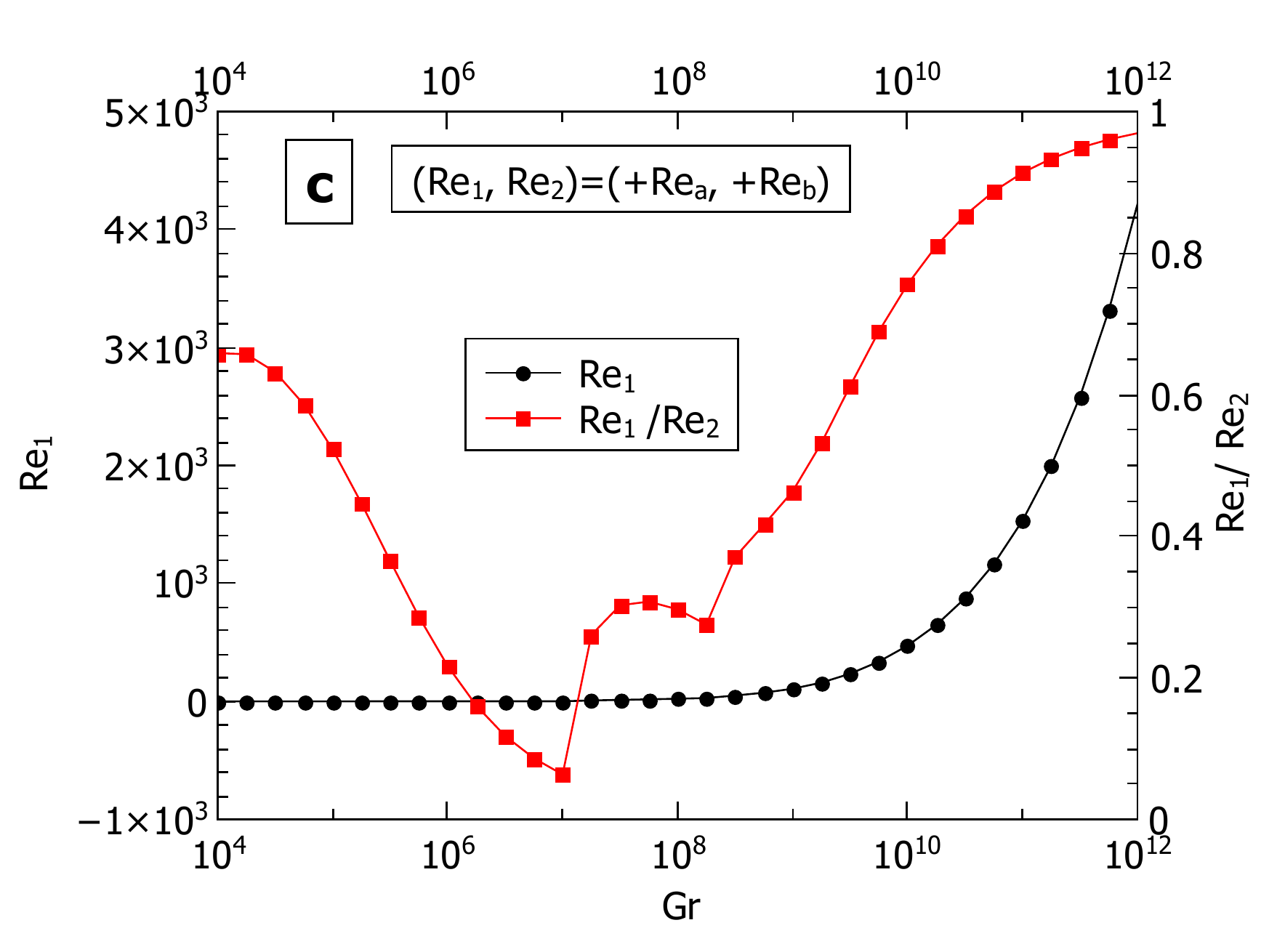}
	\end{subfigure}
	\hspace{\fill}
	\begin{subfigure}[b]{0.49\textwidth}
		\includegraphics[width=1\linewidth]{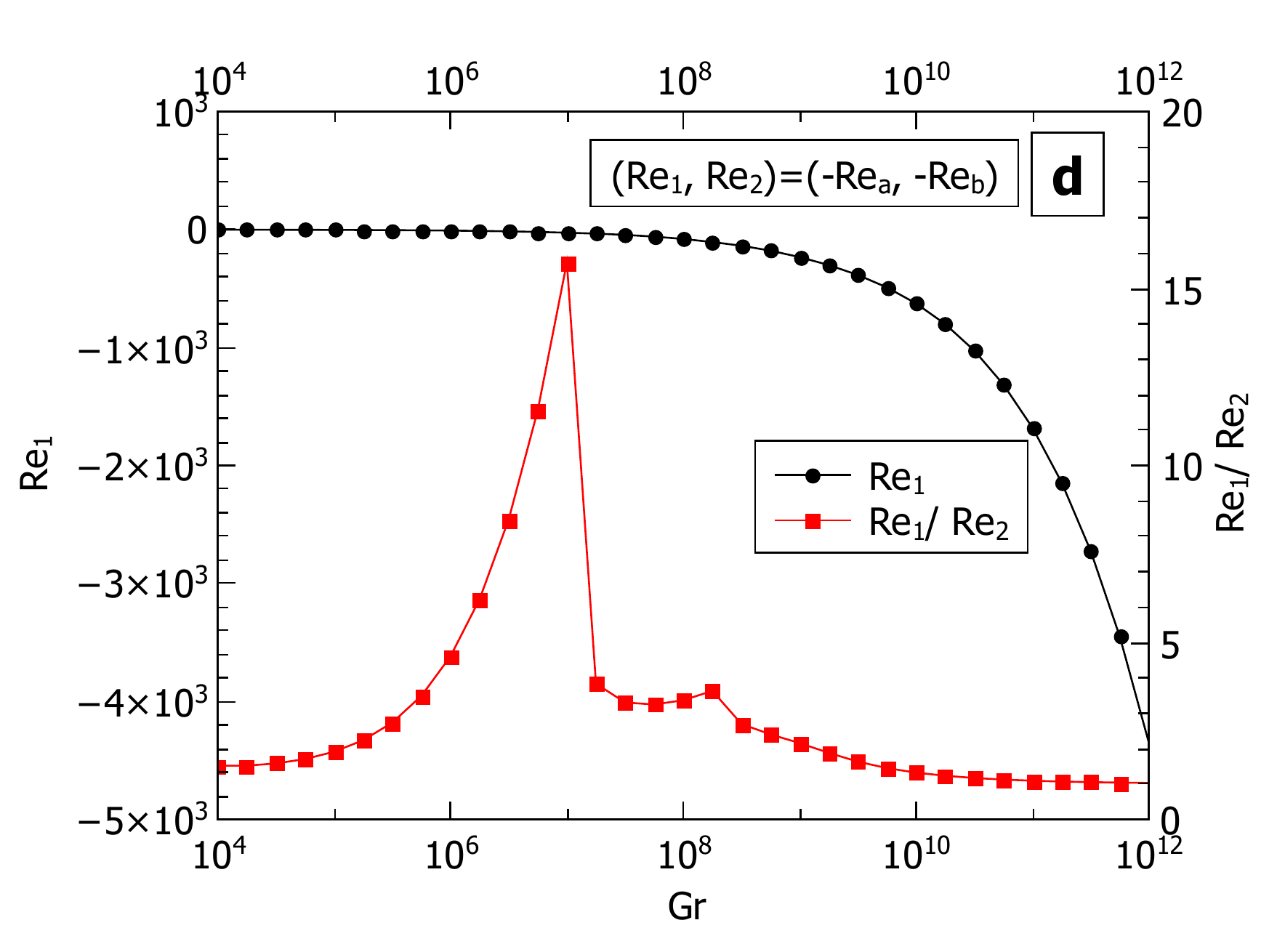}
	\end{subfigure}
	\caption{Steady state magnitude of $Re_1$ and $Re_2$ of VCNCL system for the steady state, (a) $(Re_1,Re_2)=(-Re_a,+Re_b)$, (b) $(Re_1,Re_2)=(+Re_a,-Re_b)$, (c) $(Re_1,Re_2)=(+Re_a,+Re_b)$, and (d) $(Re_1,Re_2)=(-Re_a,-Re_b)$, for $Fo=0.0001$, $As=1$, $Co_1=2846$ and $St=1000$.}
	\label{Fig16}
\end{figure}

As mentioned in section 5.1, the CNCL system has multiple steady states, but a stability map can only provide the stability of a particular case of steady-state. It was also noted in section 5.1 that the consistently observed steady states tend to lie on the lines $Re_1=Re_2$ or $Re_1=-Re_2$. Thus the VCNCL system is scoped for such steady states as represented in figure \ref{Fig16}. From figure \ref{Fig16} it can be noted that only the steady states $(Re_1,Re_2)=(-Re_a,+Re_b)$ and $(Re_1,Re_2)=(+Re_a,-Re_b)$ which lie on the line $Re_1=-Re_2$ are consistently observed and thus the stability maps of these steady states are generated. The stability maps of the other steady states can also be plotted, but the fsolve MATLAB solver which is used to compute the steady states converges to a different set of steady states based on the initial assumption provided to it, and the stability maps of steady states depend on the steady states they converge to.

\subsubsection{Stability maps of the VCNCL system}

\begin{figure}[!htb]
	\centering
	\begin{subfigure}[b]{0.49\textwidth}
		\includegraphics[width=1\linewidth]{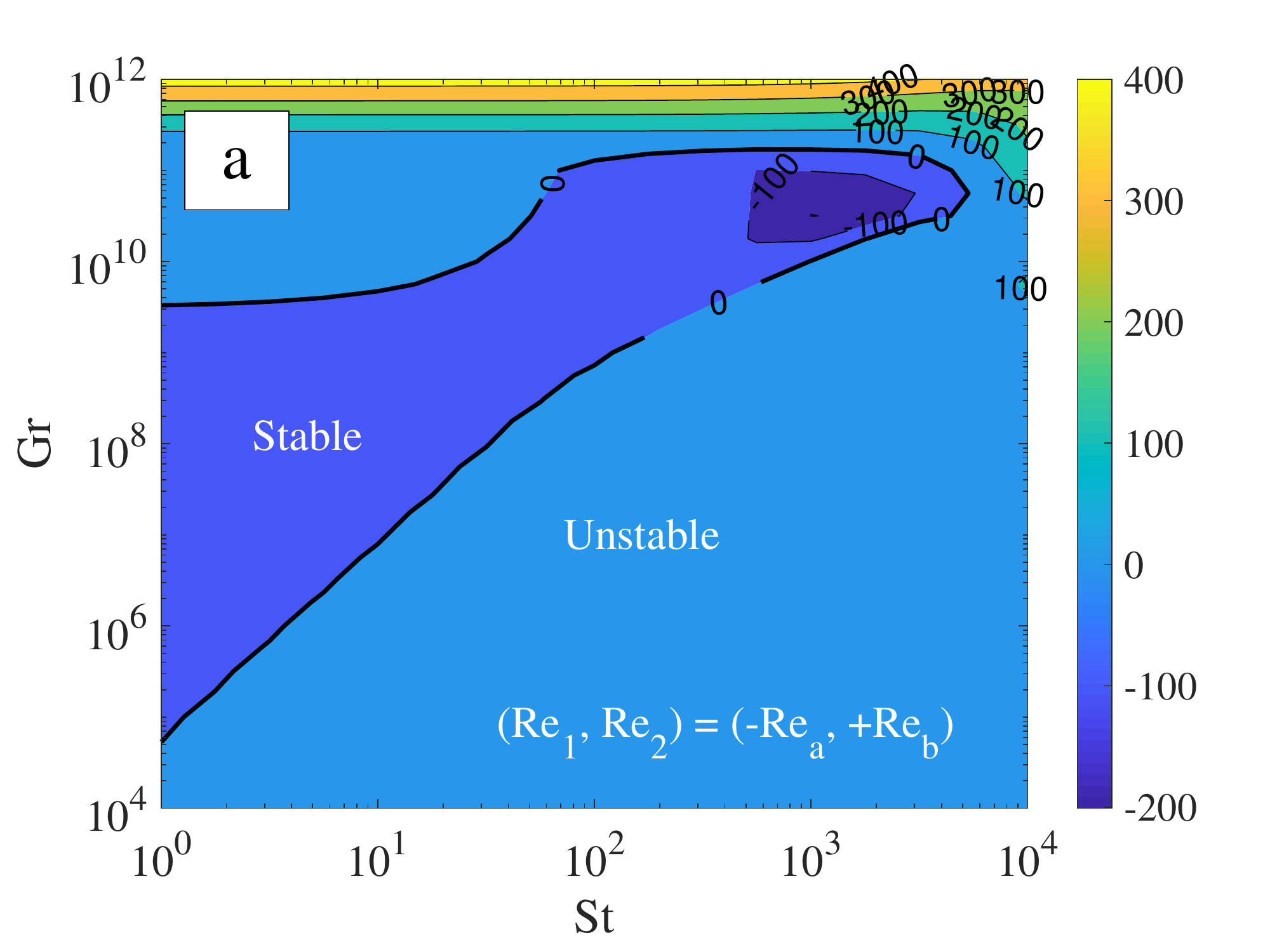}
	\end{subfigure}
	\hspace{\fill}
	\begin{subfigure}[b]{0.49\textwidth}
		\includegraphics[width=1\linewidth]{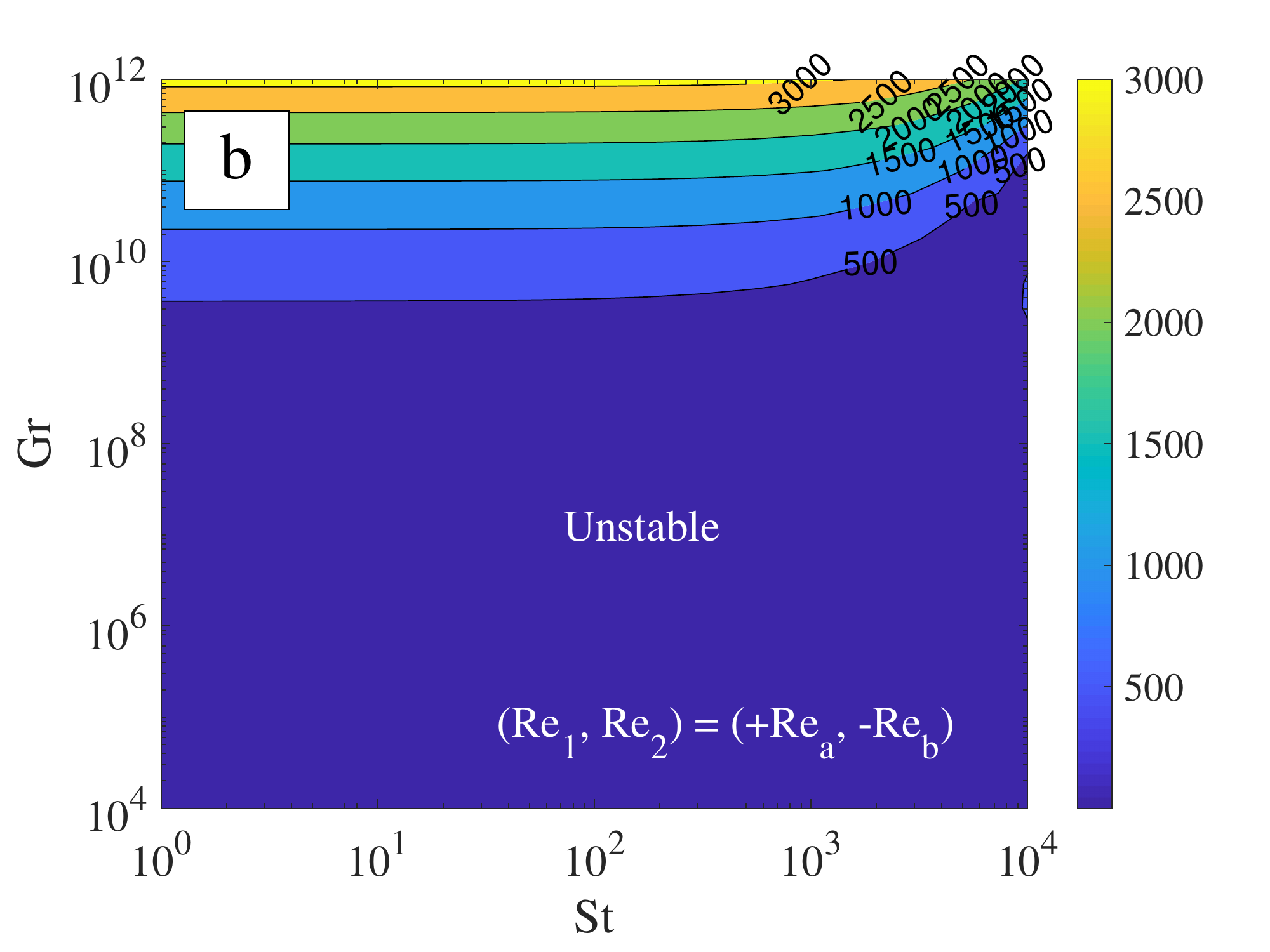}
	\end{subfigure}
	\caption{Stability map of steady state corresponding to (a) $(Re_1,Re_2)=(-Re_a,+Re_b)$ and (b) $(Re_1,Re_2)=(+Re_a,-Re_b)$, which lie on the line $Re_1=-Re_2$ in the  $Re_1 Re_2$ plane for $Fo=0.0001$, $As=1$ and $Co_1=2846$ for the VCNCL system.}
	\label{Fig17}
\end{figure}

The stability maps corresponding to steady states $(Re_1,Re_2)=(-Re_a,+Re_b)$ and $(Re_1,Re_2)=(+Re_a,-Re_b)$ which lie on the line $Re_1=-Re_2$ in the  $Re_1 Re_2$ plane are evaluated and shown in figure \ref{Fig17}. It can be noted from figure 17(b) that the steady state corresponding to $(Re_1,Re_2)=(+Re_a,-Re_b)$ which lies on the line $Re_1=-Re_2$ is completely unstable (positive eigenvalue) for the considered non-dimensional numbers. The stability map represented by figure \ref{Fig17}(a) corresponds to steady state $(Re_1,Re_2)=(-Re_a,+Re_b)$ which lies on the line $Re_1=-Re_2$. It has a distinct stability boundary which separates the stable (negative eigenvalue) from the unstable domain, thus the stability map of this particular steady state is employed for the parametric studies to determine the influence of other non-dimensional numbers on the stability of the VCNCL system.

\subsection{Stability analysis of HCNCL system}

\begin{figure}[!htb]
    \centering
    \includegraphics[width=0.42\linewidth]{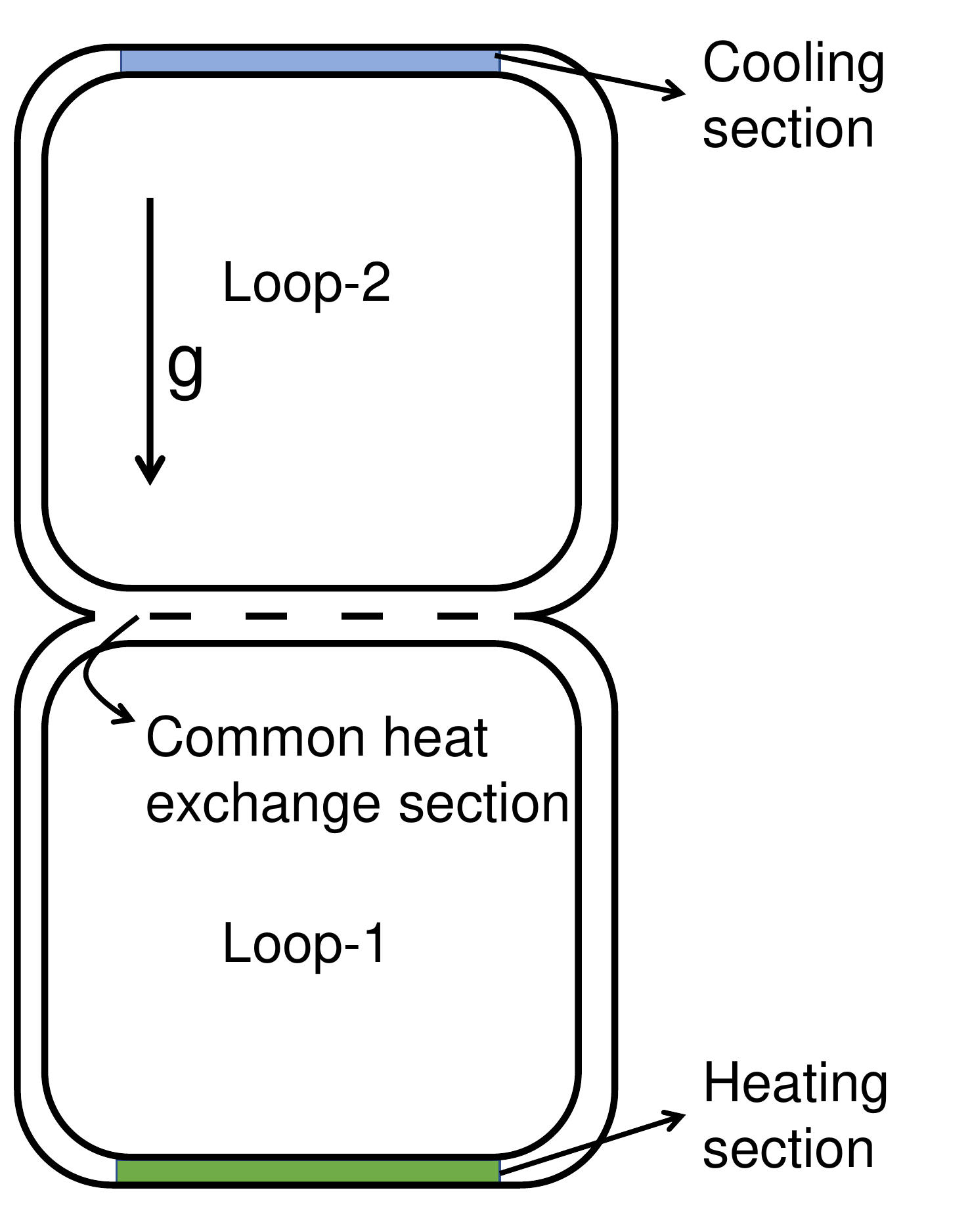}
    \caption{Geometry of the HCNCL system considered for the study.}
    \label{Fig18}
\end{figure}

Figure \ref{Fig18} presents the geometry of the HCNCL considered for the stability analysis. The peculiarity of the considered geometry is that for this particular heater-cooler configuration of the HCNCL, both parallel flow and counterflow configurations have been reported by Dass and Gedupudi \cite{dass2019}. Thus, the stability analysis of such HCNCL system with heat flux boundary conditions, square component NCLs, i.e., an aspect ratio of unity and same fluids within both the loops, can be used for comparing the stability of the parallel and counterflow configurations at the common heat exchange section. 

\newpage

\subsubsection{Steady states of the HCNCL system}

\begin{figure}[!htb]
	\centering
	\begin{subfigure}[b]{0.49\textwidth}
		\includegraphics[width=1\linewidth]{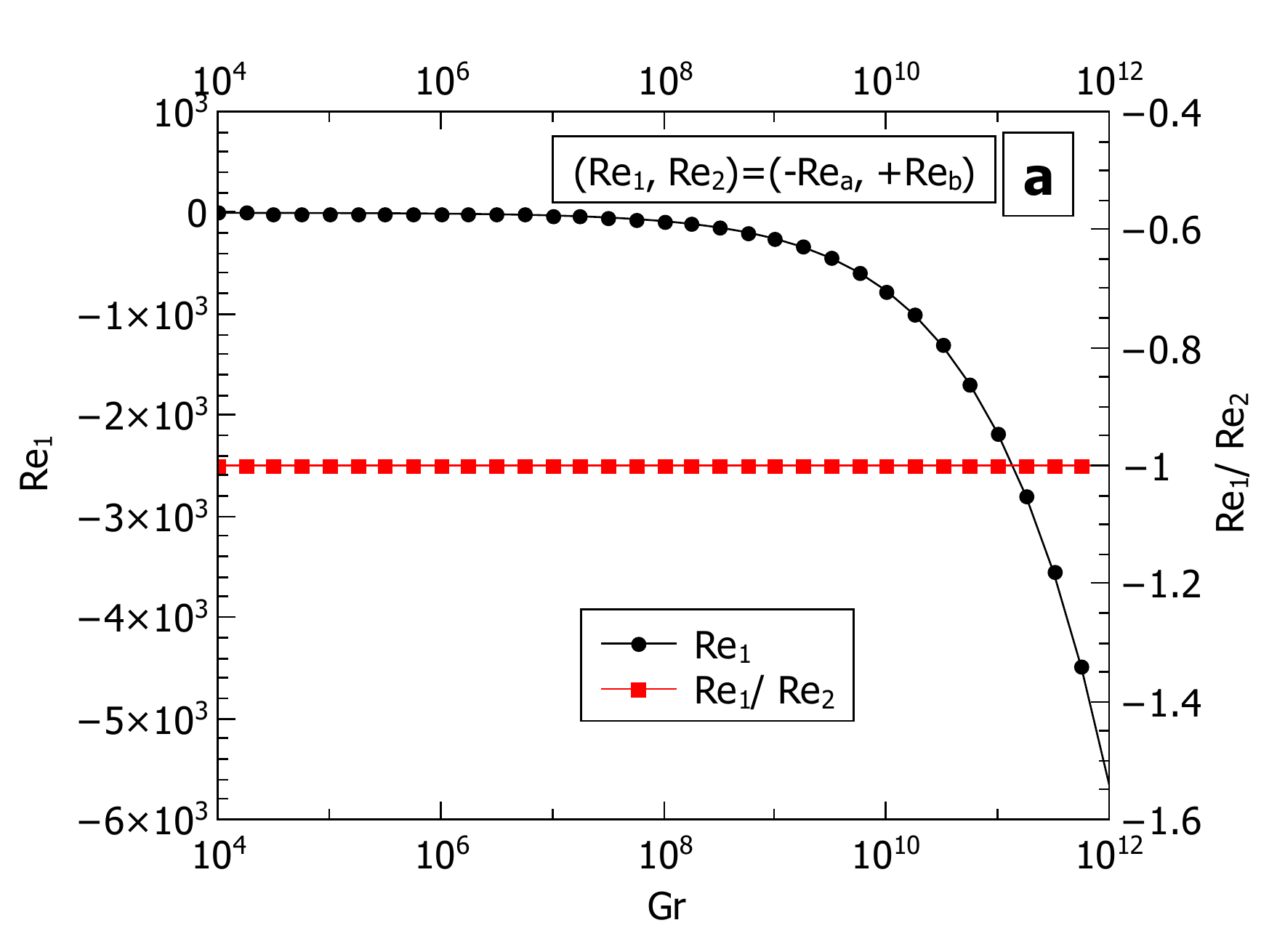}
	\end{subfigure}
	\hspace{\fill}
	\begin{subfigure}[b]{0.49\textwidth}
		\includegraphics[width=1\linewidth]{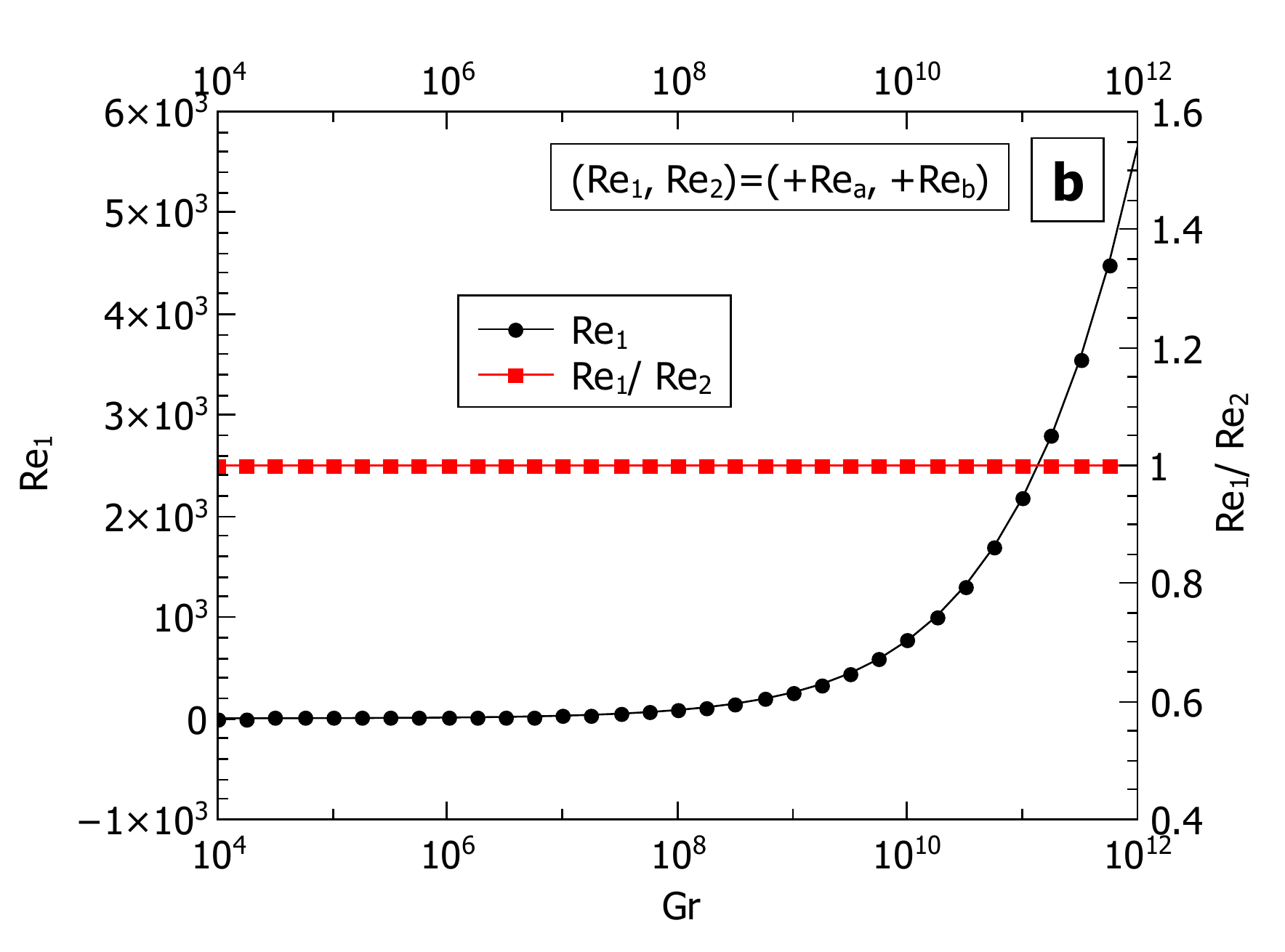}
	\end{subfigure}
	\caption{Steady state magnitude of $Re_1$ and $Re_2$ of the considered HCNCL system for the steady state, (a) $(Re_1,Re_2)=(-Re_a,+Re_b)$ which lies on the line $Re_1=-Re_2$ and (b) $(Re_1,Re_2)=(+Re_a,+Re_b)$ which lies on the line $Re_1=Re_2$, in the  $Re_1 Re_2$ plane for $Fo=0.0001$, $As=1$, $Co_1=2846$ and $St=1000$.}
	\label{Fig19}
\end{figure}

The considered HCNCL system is assessed for the presence of steady states corresponding to parallel and counterflow conditions. For the reasons presented in section 10.1.1, the study searched for steady states which lie on the lines $Re_1=Re_2$ and $Re_1=-Re_2$ in the $Re_1 Re_2$ plane. The steady-state magnitudes of $Re_1$ and $Re_2$ corresponding to the counterflow and parallel flow conditions for $Fo=0.0001$ are presented in figures \ref{Fig19}(a) and \ref{Fig19}(b) respectively. From figure \ref{Fig19}, it can be concluded that the steady-states corresponding to the parallel flow and counterflow configurations at the common heat exchange sections exist and are consistently observed, and thus the stability maps of the HCNCL corresponding to these steady states are plotted. 

\subsubsection{Stability maps of the HCNCL system}

\begin{figure}[!htb]
	\centering
	\begin{subfigure}[b]{0.49\textwidth}
		\includegraphics[width=1\linewidth]{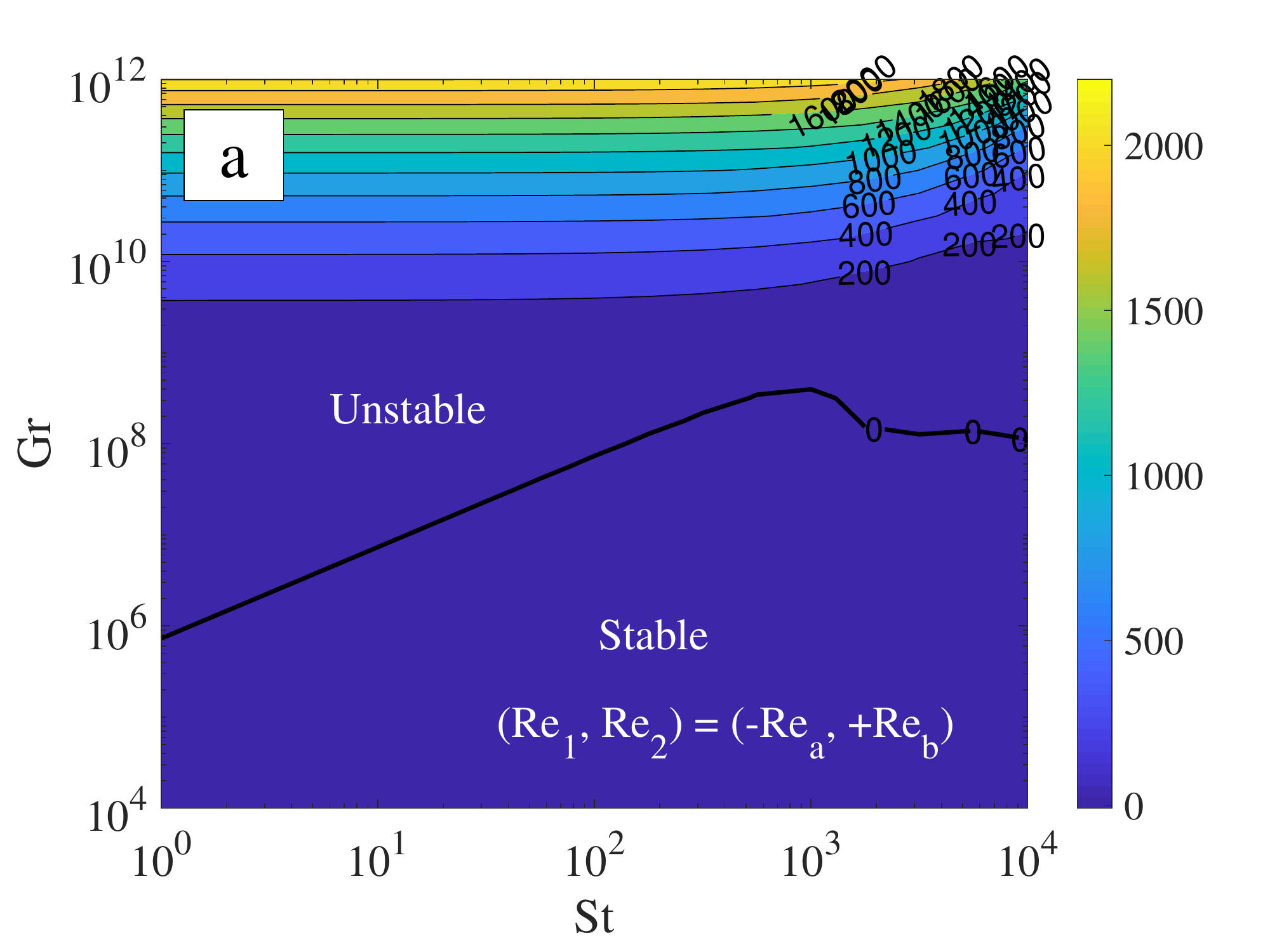}
	\end{subfigure}
	\hspace{\fill}
	\begin{subfigure}[b]{0.49\textwidth}
		\includegraphics[width=1\linewidth]{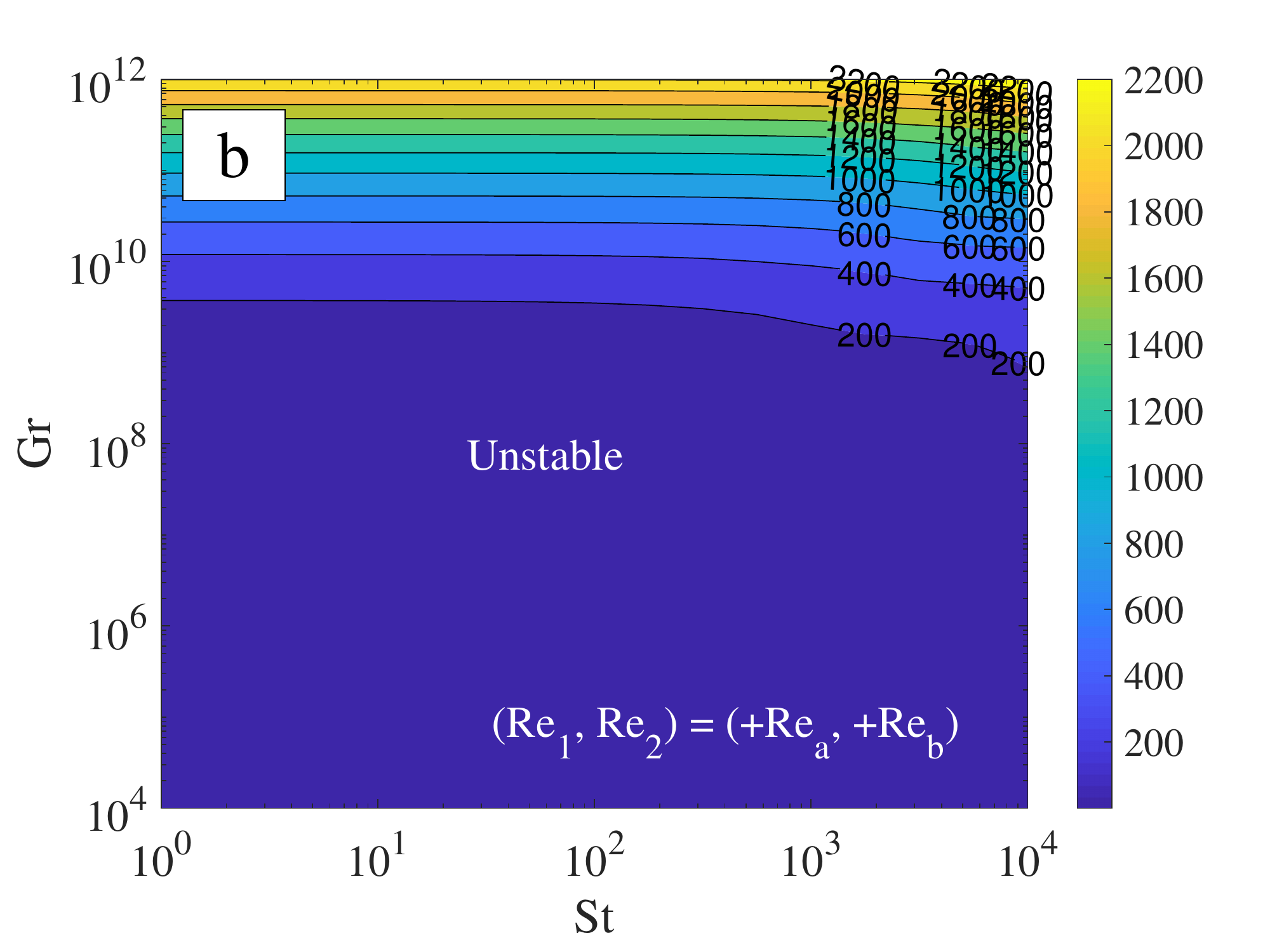}
	\end{subfigure}
	\caption{Stability maps of steady state corresponding to (a) $(Re_1,Re_2)=(-Re_a,+Re_b)$ which lies on the line $Re_1=-Re_2$ and (b) $(Re_1,Re_2)=(+Re_a,+Re_b)$ which lies on the line $Re_1=Re_2$, in the  $Re_1 Re_2$ plane for $Fo=0.0001$, $As=1$ and $Co_1=2846$ for the HCNCL system.}
	\label{Fig20}
\end{figure}

Figures \ref{Fig20}(a) and \ref{Fig20}(b) represent the comparison of the stability maps corresponding to the counterflow and parallel flow configurations at the common heat exchange section of the HCNCL system, respectively. It is observed that for the considered $Fo=0.0001$, only the counterflow configuration has a stable domain in the stability map. Therefore, for $Fo=0.0001$, only the counterflow configuration is observed at the common heat exchange section of the HCNCL for stable system operating conditions. Due to the midplane symmetry of the considered HCNCL system represented in figure \ref{Fig18}, the stability maps of the steady states corresponding to $(Re_1,Re_2)=(-Re_a,+Re_b)$, which lies on the line $Re_1=-Re_2$ and $(Re_1,Re_2)=(+Re_a,+Re_b)$, which lies on the line $Re_1=Re_2$, are identical to the stability map of steady states corresponding to $(Re_1,Re_2)=(+Re_a,-Re_b)$, which lies on the line $Re_1=-Re_2$ and $(Re_1,Re_2)=(-Re_a,-Re_b)$, which lies on the line $Re_1=Re_2$, respectively.

\subsubsection{Stability analysis of HCNCL systems for $Fo=2$}

\begin{figure}[!htb]
	\centering
	\begin{subfigure}[b]{0.49\textwidth}
		\includegraphics[width=1\linewidth]{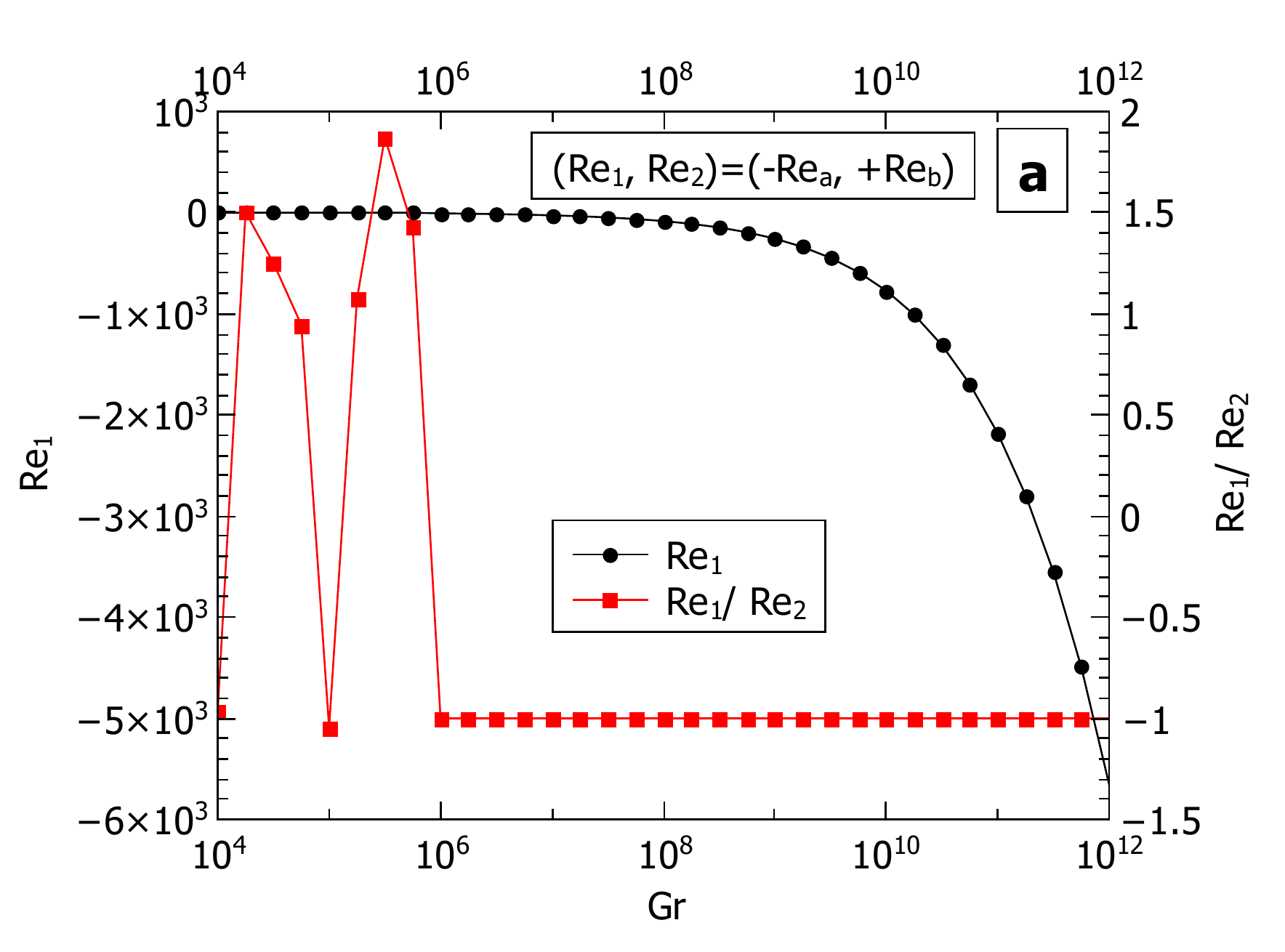}
	\end{subfigure}
	\hspace{\fill}
	\begin{subfigure}[b]{0.49\textwidth}
		\includegraphics[width=1\linewidth]{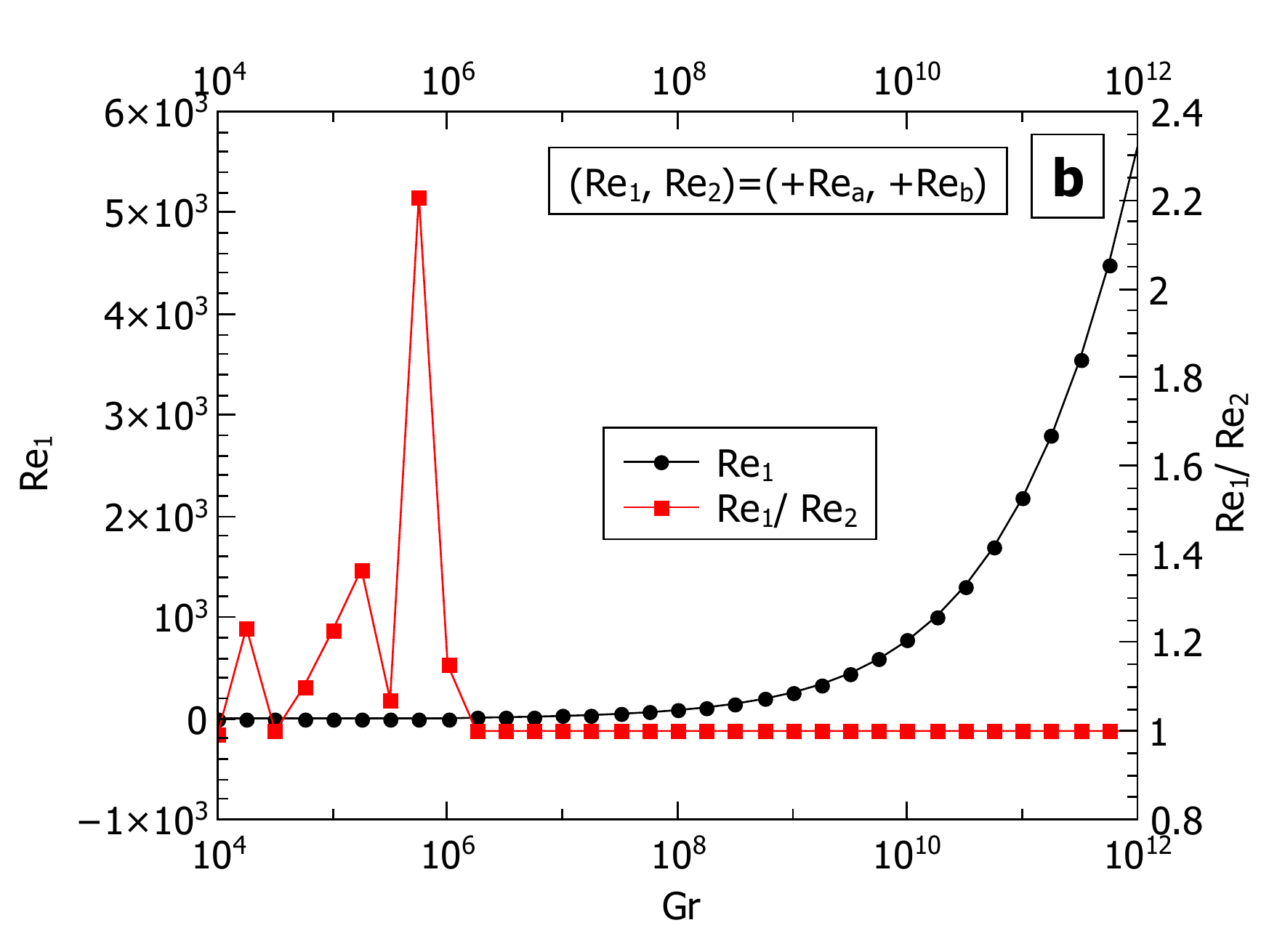}
	\end{subfigure}
	\caption{Steady state magnitude of $Re_1$ and $Re_2$ of the considered HCNCL system, (a) $(Re_1,Re_2)=(-Re_a,+Re_b)$ which lies on the line $Re_1=-Re_2$ and (b) $(Re_1,Re_2)=(+Re_a,+Re_b)$ which lies on the line $Re_1=Re_2$, in the  $Re_1 Re_2$ plane for $Fo=2$, $As=1$, $Co_1=2846$ and $St=1000$.}
	\label{Fig21}
\end{figure}

\begin{figure}[!htb]
	\centering
	\begin{subfigure}[b]{0.49\textwidth}
		\includegraphics[width=1\linewidth]{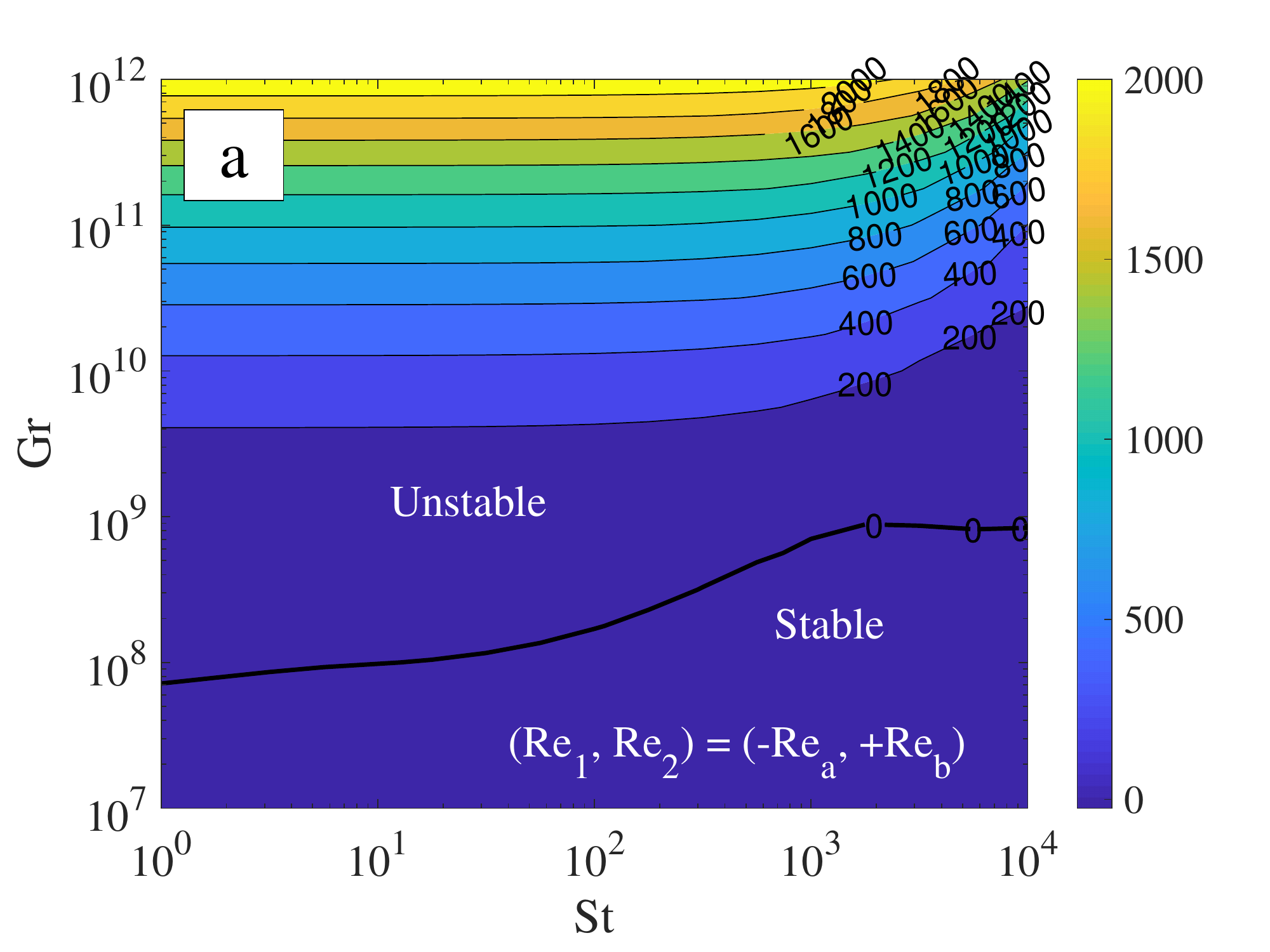}
	\end{subfigure}
	\hspace{\fill}
	\begin{subfigure}[b]{0.49\textwidth}
		\includegraphics[width=1\linewidth]{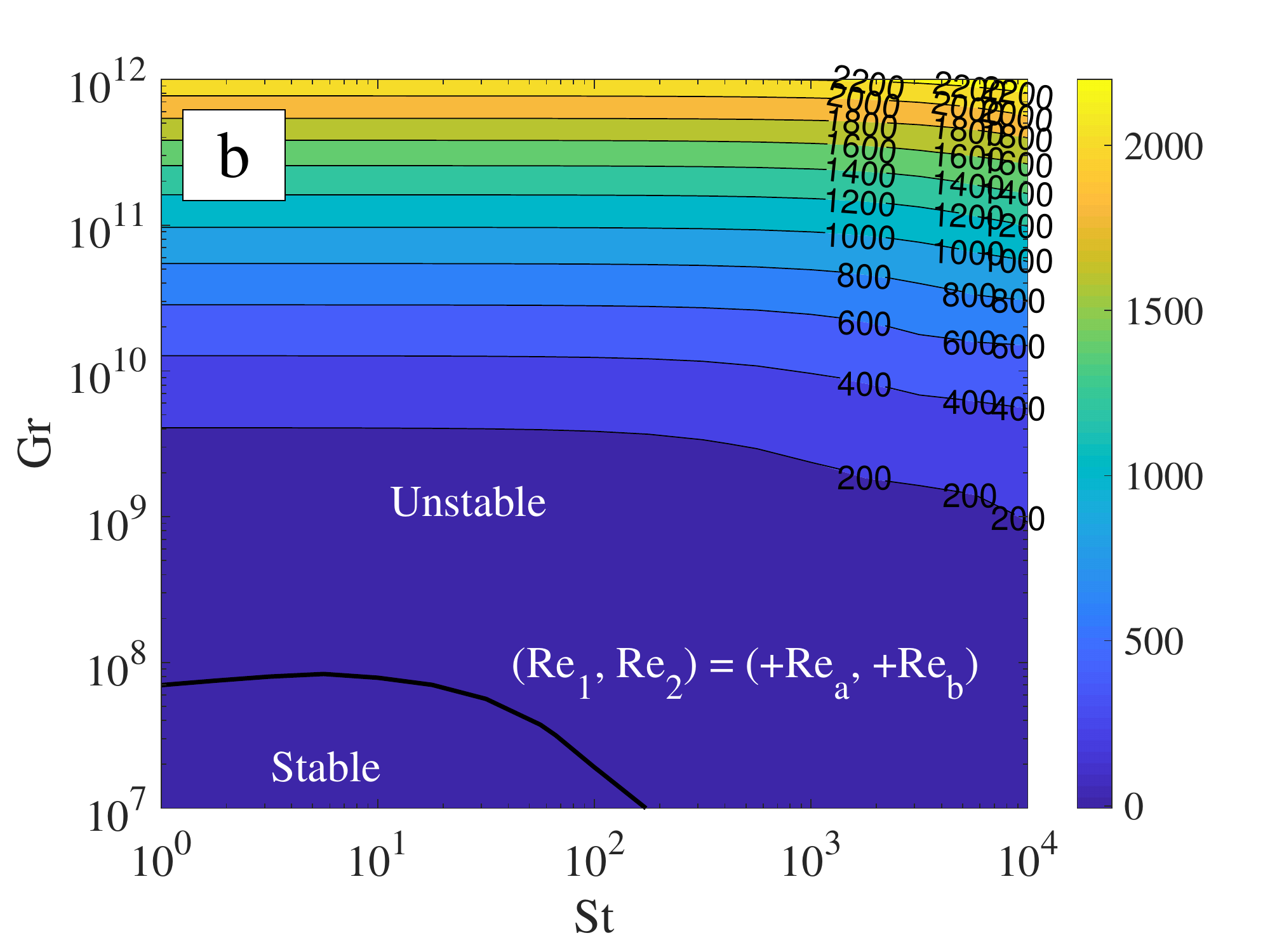}
	\end{subfigure}
	\caption{Stability maps of steady state corresponding to (a) $(Re_1,Re_2)=(-Re_a,+Re_b)$ which lies on the line $Re_1=-Re_2$ and (b) $(Re_1,Re_2)=(+Re_a,+Re_b)$ which lies on the line $Re_1=Re_2$, in the  $Re_1 Re_2$ plane for $Fo=2$, $As=1$ and $Co_1=2846$ for the HCNCL system.}
	\label{Fig22}
\end{figure}

Stable steady-state for the parallel flow configuration at the common heat exchange section is not witnessed for $Fo=0.0001$, as observed in the last section. The current section investigates the effect of a larger $Fo$ on the existence of a stable domain in a parallel flow stability map. From figure \ref{Fig21}, it can be noted that a consistent steady-state $(Re_1,Re_2)=(+Re_a,+Re_b)$  which lies on the line $Re_1=Re_2$ is observed only for $Gr>10^7$. Thus, the stability map of the HCNCL for $Fo=2$ is evaluated for $Gr>10^7$. Figure \ref{Fig22} presents the comparison of the counterflow and parallel flow stability map of the HCNCL system for $Fo=2$. It can be concluded from figure \ref{Fig22} that the counterflow configuration at the common heat exchange section has a greater domain of stable operation relative to the parallel flow configuration and that the parallel flow has a stable domain of operation for $Fo=2$.

From the analysis of stability maps of the considered HCNCL system, it can be summarised that HCNCL exhibits stable parallel and counterflow configuration at common heat exchange section only for larger magnitudes of $Fo$ and that the domain of stable operation of counterflow configuration is greater than that of the parallel flow configuration.

\subsection{Parametric study on the stability behaviour of CNCL systems}

The effects of non-dimensional numbers $Fo$, $Co_1$ and $As$ on the stability of CNCL systems are investigated. The ranges of non-dimensional numbers considered for the parametric study is as follows:
\begin{enumerate}
    \item $10^4 \leq Gr \leq 10^{12} $
    \item $10^0 \leq St \leq 10^{4} $
    \item $0 < Fo\leq 2 $
    \item $10^{-1} \leq As \leq 10^{1} $
    \item $10^2 \leq Co_1 \leq 10^{4} $
\end{enumerate}

The $Gr-St$ stability map of the VCNCL corresponding to the steady state $(Re_1,Re_2)=(-Re_a,+Re_b)$ which lies on the line $Re_1=-Re_2$ is used to conduct the parametric study with the laminar flow friction factor. Calculation of non-dimensional numbers used for the parametric study is shown in Table 3.

\newpage

\subsubsection{Effect of $Fo$  on the stability of VCNCL system}
The magnitude of $Fo$ for a particular fluid denotes the strength of axial conduction within the fluid. A fluid with $Fo \approx 0.001$ (see Table 3 for calculation) such as water has negligible axial conduction contribution. In contrast, a fluid with  $Fo \approx 2$ (see Table 3 for calculation) such as sodium has a more significant axial conduction contribution which affects the system stability dynamics. The influence of $Fo$ on the stability of the VCNCL system is presented in figure \ref{Fig23}. It can be noted that with an increase in $Fo$, there is an increase in the domain of stability. This observation is corroborated by the fact that with an increase in $Fo$, there is a decrease in the transient oscillatory behaviour of CNCL systems \cite{dass2019}. For the considered range of $Fo$, it is noted that the VCNCL system becomes unstable beyond $Gr \approx 10^{11}$. 

\subsubsection{Effect of $Co_1$ on the stability of VCNCL system}

\begin{figure}[!htb]
\begin{minipage}{0.49\linewidth}
\centering
		\includegraphics[width=1\linewidth]{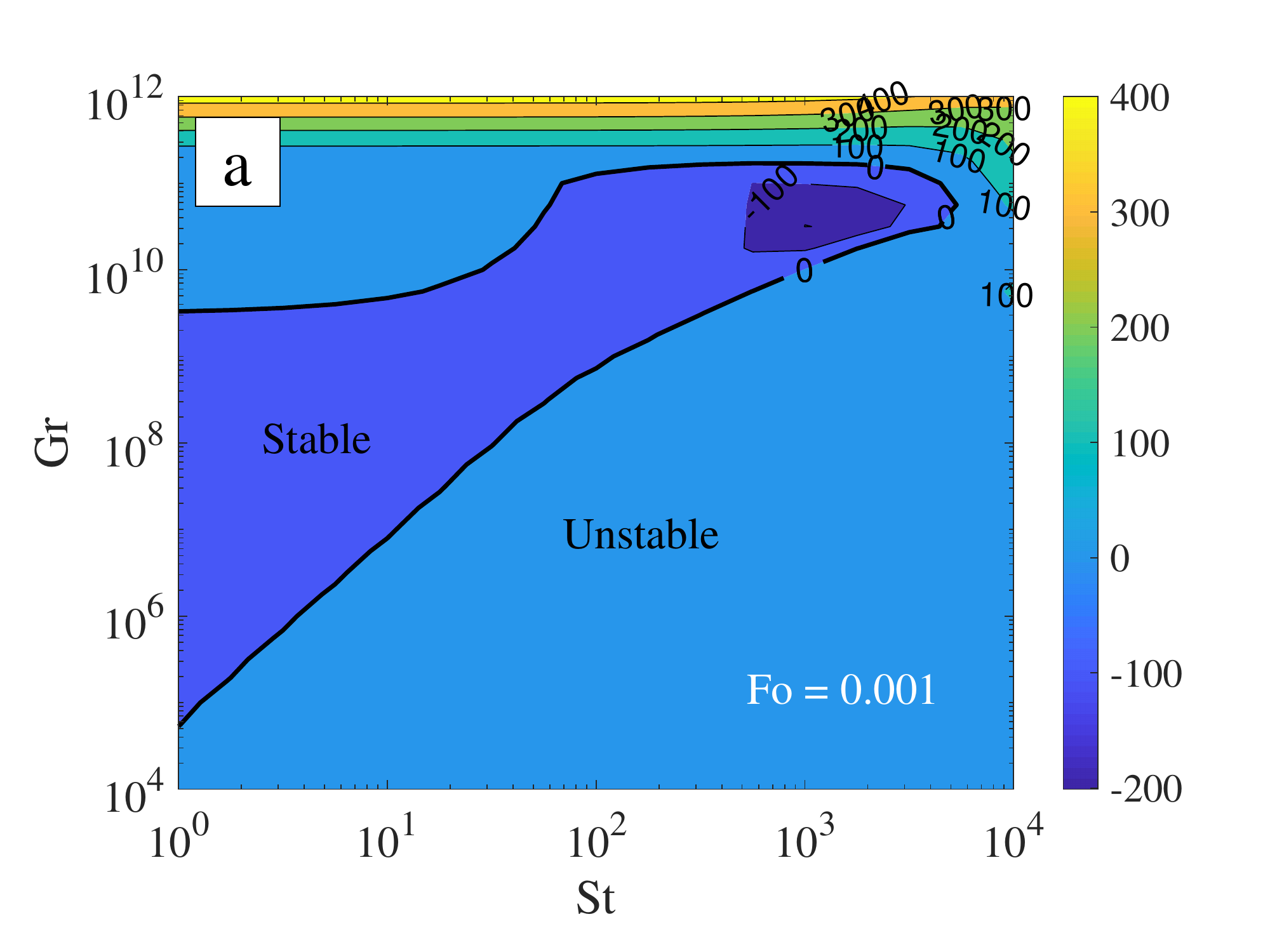}
		\includegraphics[width=1\linewidth]{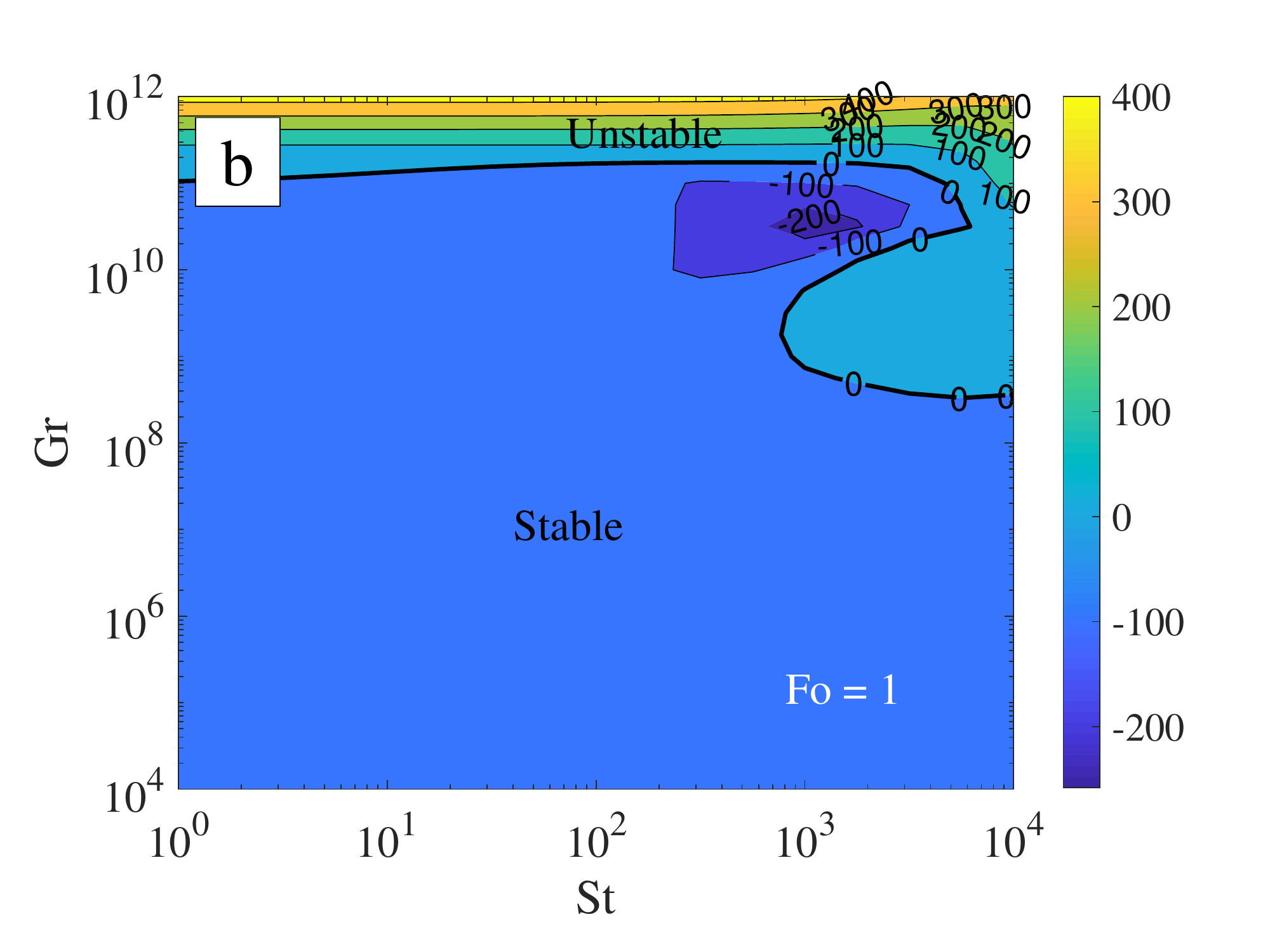}
		\includegraphics[width=1\linewidth]{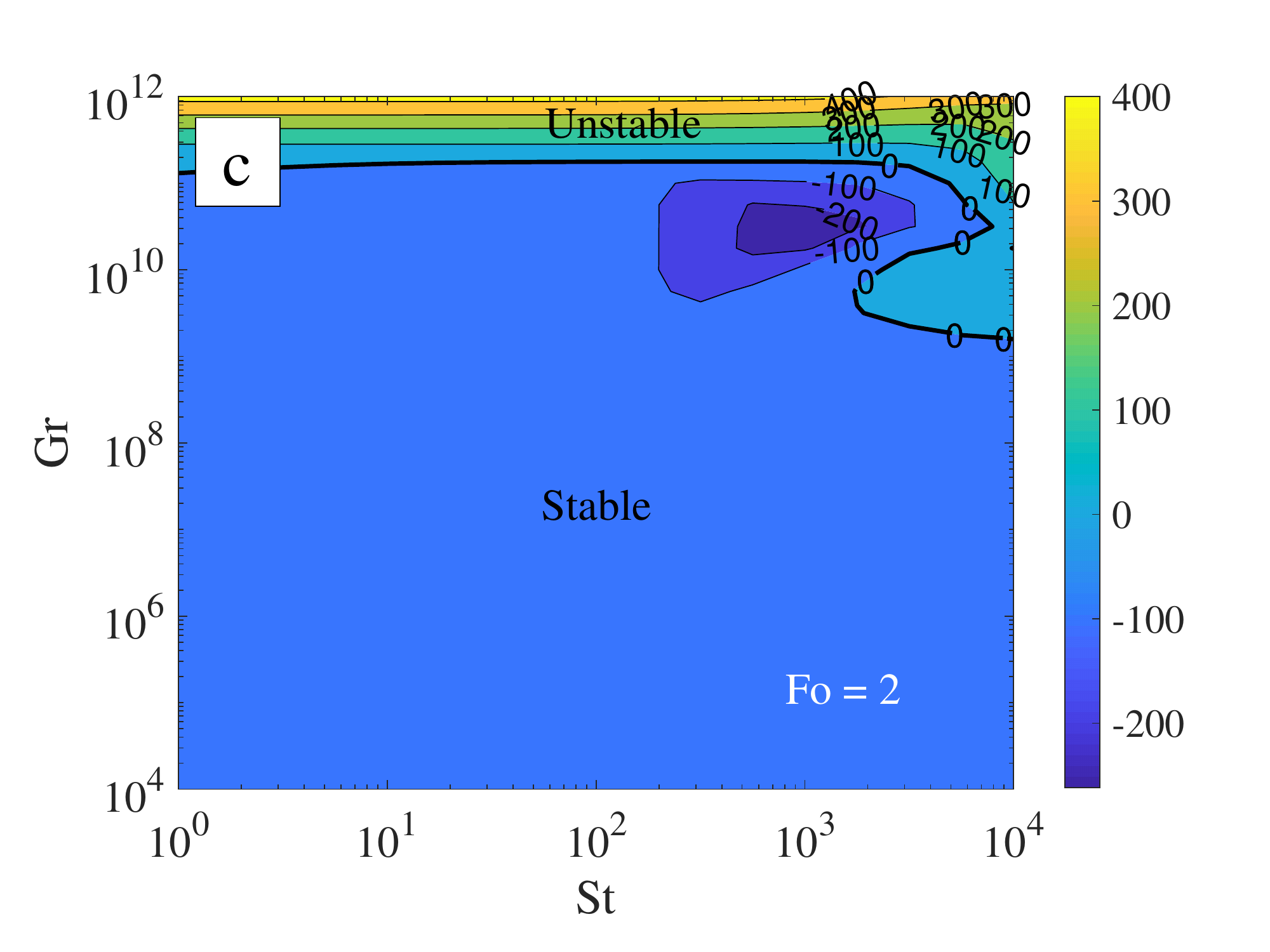}
	    \caption{Stability map of the VCNCL system corresponding to the steady state $(Re_1,Re_2)=(-Re_a,+Re_b)$ which lies on the line $Re_1=-Re_2$ with $As=1$, $Co_1= 2846$ for, (a) $Fo=0.001$, (b) $Fo=1$, and (c) $Fo=2$.}
	    \label{Fig23}
\end{minipage}
\hfill \vline \hfill
\begin{minipage}{0.49\linewidth}
\centering
		\includegraphics[width=1\linewidth]{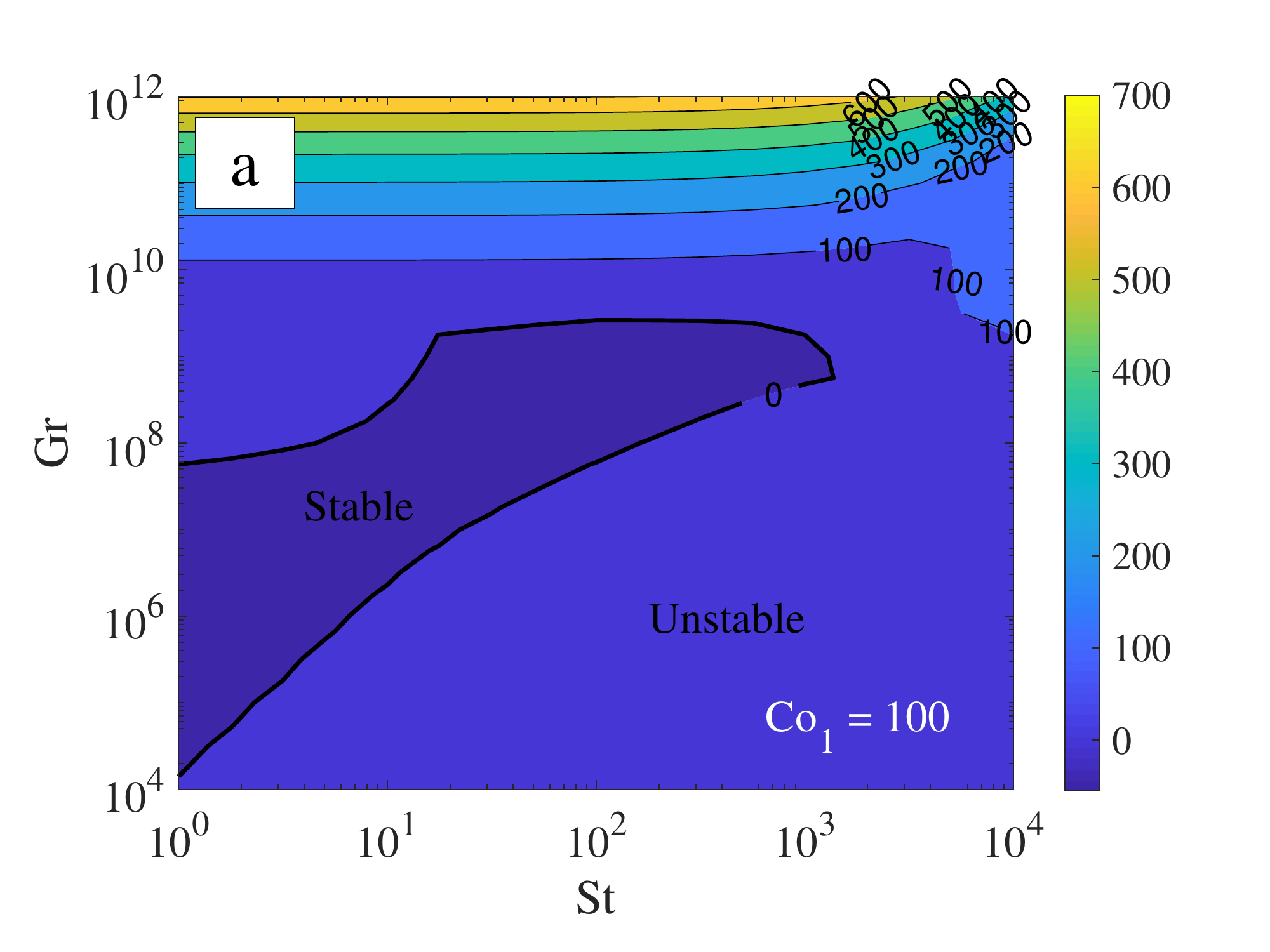}
		\includegraphics[width=1\linewidth]{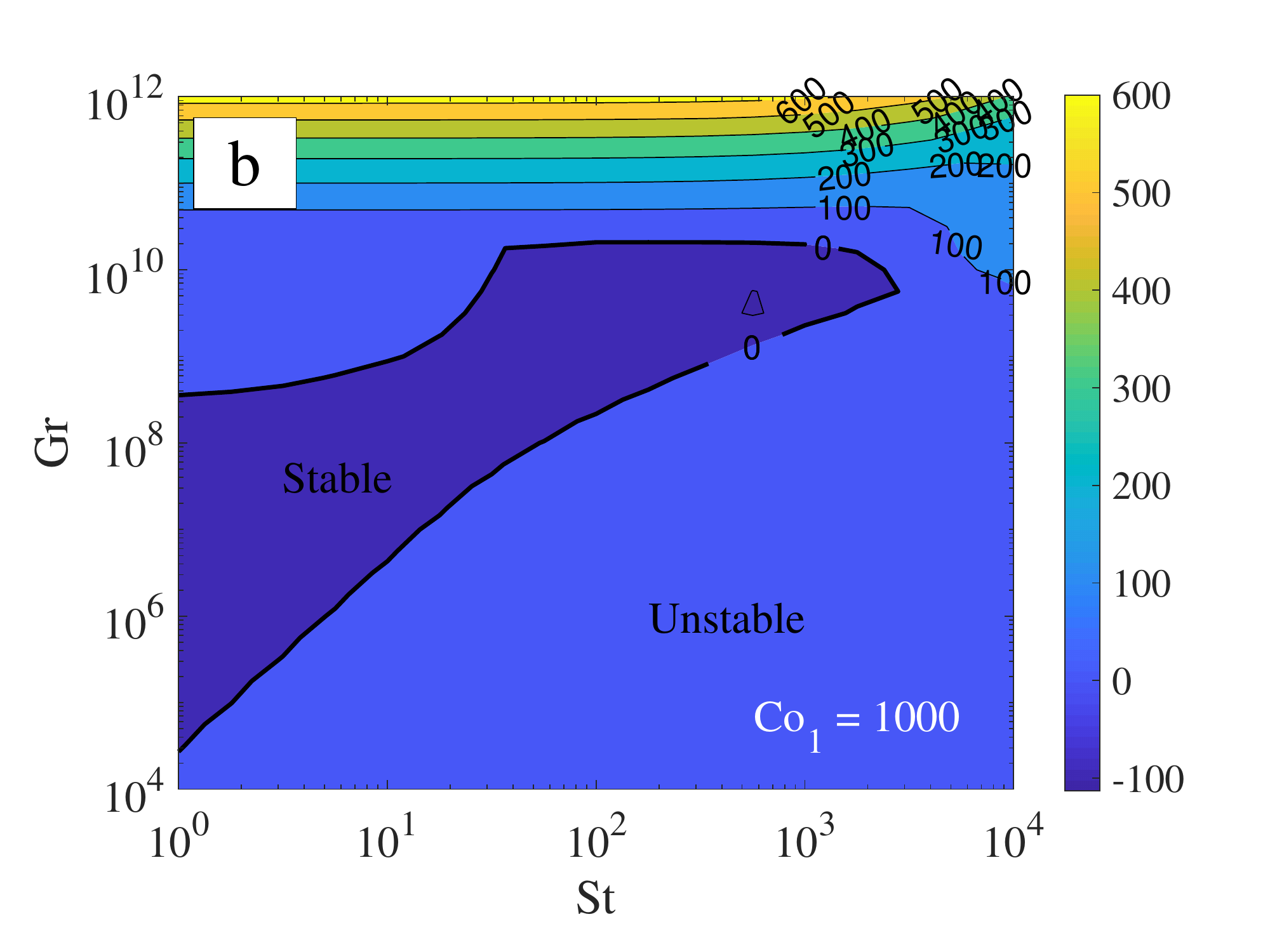}
		\includegraphics[width=1\linewidth]{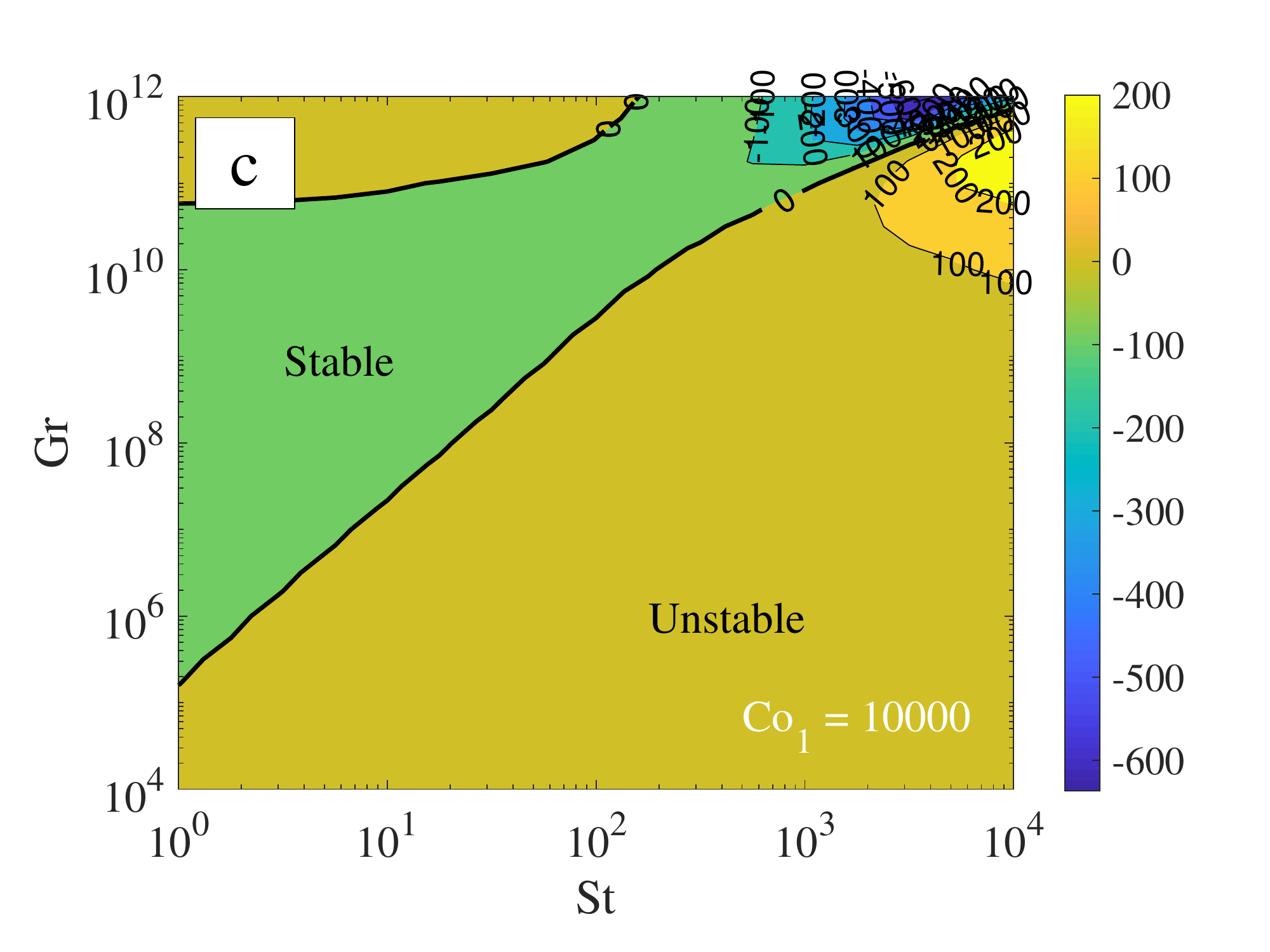}
	    \caption{Stability map of the VCNCL system corresponding to the steady state $(Re_1,Re_2)=(-Re_a,+Re_b)$ which lies on the line $Re_1=-Re_2$ with $As=1$, $Fo=0.001$ for, (a) $Co_1=10^2$, (b) $Co_1=10^3$, and (c) $Co_1=10^4$.}
	    \label{Fig24}
\end{minipage}
\end{figure}

$Co_1$ is defined as the flow resistance coefficient \cite{dass2019}, and hence with an increase in its magnitude, the effective flow resistance encountered by the fluid increases. Figure \ref{Fig24} represents the effect of an increase in $Co_1$ on the stability behaviour of VCNCL system. It is noted that with an increase in $Co_1$, the domain of stability increases, the shape of the stability boundary is unaltered, and the stability boundary shifts up. Dass and Gedupudi \cite{dass2019} have reported a decrease in the oscillatory behaviour of the CNCL system with an increase in $Co_1$, and this fact corroborates with the observed increase in the domain of stable operation with the increase in $Co_1$, as the chaotic behaviour is suppressed.

$Co_1$ is calculated as follows:

\begin{equation}
    Co_1=\frac{2b x_0}{D_h}
\end{equation}

Therefore, for a fixed $b$ (which depends on the friction factor), $Co_1 \propto	\frac{x_0}{D_h}$. Vijayan \cite{vijayan2002experimental} had also reported an upward shift in the stability boundary with an increase in $\frac{x_0}{D_h}$ for an NCL system. This observation agrees with the observed shift in the stability boundary of the CNCL system. 

\begin{table}[!htb]
	\caption{Calculation of the values of non-dimensional numbers used for the parametric study.}
	\centering
	\begin{tabular}{|l|l|l|}
		\hline
		Parameter & Water ($25 \;^{\circ}\mathrm{C}$)            & Sodium ($400 \;^{\circ}\mathrm{C}$)         \\ \hline
		$\alpha$  & $1.42 \times 10^{-7}$          & $6.59 \times 10^{-5}$        \\ \hline
		$\nu$     & $8.93 \times 10^{-7}$         & $3.23 \times 10^{-7}$        \\ \hline
		$L$       & \multicolumn{2}{l|}{$1 \; \mathrm{m}$}                       \\ \hline
		$L1$     & \multicolumn{2}{l|}{$1 \; \mathrm{m}$}                       \\ \hline
		$D_h$     & \multicolumn{2}{l|}{$0.02 \; \mathrm{m}$}                    \\ \hline
		$x_0=(L+L1)$     & \multicolumn{2}{l|}{$2 \; \mathrm{m}$}                       \\ \hline
		b         & \multicolumn{2}{l|}{14.23 (For laminar flow in square duct)} \\ \hline
		$t_0=\frac{x_0 D_h}{\nu_1}$     &      44792.83                         &      123839.01                        \\ \hline
		$Fo_i=\frac{\alpha_i t_0}{x_0^2}$      &      $1.59 \times 10^{-3}$                         &      $2.04$                        \\ \hline
		$Co_1=\frac{2bx_0}{D_h}$    &         2846                      &                2846              \\ \hline
	\end{tabular}
\end{table}

\clearpage

\clearpage 

\begin{figure}[!htb]
\begin{minipage}{0.49\linewidth}
\centering
		\includegraphics[width=1\linewidth]{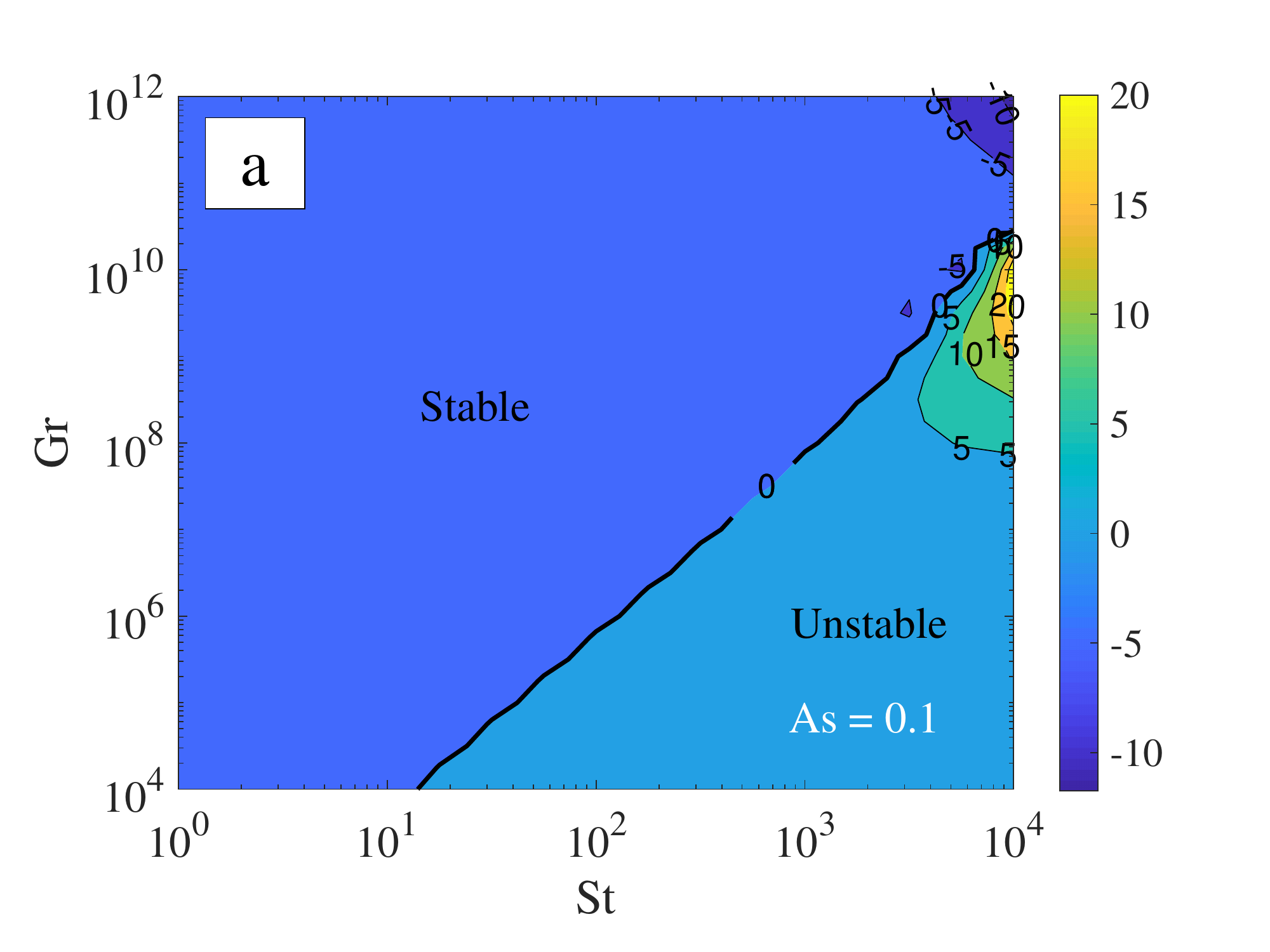}
		\includegraphics[width=1\linewidth]{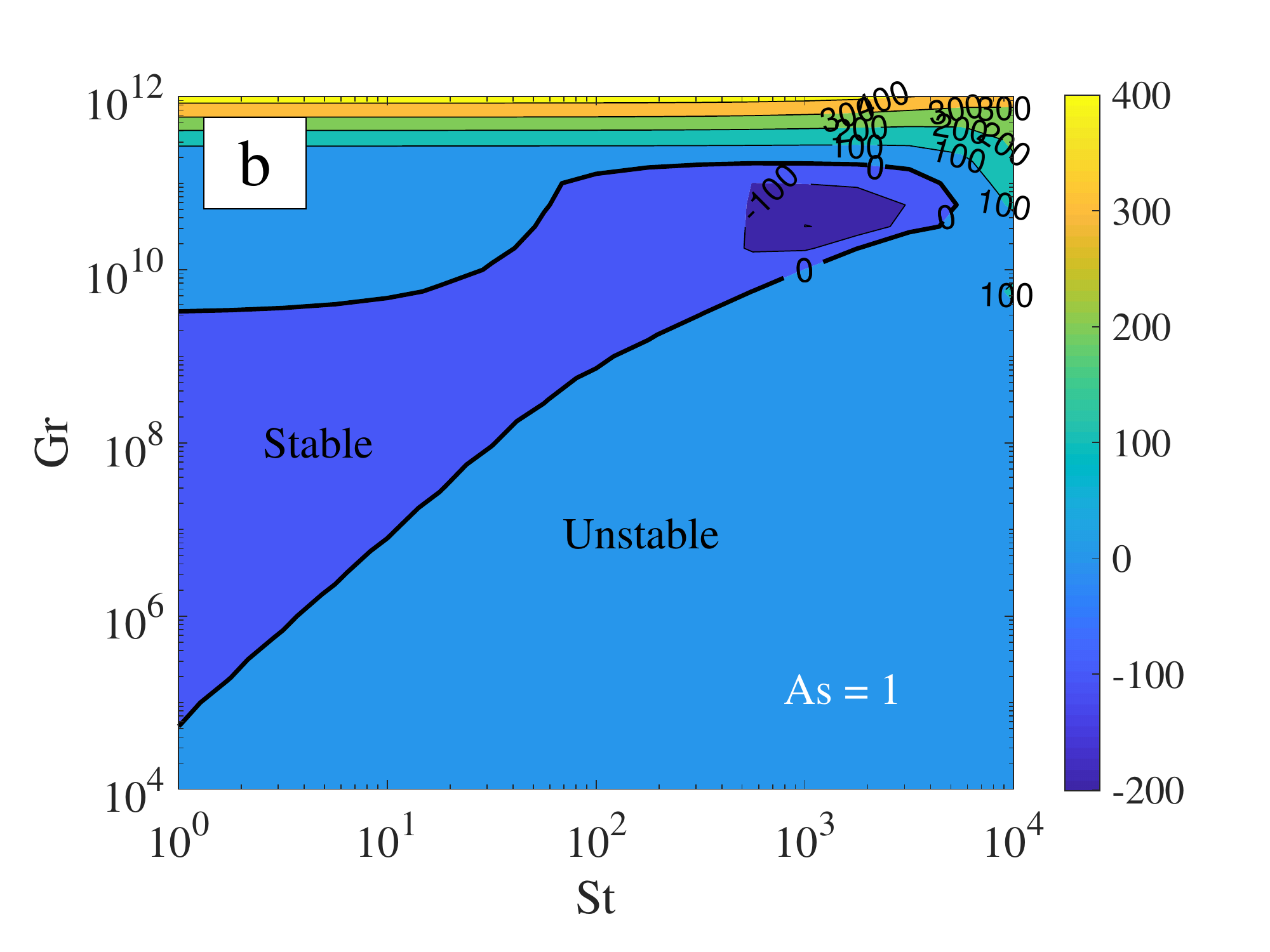}
		\includegraphics[width=1\linewidth]{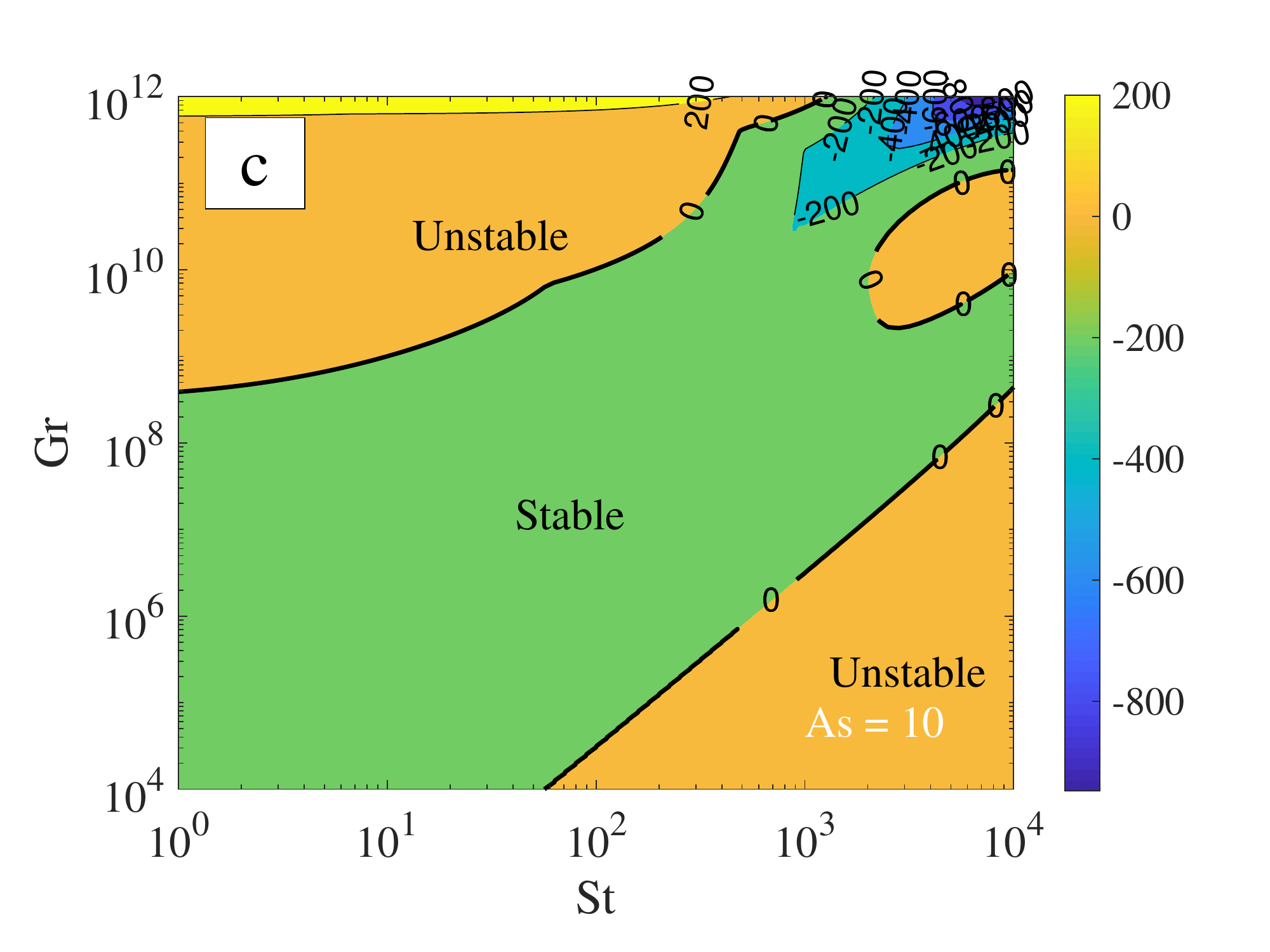}
	    \caption{Stability map of the VCNCL system corresponding to the steady state $(Re_1,Re_2)=(-Re_a,+Re_b)$ which lies on the line $Re_1=-Re_2$ with $Fo=0.001$, $Co_1= 2846$ for, (a) $As=0.1$, (b) $As=1$, and (c) $As=10$.}
	    \label{Fig25}
\end{minipage}
\hfill \vline \hfill
\begin{minipage}{0.49\linewidth}
\centering
		\includegraphics[width=1\linewidth]{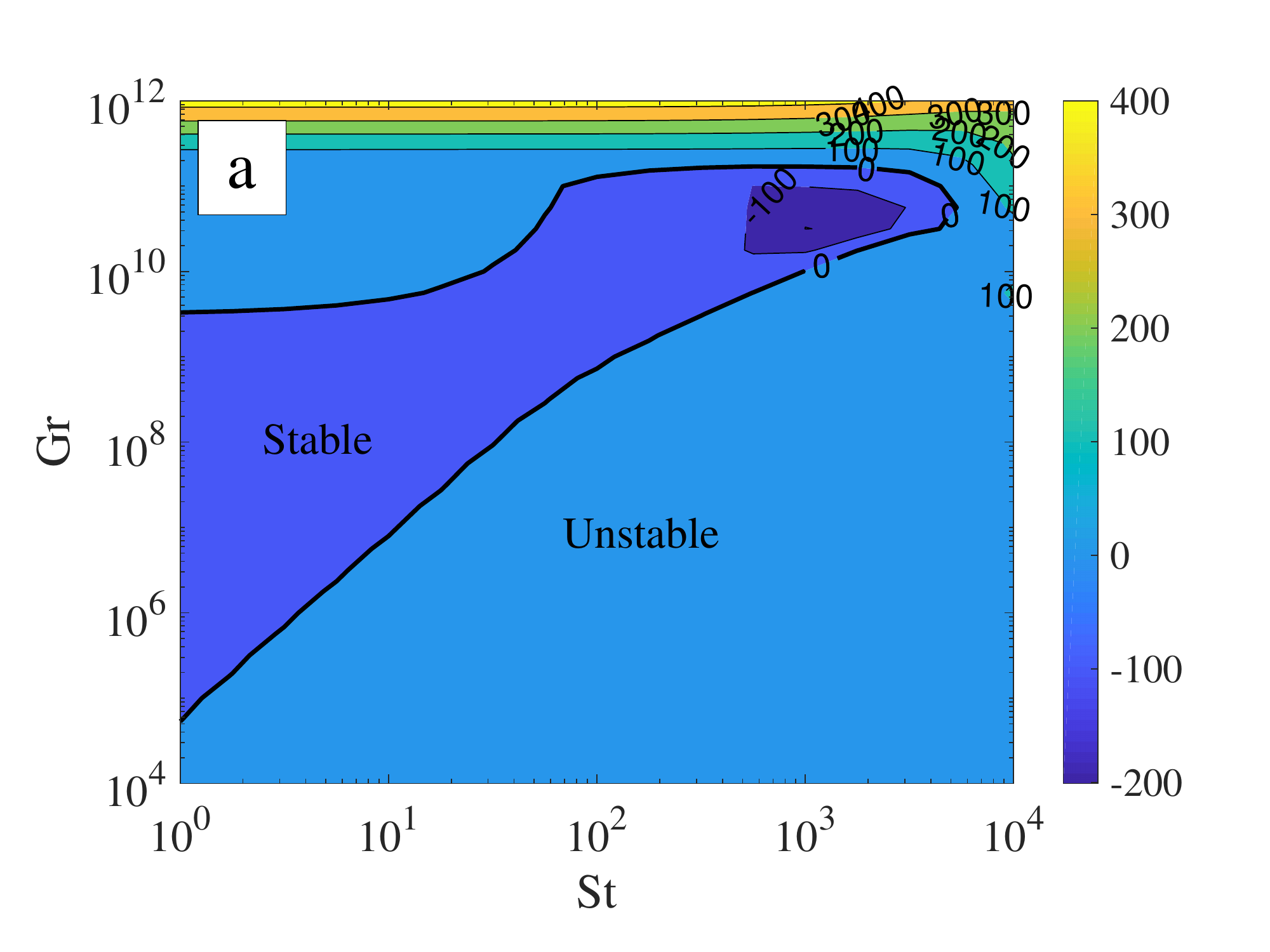}
		\includegraphics[width=1\linewidth]{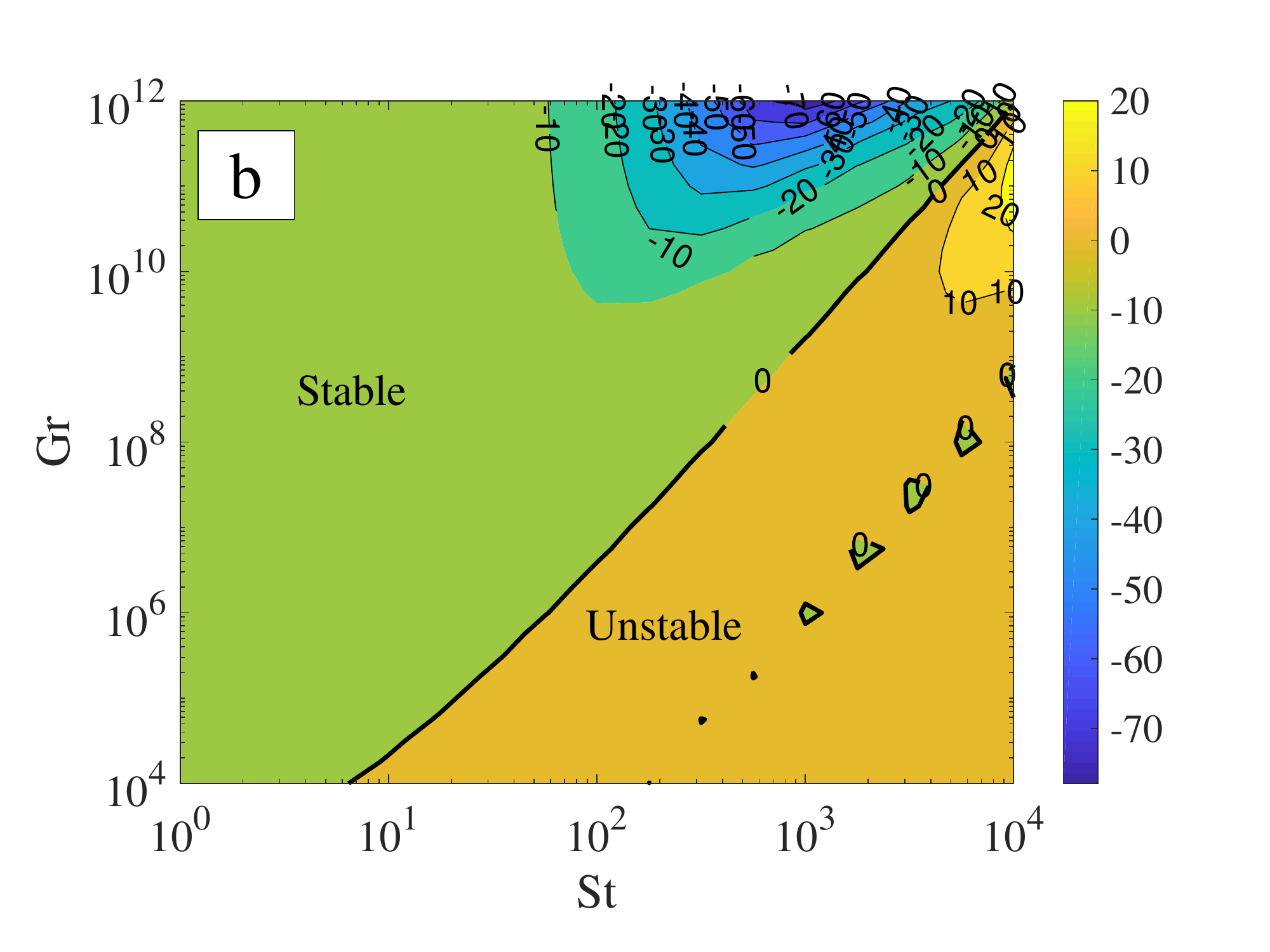}
	    \caption{Stability map of the VCNCL system corresponding to the steady state $(Re_1,Re_2)=(-Re_a,+Re_b)$ which lies on the line $Re_1=-Re_2$ for $As=1$, $Fo=0.001$, $Co_1=2846$ employing (a) Laminar friction factor and (b) Turbulent friction factor.}
	    \label{Fig26}
\end{minipage}
\end{figure}

\subsubsection{Effect of $As$ on the stability of VCNCL system}

The effect of an increase in the aspect ratio ($As$), of the CNCL system on the stability behaviour of VCNCL system is presented in figure \ref{Fig25}. It is observed that VCNCL system with $As=1$ has the smallest region of stable operation relative to the other considered cases ($As=0.1\; \mathrm{and} \; 10$). This observation is in agreement with the behaviour of an NCL system, as reported by Cammarata et al. \cite{cammarata2003stability}.

\subsection{Effect of flow regime on the stability of VCNCL system}

The shear stress experienced by the fluid is given by the expression:

\begin{equation}
    f_F=\frac{b}{Re^d}
\end{equation}

Here, the magnitudes of $b$ and $d$ decide the flow regime. The laminar friction factor is calculated using equation (47) for a square duct with $b=14.23$ and $d=1$. The turbulent friction factor is calculated using equation (47) with $b=0.316/4$ and $d=0.25$. From figure \ref{Fig26}, it can be noted that the stability map determined using the turbulent friction factor has a larger stable domain of operation relative to the laminar friction factor.

\subsection{Sharp point on the stability map}

A closer look at the stability boundary of the VCNCL system corresponding to the steady-state $(Re_1,Re_2)=(-Re_a,+Re_b)$ which lies on the line $Re_1=-Re_2$ for $As=1$, $Fo=0.001$, $Co_1=2846$ at the point denoted by $Gr \approx 1.758 \times 10^{11}$ and $St \approx 74$ in figure \ref{Fig27}(a) reveals two sharp points (SP) . These sharp points are visible not due to the lack of adequate resolution of the stability contour plot, but due to the change in the transient behaviour of the VCNCL system close to those points. 

The transient behaviour of the points denoted by the ellipses in figure \ref{Fig27}(a) are represented in figure \ref{Fig27}(b) and the transient behaviour of the points denoted by the rectangles in figure \ref{Fig27}(a) are represented in figure \ref{Fig27}(c). From figure \ref{Fig27}(b), it can be noted that with an increase in the considered $Gr$ from $10^9$ to $10^{12}$ for $St=50$, the transient behaviour of the CNCL system changes from converging to a particular steady-state to a diverging oscillatory flow. Furthermore, from figure \ref{Fig27}(c), it can be noted that with an increase in the considered $Gr$ from $10^9$ to $10^{12}$ for $St=350$, the transient behaviour of the CNCL system changes from chaotic to a flow which converges to a particular steady or quasi-steady-state and back to chaotic flow. The change in the transient behaviour as the system progresses from having a steady-state flow to a diverging transient flow for $St$ approximately less than 74, and from a steady-state flow to a chaotic flow for $St$ approximately greater than 74 results in the sharp points in the stability map.

From figures \ref{Fig27}(b) and \ref{Fig27}(c) it can be noted that the stability boundary shown in figure \ref{Fig27}(a), obtained from the linear stability analysis, accurately captures the transition from stable (flow which converges to a steady-state or a quasi-steady-state) to unstable (flow which has a diverging transient behaviour or a flow which exhibits chaotic oscillatory behaviour).

\begin{figure}[!htb]
	\centering
	\begin{subfigure}[b]{0.55\textwidth}
		\includegraphics[width=1\linewidth]{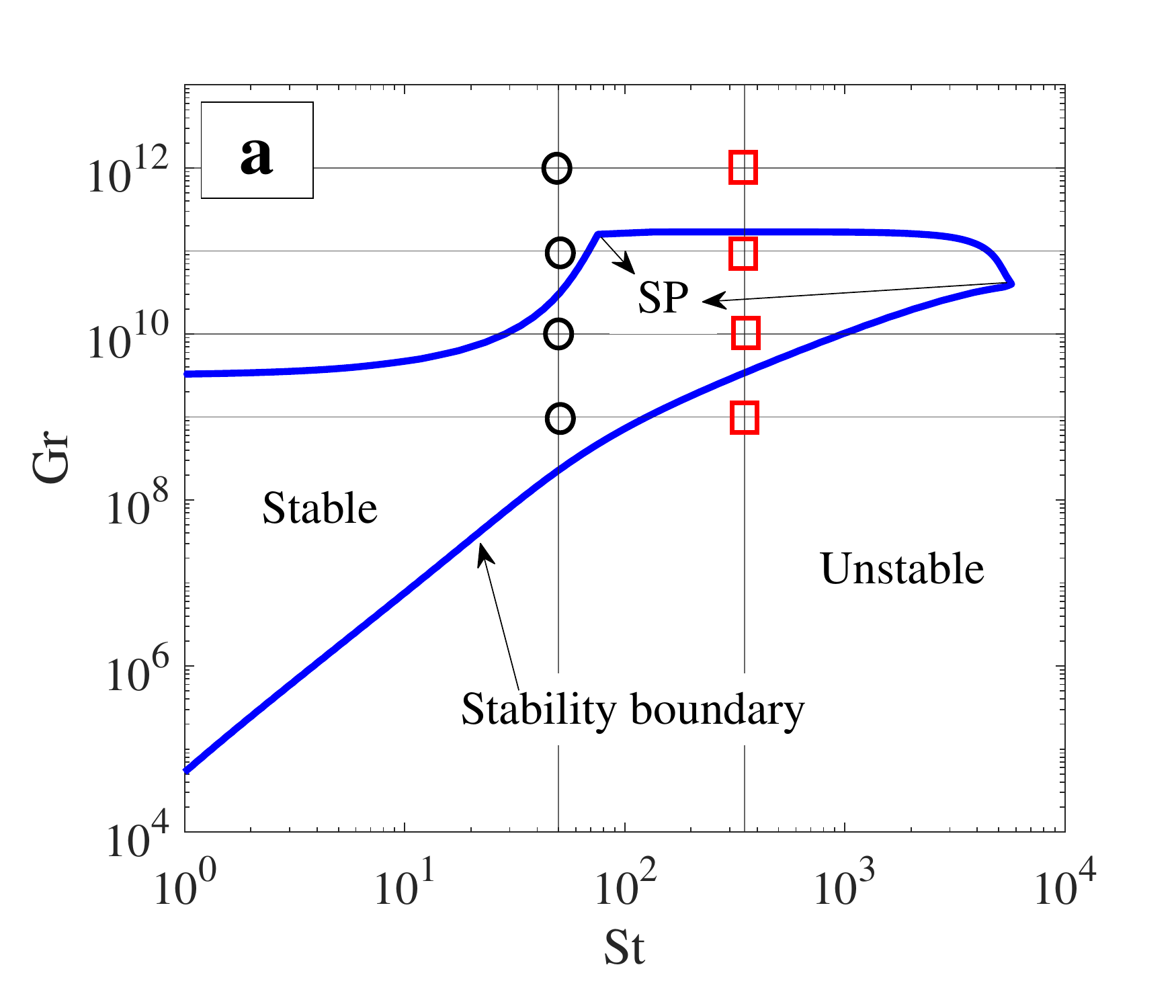}
	\end{subfigure}
	\end{figure}
	\clearpage
	\begin{figure}[!htb]
	\centering
	\ContinuedFloat
	\begin{subfigure}[b]{0.7\textwidth}
		\includegraphics[width=1\linewidth]{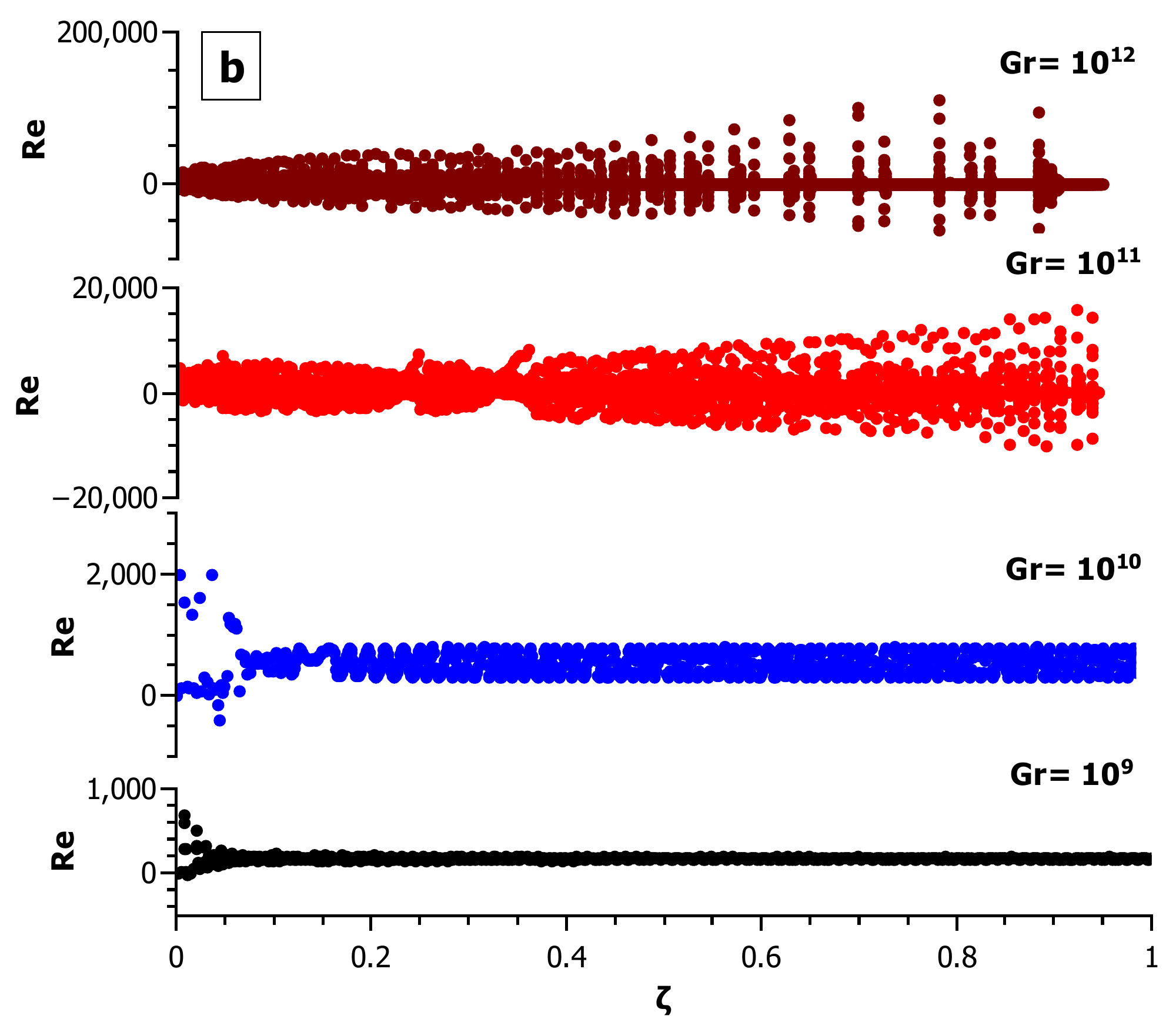}
	\end{subfigure}
	\hspace{\fill}
	\begin{subfigure}[b]{0.7\textwidth}
		\includegraphics[width=1\linewidth]{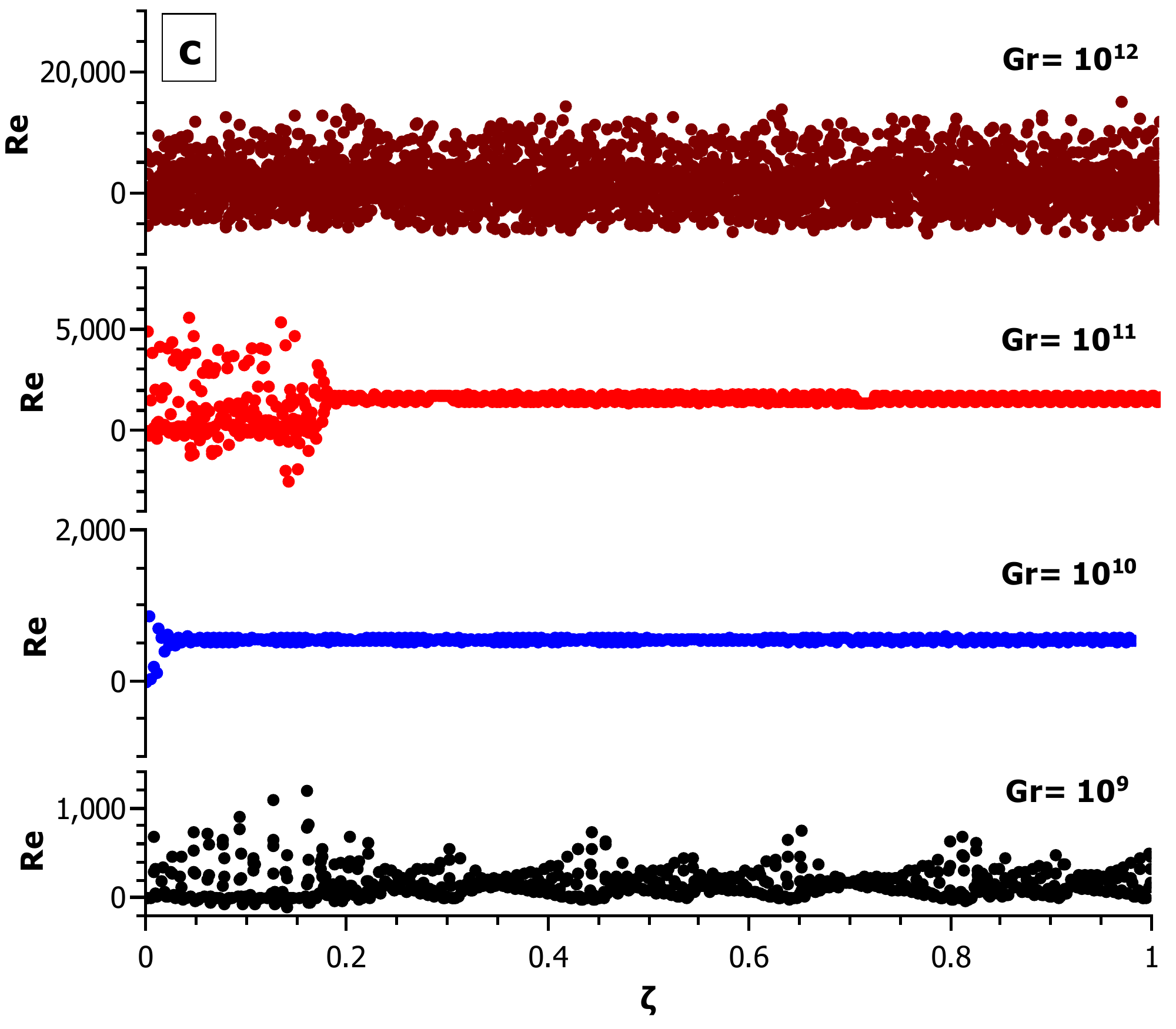}
	\end{subfigure}
	\caption{(a) Stability boundary of the VCNCL system corresponding to the steady state $(Re_1,Re_2)=(-Re_a,+Re_b)$ which lies on the line $Re_1=-Re_2$ for $As=1$, $Fo=0.001$, $Co_1=2846$ indicating the points of intersection of lines of constant $Gr=10^v$ with line of constant $St=50$ represented by circles and line of constant $St=350$ represented by squares, where $v={9,10,11,12}$, (b) Transient plot of $Re$ versus $\zeta$ for the points represented by circles in figure \ref{Fig27}(a), and (c) Transient plot of $Re$ versus $\zeta$ for the points represented by squares in figure \ref{Fig27}(a).} 
	\label{Fig27}
\end{figure}

\subsection{Assessment of the linearization approximation for stability prediction}

\begin{figure}[!htb]
    \centering
    \includegraphics[width=0.8\linewidth]{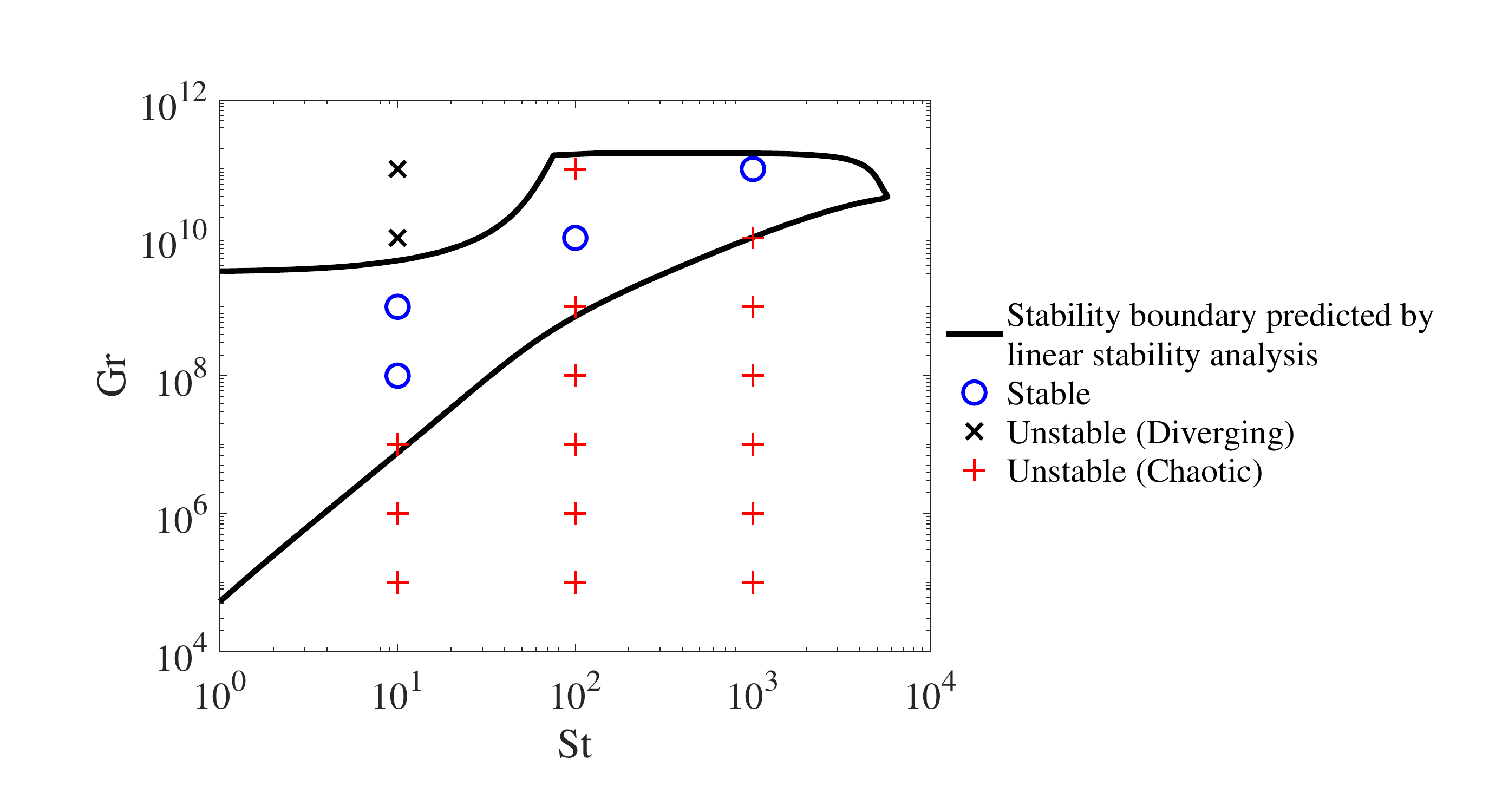}
    \caption{ Comparison of the linear stability map with the empirical stability map (represented by markers) for a VCNCL system corresponding to the steady state $(Re_1,Re_2)=(-Re_a,+Re_b)$ which lies on the line $Re_1=-Re_2$ for $As=1$, $Fo=0.001$, $Co_1=2846$.}
    \label{Fig28}
\end{figure}

The current paper employs the linear stability analysis to predict the stability boundary of the CNCL system. However, the linear stability analysis is just an approximation of the system, which is, in fact, non-linear. Therefore, to capture the non-linear effects in the stability map, the transient behaviour of the interior points of the $Gr-St$ stability map are observed and the system is classified as stable when it tends to a steady-state or exhibits a quasi-steady state and unstable if it exhibits chaotic or diverging transient behaviour. The transient behaviour of several interior points is noted and is utilised to construct the empirical stability map denoted by markers in figure \ref{Fig28}, which captures all the non-linear effects. From figure \ref{Fig28}, it is observed that the stable and unstable domains demarcated by the stability boundary predicted using linear stability analysis matches well the empirical stability map (non-linear stability map), indicating that the linear stability analysis approach is suitable to determine the stability boundary of the CNCL system with reasonable accuracy. Vijayan et al. \cite{vijayan2007steady} have observed an excellent agreement between the prediction of stability boundary employing linear and non-linear stability analysis for an NCL system. Cammi et al. \cite{cammi2017stability} also observed a good agreement with a slight deviation in the stability boundary predicted employing linear and non-linear stability analysis for the NCL system. Therefore, the assessment made in this section confirms that the linear stability analysis can be employed to obtain the stability maps of CNCL system with good accuracy.

\subsection{Prediction of NCL system stability from 1-D CNCL model}

The present section describes the modifications to be made to the 1-D CNCL model to predict NCL system stability with the following objectives:
\begin{enumerate}
	\item To demonstrate that the 1-D CNCL model is more general and hence practically relevant.
	\item To assess the influence of the thermal coupling at the common heat exchange section.
	\item To provide an additional and thorough validation of the ability of the 1-D CNCL model to accurately predict the stability boundary.
\end{enumerate}

In the preceding sections the stability of CNCL systems with identical fluids in both loops has been presented in detail. The present section demonstrates that the 1-D CNCL model is capable of predicting the stability of CNCL systems with different fluids in the two loops. To approximate the NCL system with horizontal heater with heat flux boundary condition and horizontal cooler with constant wall temperature boundary condition using the 1-D CNCL model, the 1-D CNCL model of the HCNCL system represented in figure \ref{Fig18} is considered and the specific heat and thermal diffusivity of the fluid in Loop 2 is set to a large value i.e. $C_{p,2} \to \infty$ and $\alpha_2 \to \infty$. This leads to the following changes in the non-dimensional numbers corresponding to Loop 2:

\begin{equation}
	Gr_2 \to 0 \;, St_2 \to 0\;, Co_2 \to \infty\;, Fo_2 \to \infty
\end{equation}

Employing the conditions listed in equation (48) results in the fluid of Loop 2 acting as an infinite heat sink of constant uniform temperature for Loop 1, thus approximating the desired NCL system. This demonstrates the versatility of the 1-D model which can be modified and  adapted and hence showcases its generality and relevance. 

Utilising the conditions listed in equation (48), the stability map of the NCL system with horizontal heater and cooler is determined. The stability boundary prediction of an NCL system with such a heater cooler arrangement was done by Vijayan et al. \cite{vijayan2007steady}. The following alterations apart from the conditions listed in equation (48) are made to the 1-D CNCL model with a square cross-section to match the system described by Vijayan et al. \cite{vijayan2007steady}:
\begin{enumerate}
	\item The laminar friction factor of a circular duct is used ($b=16$, $d=1$).
	\item The dimensions are modified to $L=1.14 \; \mathrm{m}$, $L1=2.1 \;\mathrm{m}$, $D_h=6\;\mathrm{mm}$ and reduced heater and cooler length of $0.8 \; \mathrm{m}$. The bend losses are neglected and $Fo_1=0$.
	\item The common heat exchange section is modified from a flat plate heat exchanger with square cross section to a tube of circular cross section with constant wall temperature boundary. This results in an increase in heat exchange area and hence and an increase in $St_1$. The magnitude of increase in $St_1$ is calculated from the ratio of heated area to cross-sectional area of the circular duct with constant wall temperature boundary condition to that of a square duct with flat plate heat exchange section. This results in a fourfold increase in the magnitude of $St_1$. Thus, by modifying the flat plate common heat exchange section with square cross-section to a tube of circular cross section with constant wall temperature boundary condition, the $St_1$ for the tube is defined as: $St_1=(4Ut_0)/(\rho_1 Cp_1 D_h)$.
\end{enumerate}

\begin{figure}[!htb]
	\centering
	\begin{subfigure}[b]{0.49\textwidth}
		\includegraphics[width=1\linewidth]{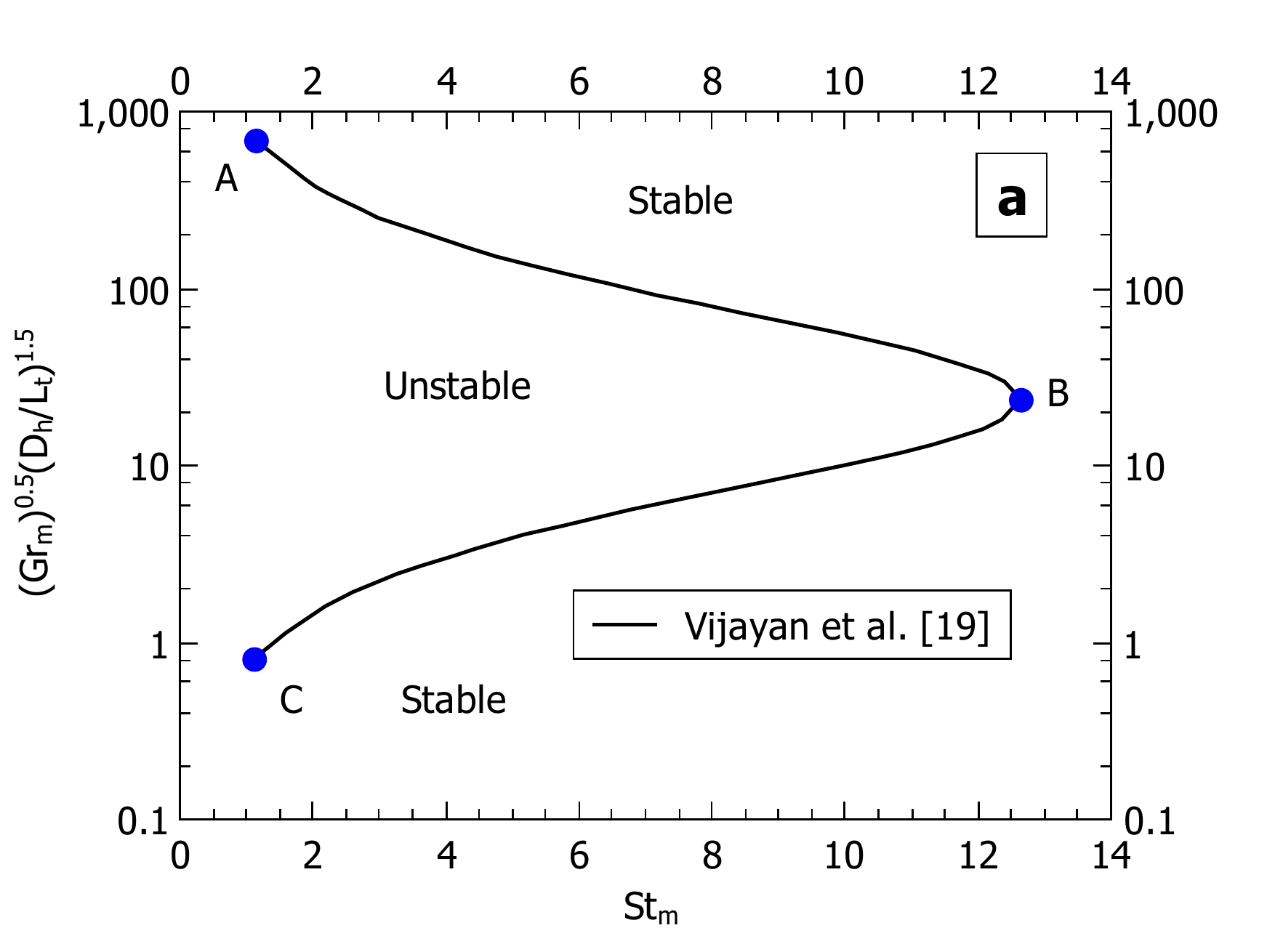}
	\end{subfigure}
	\hspace{\fill}
	\begin{subfigure}[b]{0.49\textwidth}
		\includegraphics[width=1\linewidth]{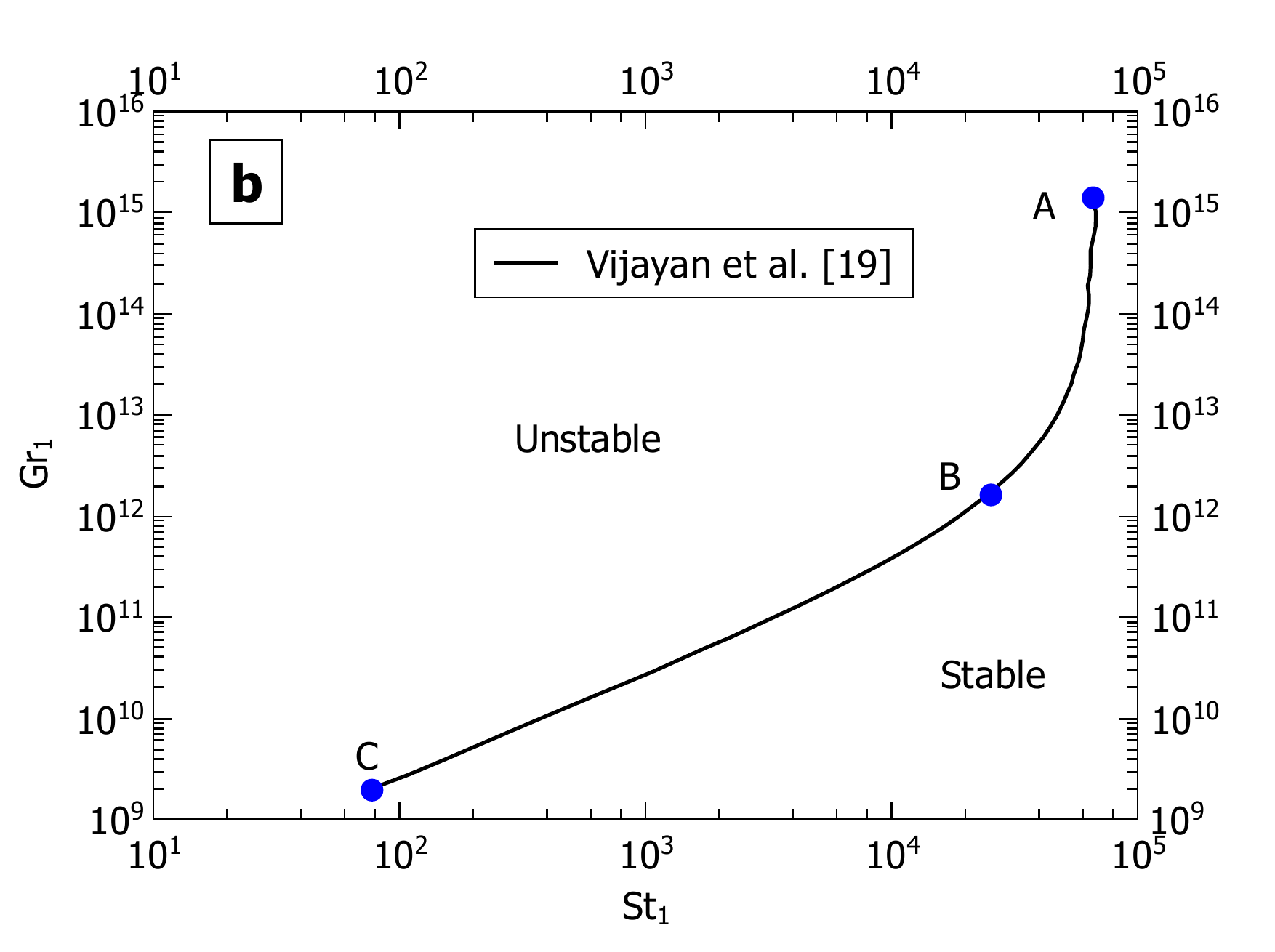}
	\end{subfigure}
	\caption{Mapping of stability boundary predicted by Vijayan et al. \cite{vijayan2007steady} with the laminar friction factor from the $((Gr_m)^{0.5}(D_h/L_t)^{1.5},St_m)$ domain to the $(Gr_1,St_1)$ domain.}
	\label{Fig29}
\end{figure}

\begin{figure}[!htb]
	\centering
	\begin{subfigure}[b]{0.49\textwidth}
		\includegraphics[width=1\linewidth]{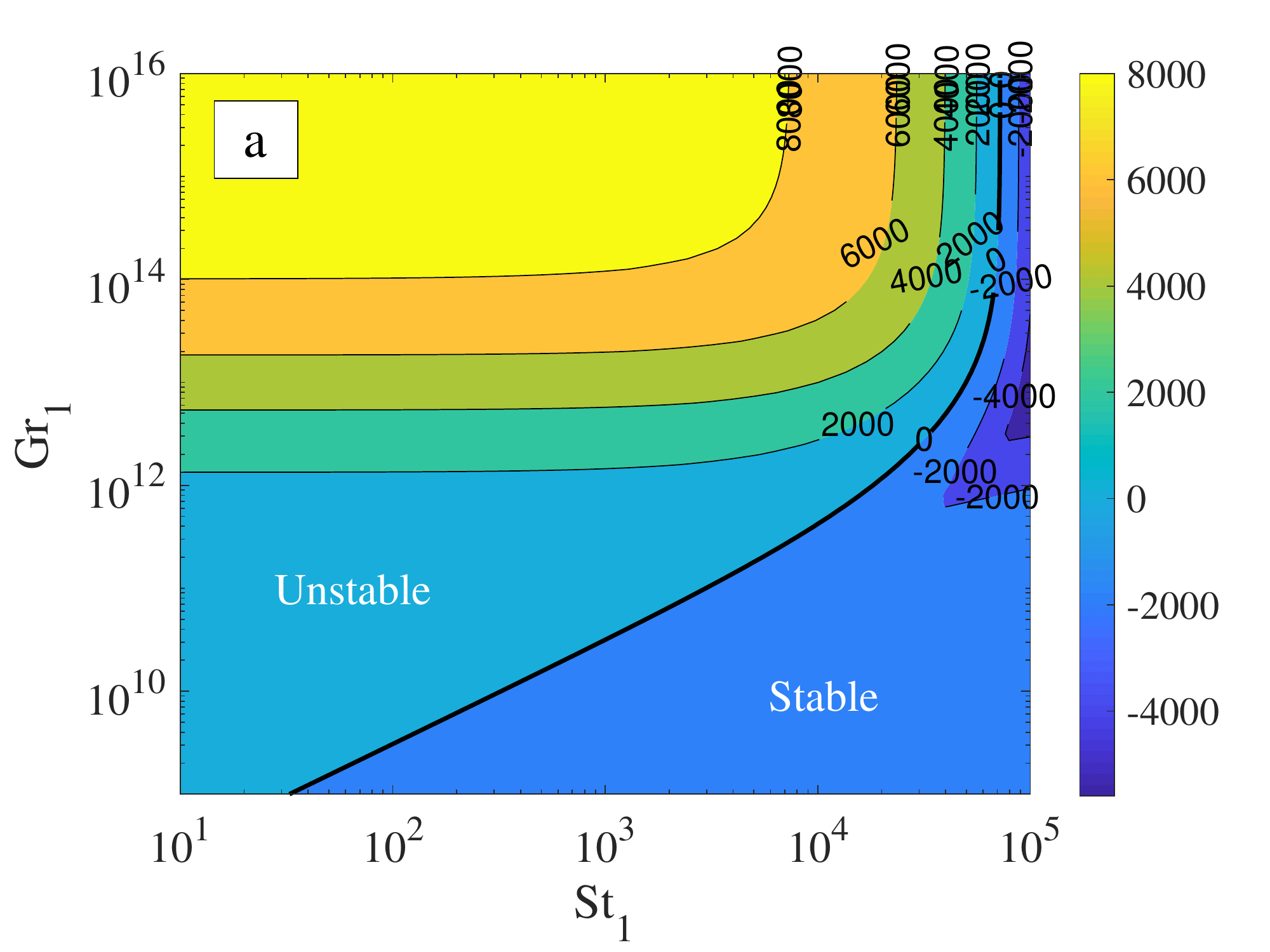}
	\end{subfigure}
	\hspace{\fill}
	\begin{subfigure}[b]{0.49\textwidth}
		\includegraphics[width=1\linewidth]{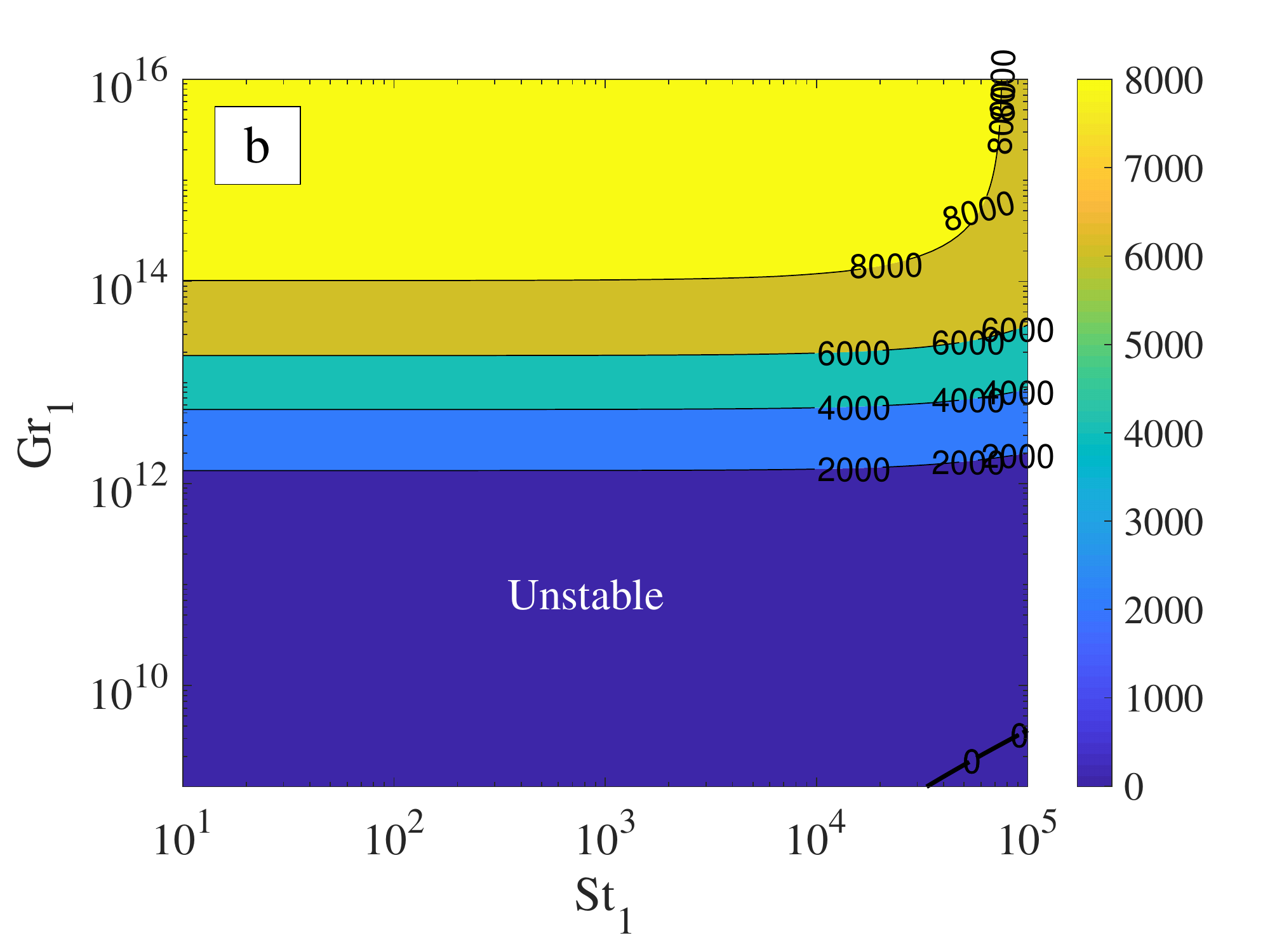}
	\end{subfigure}
	\hspace{\fill}
	\begin{subfigure}[b]{0.49\textwidth}
	\includegraphics[width=1\linewidth]{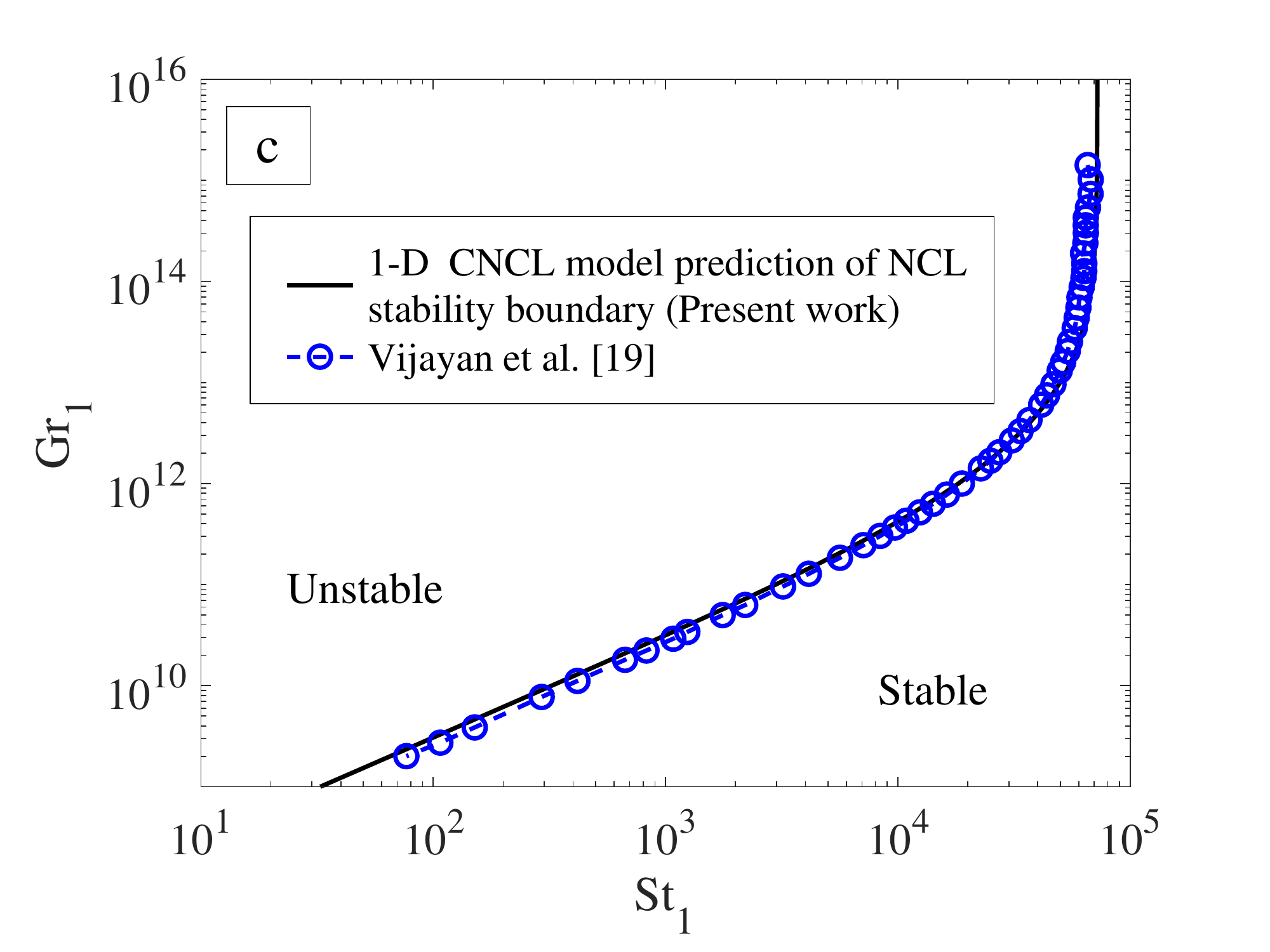}
	\end{subfigure}
	\caption{Assessment of NCL system stability, (a) stability map prediction of the NCL system with 1-D CNCL model, (b) influence of thermal coupling at the common heat exchange section on the CNCL system stability, and (c) verification of the stability boundary predicted by the 1-D CNCL model with the existing literature. }
	\label{Fig30}
\end{figure}

Figure \ref{Fig29} shows the mapping of stability boundary from the $((Gr_m)^{0.5}(D_h/L_t)^{1.5},St_m)$ domain to the $(Gr_1,St_1)$ domain. The points A, B and C are used to represent equivalent points in each of the domains. It is observed that the mapping changes the shape of the stability boundary. Figure \ref{Fig30}(a) shows the stability map predicted with the aforementioned modifications to model the NCL system with 1-D CNCL model. An excellent agreement is observed in the stability boundary of the NCL predicted by the 1-D CNCL model with the results from Vijayan et al. \cite{vijayan2007steady} as shown in figure \ref{Fig30}(c), indicating the accuracy of the 1-D CNCL model. 

Comparing the stability map of the HCNCL with $Gr_2=Gr_1 ;\; St_2=St_1 ;\; Co_2=1 \; \mathrm{and}\; Fo_2=Fo_1$ shown in figure \ref{Fig30}(b), with the stability map of the NCL (HCNCL with the conditions described by equation (48)) as shown in figure \ref{Fig30}(a), the influence of thermal coupling at the heat exchange section can be determined. The common heat exchange section of the CNCL system demonstrates the thermal coupling of fluids which are propelled by buoyancy forces, i.e., it is a two way thermal coupling (any change in behaviour in the first loop influences the second loop and vice versa) and employing conditions in equation (48) reduces it to a one way coupling (any change introduced in Loop 1 has no influence on Loop 2). Dass and Gedupudi \cite{dass2019} have reported that if $Co_2 >>1$ or $Co_2 <<1$, the loops of the CNCL system are effectively decoupled. This confirms the fact that the conditions listed in equation (48) decoupled the CNCL and approximatated the CNCL to an NCL with constant wall temperature boundary condition. Comparing figures \ref{Fig30}(a) (NCL system stability map) and \ref{Fig30}(b) ( HCNCL system for counterflow arrangement at the common heat exchange section), it is noted that the two way thermal coupling present in the HCNCL system leads to a decrease in the domain of stable operation. Hence, as the practical systems such as PRHRS and LMFBR are better understood via the CNCL system, the stability map of the CNCL system provides a much better estimate of the stability of such systems.

\section{Conclusions}
The linear stability analysis of the CNCL system employing Fourier series based 1-D model has been carried out. A 3-D CFD study of the CNCL systems has also been performed and it shows that the CNCL system is a dynamical system, i.e., it exhibits chaotic oscillatory behaviour under specific conditions. The 3-D CFD study has also been employed to evaluate the reliability of the 1-D CNCL model to predict observed non-periodic oscillatory behaviour. The 1-D model and the 3-D CFD methodology have also been validated with the literature data. The important conclusions which can be drawn from the present study for a CNCL system are:
\begin{enumerate}
    \item The CNCL system is a dynamical system and exhibits chaotic behaviour under specific conditions. The Fourier series based 1-D model of the CNCL is capable of predicting the conditions for which the system displays chaotic behaviour as observed from the good agreement with the 3-D CFD predictions.
    \item The oscillatory behaviour of the CNCL system is a direct result of generations of hot and cold fluid packets within the system. The CNCL system velocity at the peak or valley of the transient oscillatory behaviour is a result of the location of the hot or cold fluid packets in the vertical or horizontal limbs of the CNCL, respectively. 
    \item The CNCL system has multiple steady states, and these steady states can be represented entirely or calculated from the steady-state magnitudes of $Re_1$ and $Re_2$. For the CNCL system considered in the present study, it is noted that symmetric steady-state solutions in the $Re_1Re_2$ plane along the line $Re_1=-Re_2$ are consistently observed, and thus the stability maps of these steady-states are presented in the current work.
    \item The stability map of the CNCL system for a particular steady-state identifies the stable (negative eigenvalue) and unstable (positive eigenvalue) and the stability boundary ( null eigenvalue) which demarcates the stable and unstable domains. The CNCL system is classified as stable if it exhibits stable or quasi-steady (periodic oscillatory behaviour with small amplitudes) convective flow and the system is classified as unstable it exhibits divergent or chaotic transience. 
    \item The stability boundary of the CNCL system for a considered set of non-dimensional parameters is constructed using the contour map of the eigenvalues and, thus is dependent on the resolution (no of points used to construct the contour map) of the study and also on the number of Fourier nodes to which the CNCL system is truncated. From the present study, it is identified that a CNCL system truncated to 5 Fourier nodes is adequate to characterise the dynamic behaviour of the CNCL system completely.
    \item The HCNCL system with the heater and cooler on the horizontal limbs is a configuration which displays both counterflow and parallel flow configuration at steady-state.   The counterflow arrangement at the common heat exchange section of the HCNCL has a more significant stable domain of operation relative to the parallel flow arrangement. The parallel flow configuration at the common heat exchange section is observed only for $Fo>0$.
    \item The counterflow arrangement at the common heat exchange section of the VCNCL and HCNCL systems has signficantly different stability maps indicating the strong influence the orientation and the heater cooler arrangement have on the system stability.
    \item The increase in $Fo$ and $Co_1$ result in an increase of stable domain in the $Gr-St$ stability map, whereas the domain of stability decreases from $As=0.1$ to $As=1$ and increases from $As=1$ to $As=10$ for the steady-state with symmetric counterflow configuration at the common heat exchange section of the VCNCL system.
    \item The linear stability map is in good agreement with the empirical stability map, which represents the actual non-linear behaviour of the CNCL system. Therefore, the usage of linear stability analysis to obtain accurate stability maps of the CNCL system is justified.
    \item The 1-D CNCL model is capable of predicting the stability maps of CNCL systems with two different fluids in the two component loops and is thus a more general and practically relevant system. The NCL system stability predicted after employing suitable modifications is in good agreement with the available literature indicating the accuracy of the 1-D model. The influence of the additional thermal coupling introduced in the CNCL system relative to the NCL system leads to a decrease in the domain of stability.
    \item The influence of wall effects on the system stability is not considered in the present work and will be a part of the future study.
\end{enumerate}

\section*{\hfil \Large Nomenclature \hfil}

\begin{tabular}{ll}
	$C_{p}$ & Specific heat ($\mathrm{J/kg\;K}$)\\
	$D_h$ & Hydraulic diameter of both Loop 1 \&\ 2 ($\mathrm{m}$)\\
	$F_i$ & Implicit form of non-dimensional momentum equation of Loop $i$ as functions of $Re_1$ and $Re_2$\\
	$g$ & Acceleration due to gravity ($g=\mathrm{9.81\;m/s^2}$)\\
	$L$ & CNCL height used for 1-D model ($\mathrm{m}$)\\
	$L1$ &  CNCL width used for 1-D model ($\mathrm{m}$)\\
	$L_t$ &  Total loop length of the constituent NCL of the CNCL system ($\mathrm{m}$)\\
	$n$ & Number of bends on the component NCL of the CNCL system\\
	$N$ & Largest Fourier node considered in the expanded ODE stencil\\
	$Q^{\prime\prime}$ & Heat flux ($\mathrm{W/m^2}$)\\
	$R_d$ & Radius of curvature ($\mathrm{m}$)\\
	$T_{1}$ & Temperature of Loop 1 ($\mathrm{K}$)\\
	$T_{2}$ & Temperature of Loop 2 ($\mathrm{K}$)\\
	$t$ &  Time ($\mathrm{s}$)\\
	$T_{0}$ & Reference temperature of Loop 1 \&\ 2 ($\mathrm{K}$)\\
	$U$ & Overall heat transfer coefficient at the heat exchanger section ($\mathrm{W/(m^2K)}$)\\
\end{tabular}

 \section*{Greek letters}

\begin{tabular}{ll}
    $\alpha$ & Thermal diffusivity ($\mathrm{m^2/s}$)\\
	$\beta$ & Coefficient of thermal expansion ($\mathrm{1/K}$)\\
	$\kappa$ & Thermal conductivity ($\mathrm{W/(mK)}$)\\
	$\mu$ & Dynamic viscosity ($\mathrm{Pas}$)\\
	$\rho$ & Density ($\mathrm{kg/m^3}$)\\
	$\tau$ & Wall shear stress exerted on fluid ($\mathrm{Pa}$)\\
	$\omega_{1}$ &  Fluid velocity of Loop 1 ($\mathrm{m/s}$)\\
	$\omega_{2}$ &  Fluid velocity of Loop 2 ($\mathrm{m/s}$)\\
 \end{tabular}

\section*{Constants}

\begin{tabular}{ll}
	$b$ & 14.23 (for fully developed flow in laminar regime for a square duct)\\
	$d$ & 1 (for fully developed flow in laminar regime for a square duct) \\
	$\Delta T$ & ( $\Delta T_i=(4Q^{\prime\prime}t_0)/(\rho_i Cp_i D_h) $) \\
	$to$ & ( $t_0={x_0D_h}/{\nu_1} $) \\
	$x_0$ & ( $x_0=(L+L1)$) \\
\end{tabular}

\section*{Piece-wise functions}
\begin{tabular}{ll}
	
	$f(x)$ & Function which represents the geometry of the loop\\
	$h_1 (x)$ & Function which represents the heating section location\\
	$h_2 (x)$ & Function which represents the cooling section location \\
	$\lambda (x)$ & Function which represents the location of thermal coupling on the CNCL \\
\end{tabular}

\section*{Fourier coefficients}
\begin{tabular}{ll}
	
	$\theta_{1,k}$ & $k^{th}$ Fourier node of $\theta_1(s,\zeta)$\\
	$\theta_{2,k}$ & $k^{th}$ Fourier node of $\theta_2(s,\zeta)$\\
	$h_{1,k}$ & $k^{th}$ Fourier node of $h_1(s)$\\
	$h_{2,k}$ & $k^{th}$ Fourier node of $h_2(s)$\\
	$\lambda_{k}$ & $k^{th}$ Fourier node of $\lambda (s)$\\
	$f_{k}$ & $k^{th}$ Fourier node of $f(s)$\\
\end{tabular}

\section*{Matrix}
\begin{tabular}{ll}
	
	$A$ & Single column matrix denoting the variables on the L.H.S of the truncated ODE stencil\\
	$A^{\prime}$ & Single column matrix denoting the variables on the L.H.S of the expanded ODE stencil \\
	$B$ & Single column matrix denoting the expressions on the R.H.S of the truncated ODE stencil \\
	$B^{\prime}$ & Single column matrix denoting the expressions on the R.H.S of the expanded ODE stencil \\
	$B^{\prime}_J$ & Jacobian of matrix $B^{\prime}$ \\
\end{tabular}

\section*{Non-dimensional numbers}
\begin{tabular}{ll}
    $f_{F}$ & Fanning friction factor ($f_{F}=b/Re^d$)\\
    $Gr_m$ & Modified Grashof number (Vijayan et al. \cite{vijayan2002experimental}) \\
    $K$ & 	Bend losses coefficient  \\
    $N_g$ & Geometric parameter (Vijayan et al. \cite{vijayan2002experimental}) \\
    $r$ & Ratio of Rayleigh number to critical Rayleigh number (Lorenz \cite{lorenz1963deterministic})\\
    $r_c$ & Critical value of $r$ for steady convection (Lorenz \cite{lorenz1963deterministic})\\
    $\sigma$ & Prandtl number (Lorenz \cite{lorenz1963deterministic})\\
\end{tabular}
    
\section*{Non-dimensional parameters}

\begin{tabular}{ll}
    $As$ & Aspect ratio ($As=L/L1$) \\
	$Co_1$ & Flow resistance coefficient  ($Co_1=\frac{2bx_0}{D_h}$)\\
	$Co_2$ & Thermal coupling sensitivity coefficient  ($Co_2=\frac{\Delta T_1}{\Delta T_2}$)\\
	$Fo$ & Fourier number ($Fo_i=\frac{\alpha_it_{0}}{x_0^2}$)\\
	$Gr$ & Grashof number ($Gr_i=\frac{g \beta_i \Delta T_i x_0 D_h t_{0}}{(L+L1) \nu_i}$)\\
	$Re$ & Reynolds number ($Re_i=\frac{\omega_iD_h}{\nu_i}$)\\
	$St$ & Stanton number ($St_i=\frac{Ut_{0}}{\rho_i Cp_i D_h}$)\\
	$s$ & Non-dimensional length ( $s={x}/{x_0}$)\\
	$\theta$ & Non-dimensional temperature ($\theta_i={T_i-T_0}/{\Delta T_i}$)\\
	$\zeta$ & Non-dimensional time ($\zeta={t}/{t_0}$)\\
	
\end{tabular}

\section*{Subscripts}

\begin{tabular}{ll}
	$0$ & Any parameter at time $t=0$ $\mathrm{s}$ \\
	$1$ & Any parameter referring to Loop 1 \\
	$2$ &  Any parameter referring to Loop 2 \\
	$Avg$ & Average value of the parameter\\
	$k$ & $k^{th}$ Fourier node  ($-\infty \leq k \leq \infty$)\\
	$l$ & $l^{th}$ Fourier node  ($-k \leq l \leq k$)\\
	$i$ & Refers to subscript `1' or subscript `2' according to relevance\\
	$I$ & Imaginary part of the complex number\\
	$R$ & Real part of the complex number\\
	$ss$ & Steady state \\
\end{tabular}

\section*{Abbreviations}

\begin{tabular}{ll}
	$CFD$ &  Computational Fluid Mechanics \\
	$CNCL$ & Coupled Natural Circulation Loop \\
	$FDM$ & Finite Difference Methods \\
	$FFT$ & Fast Fourier Transform \\
	$HCNCL$ & Horizontal Coupled Natural Circulation Loop \\
	$LMFBR$ & Liquid Metal Fast Breeder Reactor \\
	$NCL$ & Natural Circulation Loop \\
	$ODE$ & Ordinary Differential Equation \\
	$PDE$ & Partial Differential Equation\\
	$PRHRS$ &  Passive Residual Heat Removal system\\
	$VCNCL$ & Vertical Coupled Natural Circulation Loop \\
\end{tabular}





\bibliographystyle{model1-num-names}
\bibliography{sample}







\end{document}